\documentclass[aps, prb, superscriptaddress]{revtex4-2}
\usepackage{bm, amsmath, amsfonts, amssymb, mathtools, braket}
\usepackage{multirow}
\usepackage{graphicx}
\usepackage{float, color, xcolor}
\usepackage{physics}
\usepackage{subfig}

\usepackage{bbm}
\usepackage{tabularx}

\usepackage[
pagebackref=false,
colorlinks=true,
linkcolor=blue,
urlcolor=blue,
filecolor=black,
citecolor=red,
pdfstartview=FitV,
pdftitle={},
pdfauthor={},
pdfsubject={},
pdfkeywords={},
pdfpagemode=None,
bookmarksopen=true
]{hyperref}

\newcommand{\kk}[1]{#1}

\begin{document}

\title{Non-linear sigma models for non-Hermitian
random matrices in symmetry classes AI$^{\dagger}$ and AII$^{\dagger}$}

\author{Anish Kulkarni}
\email{anishk@princeton.edu}
\affiliation{Department of Physics, Princeton University, Princeton, New Jersey 08544, USA}

\author{Kohei Kawabata}
\affiliation{Institute for Solid State Physics, University of Tokyo, Kashiwa, Chiba 277-8581, Japan}

\author{Shinsei Ryu}
\affiliation{Department of Physics, Princeton University, Princeton, New Jersey 08544, USA}

\date{\today}

\begin{abstract}
Symmetry of non-Hermitian matrices
underpins many physical phenomena. 
In particular, chaotic open quantum systems exhibit
universal bulk spectral correlations classified on the basis of time-reversal
symmetry$^{\dagger}$
(TRS$^{\dagger}$), coinciding with those of non-Hermitian random matrices in the same symmetry class. 
Here, we analytically study the spectral
correlations of non-Hermitian random matrices in the presence of 
TRS$^{\dagger}$ with signs 
$+1$ and $-1$, corresponding to symmetry classes AI$^{\dagger}$
and AII$^{\dagger}$, respectively. 
Using the fermionic replica non-linear sigma model approach, we derive $n$-fold integral expressions for
the $n$th moment of the one-point and two-point characteristic polynomials, valid 
for 
any matrix dimension.
We also study,
in the limit of 
large matrix dimensions, 
the replica limit $n\to 0$
to derive 
the density of states and level-level correlations of non-Hermitian random
matrices with TRS$^{\dagger}$.
We compare our analytical findings with numerical results.
\end{abstract}

\maketitle

\tableofcontents

\section{Introduction}

Random matrix theory (RMT), originally developed by Wigner to describe the energy levels of heavy atomic nuclei~\cite{Wigner-51, Wigner-58}, has since evolved into a versatile tool for understanding complex systems~\cite{Akemann:2011csh}. 
Its applications span numerous subfields of physics, including quantum chaos, Anderson localization, quantum chromodynamics (QCD), statistical mechanics, quantum information, and quantum gravity. 
Beyond physics, RMT has found widespread use in disciplines such as number theory, biological systems, and data science, among others.
At its core, RMT studies the statistical properties of matrices with randomly chosen elements, providing insight into the behavior of large, interacting systems. 
In many cases, the eigenvalue distributions of these random matrices exhibit universal properties, meaning that they are largely independent of the specific details of the matrix ensemble. 
This universality is one of the most compelling features of RMT, making it applicable to diverse problems.

Non-Hermitian RMT extends the powerful framework of the traditional RMT to the study of matrices that lack Hermiticity~\cite{Byun-Forrester-book}.
While Hermitian matrices are central in quantum mechanics, non-Hermitian matrices emerge naturally in a wide array of physical systems where dissipation, gain, 
or non-conservative interactions play a role~\cite{Konotop-review, Christodoulides-review}. 
These include open quantum systems, non-equilibrium statistical mechanics, disordered materials, and biological and sociological systems.
The study of non-Hermitian random matrices began in earnest with the work of Ginibre~\cite{Ginibre-65}, who introduced ensembles of non-Hermitian random matrices. 
Unlike Hermitian matrices, eigenvalues of non-Hermitian matrices generally spread over the complex plane, leading to rich and intricate spectral structures. 
One of the most remarkable features of non-Hermitian RMT is the emergence of complex spectral distributions, such as Girko’s circular law~\cite{Girko-85}, which describes the eigenvalue distribution for large non-Hermitian matrices with independent, identically distributed entries.
Other classic works on non-Hermitian RMT are found, for example, in Refs.~\cite{Verbaarschot:1985jn, Sokolov:1988ata, Sokolov:1988df, Haake-92, Stephanov_1996, Fyodorov_1997, Kogut:2000ek}.
The relevance of non-Hermitian RMT has grown considerably in recent decades, particularly with the recognition of non-Hermitian physics in various fields.
For example, beyond the Hermitian regime, the physics of Anderson localization in non-Hermitian systems has attracted growing interest~\cite{Hatano-Nelson-96, Efetov-97, Feinberg-97, Efetov-97B, Hatano-Nelson-97, Brouwer-97, Feinberg-99, Longhi-19, 
Zeng-20, Tzortzakakis-20, Huang-20, KR-21, Claes-21, Luo-21L, Luo-21B, Luo-22R, Liu-Fulga-21}.
Additionally, topological phenomena intrinsic to non-Hermitian systems have been extensively explored, offering new insights into the behavior of these systems~\cite{BBK-review, Okuma-Sato-review}.

One of the key impacts of non-Hermiticity is the enrichment of symmetry classification, expanding 
the traditional 10-fold classification~\cite{AZ-97, Evers-review, CTSR-review, Haake-textbook} to a more complex 38-fold scheme~\cite{Bernard-LeClair-02, KSUS-19}.
This expanded classification is crucial not only for non-Hermitian RMT but also for understanding chaotic behavior in open quantum systems~\cite{Grobe-88, Grobe-89, Xu-19, Hamazaki-19, Denisov-19, Can-19PRL, Can-19JPhysA, Hamazaki-20, Akemann-19, Sa-20, Wang-20, Xu-21, GarciaGarcia-22PRL, JiachenLi-21, Cornelius-22, GarciaGarcia-22PRX, Prasad-22, Sa-22-SYK, Kulkarni-22-SYK, GarciaGarcia-22PRD, GJ-23, Xiao-22, Shivam-22, Ghosh-22, Sa-23, Kawabata-23, GJ-24-ETH, Kawabata-23SVD, Xiao-24}.
Similar to their Hermitian counterparts,
spectral properties of 
non-Hermitian random matrices encode information about the dynamics of the associated physical system,
many of which are expected to be universal. 
They are insensitive to the details of the ensemble distribution and classified based on the symmetry of the matrices in the ensemble. 

\begin{figure}
    \centering
    \subfloat{\includegraphics[height = 0.27 \textwidth]{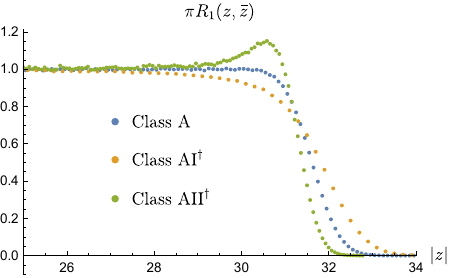}}
    \hspace*{30 pt}
    \subfloat{\includegraphics[height = 0.27 \textwidth]{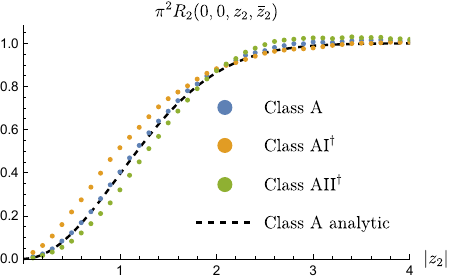}}
    \caption{
    \label{fig:R1R2-all}
    Numerical calculations for one-point (left) and two-point (right) functions for the three universality classes. We sample matrices of size $10^3\times 10^3$ for classes A and AI$^\dagger$ and size $\left( 2 \times 10^3 \right) \times \left( 2 \times 10^3 \right)$ for class AII$^\dagger$. To numerically obtain $R_1$, we simply compute the histogram of the sampled spectra in each class. The numerical computation of $R_2$ is more subtle. To analyze bulk correlations, we must choose eigenvalues away from the edge of the spectrum. Our prescription is to choose eigenvalues within a disc centered at the origin with $ 2/3^{\text{rd}}$ the radius of the spectrum. In this region, $R_2$ depends only on the distance $|z_1-z_2|$. Therefore, the Mathematica library function \texttt{PairCorrelationG} can be used, which gives $\rho^{-2} R_2(\sqrt \rho |z_1 - z_2|)$, where $\rho$ is the density of eigenvalues in this region. We then average this quantity over $10^3$ samples in each class. In the above plot, to compare correlations across different classes, we have scaled the data such that $\rho$ is identical for each class. For comparison, we also plot the known analytical result for the bulk two-point correlation function in class A~\cite{mehta2004random}. }
\end{figure}

For non-Hermitian matrices, time-reversal symmetry$^{\dag}$ (TRS$^{\dag}$) plays a particularly important role \cite{Hamazaki-20}. It is defined by the relation  ${\cal T} H {\cal T}^{-1} = H^\dagger$,
where $H$ is a non-Hermitian matrix and 
${\cal T}$ is an anti-unitary operator. We distinguish two symmetry classes of non-Hermitian random matrices, AI$^{\dag}$ and AII$^{\dag}$ \cite{KSUS-19}, 
which respect TRS$^{\dag}$ with 
${\cal T}^2 = +1$ and
${\cal T}^2 = -1$, respectively.
The presence of  TRS$^{\dag}$  
induces correlations
among nearby complex eigenvalues.
This should be contrasted with
other discrete symmetries such as 
TRS, particle-hole symmetry (PHS) and PHS$^{\dag}$---they 
relate complex eigenvalues with their 
complex-conjugate or opposite-sign
partners.
As a consequence, 
TRS$^{\dag}$ 
alters level statistics in the bulk of the complex spectrum, 
while 
TRS, PHS, and PHS$^{\dag}$
primarily affect spectral statistics 
around (near)  symmetric lines or points.
Moreover, just as the bulk spectral correlations of Hermitian random matrices in symmetry classes A, AI, and AII are known to be universal, the bulk spectral correlations of non-Hermitian random matrices in symmetry classes A, AI$^{\dag}$, 
and AII$^{\dag}$ are also expected to exhibit universal behavior.
While numerical evidence supports this universality \cite{Hamazaki-20}, a rigorous proof remains an open challenge.
In this work, we take a step forward in this direction by analytically computing spectral correlations in the Gaussian ensemble for these symmetry classes.
Figure \ref{fig:R1R2-all} 
shows numerically calculated 
density of states and two-point correlation functions 
for classes A, AI$^{\dag}$ and
AII$^{\dag}$.
The goal of our work is 
to develop analytical understanding of these quantities.

Non-linear $\sigma$ models (NL$\sigma$Ms) have long been employed to calculate correlation functions in disordered systems~\cite{Efetov-textbook}.
Subsequently, field-theoretic treatments using the replica trick have been successfully used, even for non-Hermitian matrices. 
In Ref.~\cite{Nishigaki-02},
this method was used to calculate the density of states for non-Hermitian random matrices in classes A, AI, and AII. 
In this work, we further develop replica space NL$\sigma$Ms for classes AI$^\dagger$ and AII$^\dagger$, which are relevant to the threefold universality classes of the bulk spectral correlations.
Specifically, we consider the $n$th moments 
of the $k$-point characteristic polynomials 
$Z_n^{(k)}(z_1,\bar{z}_1,\cdots, z_k, \bar{z}_k)$, 
defined in Eq.~\eqref{eq:def-characteristic-polynomial}, for Gaussian non-Hermitian random matrices in classes AI$^\dagger$ and AII$^\dagger$. 
We also study the density of states, 
$R_1(z, \bar{z})$, and the two-point correlation function, 
$R_2(z_1,\bar{z}_1,z_2, \bar{z}_2)$, of the complex spectrum, derived from these polynomials via the replica limit $n\rightarrow 0$.

Our main results are summarized as follows. 
We derive general replica space matrix integrals for 
$Z_n^{(k)}(z_1,\bar{z}_1,\cdots, z_k,\bar{z}_k)$ in both symmetry classes AI$^{\dag}$ [Eq.\ \eqref{eq:AI-dag-Zn-general-k}] 
and AII$^{\dag}$ [Eq.\ \eqref{eq:AII-dag-Zn-general-k}]. 
We then study the cases $k=1$ and $k=2$ in detail.
For non-Hermitian random matrices of size $N$, we obtain exact expressions for 
$Z_n^{(1)}(z, \bar{z})$ with arbitrary $N$ 
[Eq.\ \eqref{eq:AI-dag-1pt-microscopic}
for class AI$^{\dag}$
and
Eq.\ \eqref{eq:AII-dag-1pt-microscopic}
for class AII$^{\dag}$]
and for 
$Z_n^{(2)}(z_1,\bar{z}_1, z_2, \bar{z}_2)$ in the limit $N\rightarrow \infty$ in both symmetry classes
[Eq.\ \eqref{eq:AI-dag-2pt-microscopic}
for class AI$^{\dag}$
and
Eq.\ \eqref{eq:AII-dag-2pt-microscopic}
for class AII$^{\dag}$]. 
These expressions are in the form of $n$-fold or $2n$-fold integrals over replica space singular values or cosine-sine values. 
We confirm these exact results by comparing them with numerically computed ensemble averages of characteristic polynomials.
Furthermore, 
following the approach in Refs.\ \cite{Kamenev_1999a, Kamenev_1999b, Yurkevich_Lerner_1999, Nishigaki-02}, 
we perform the replica limit $n\rightarrow 0$ of these $n$-fold integrals to obtain closed form expressions for the spectral distributions 
$R_1(z, \bar{z})$ 
[Eqs.\ \eqref{AId DOS analytical}
and \eqref{DOS AIIdag}]
and
$R_2(z_1, \bar{z}_1,z_2, \bar{z}_2)$ 
[Eqs.\ \eqref{two-pt AIdag} and \eqref{two-pt AIIdag}]
in the regimes $|z| \ll \sqrt N$ and $
|z_1-z_2| \ll \sqrt N$, respectively. 
We compare our analytical findings and numerical results.
For the density of states, 
our results  
show reasonable agreement with the numerics 
within the range of their validity.
However, for the two-point function,
our results substantially deviate
from the numerics 
in the regime $|z_1-z_2| \lesssim \sqrt{g}$
where level repulsion among complex eigenvalues is prominent.  
To obtain more accurate results, 
refining our current approach or adopting a more systematic method for taking the replica limit may be necessary.

\section{NL$\sigma$M for class AI$^\dagger$}
    \label{sec:NLSM AIdag}
    
\subsection{Replica space matrix integral for characteristic polynomials}
    \label{sec: AIdag - replica}

Non-Hermitian matrices in class AI$^\dagger$ are defined to respect TRS$^\dagger$ with sign $+1$:
\begin{equation}
    H^\dagger = \mathcal T H \mathcal T^{-1} = \bar H,
\end{equation}
where we choose the symmetry operator $\mathcal T$ to be complex conjugation $\mathcal K$, and $\bar{H}$ denotes complex conjugate of $H$.
This is equivalent to $H=H^T$, where ${H}^T$ denotes the transpose of $H$.  
We thus consider a Gaussian ensemble of symmetric $N\times N$ complex matrices $H$:
\begin{align}
\label{eq:AIdag-gaussian}
P(H) dH &= dH \exp[-g^{-1}\text{tr } H ^\dagger H] \nonumber
\\
&=\mathcal N_
{\text{AI}^\dagger} \prod_{i=1}^N dH_{ii}dH^*_{ii} \prod_{1\leq i<j \leq N}dH_{ij}dH^*_{ij} \exp[-g^{-1}\left(\sum_{i=1}^N |H_{ii}|^2 + 2\sum_{1\leq i<j \leq N} |H_{ij}|^2\right)]
\end{align}
where $g$ parameterizes the width of 
the Gaussian and the normalization $\mathcal N_{\text{AI}^\dagger}$ is defined by setting $\int dHP(H) = 1$.
Our main objects of interest are the $n$th moments of $k$th-order characteristic polynomial of $H$, defined as 
\begin{align}
    \label{eq:def-characteristic-polynomial}
    Z^{(k)}_n(z_1,\bar{z}_1, \cdots, z_k,\bar{z}_k)
    = \left\langle \prod_{l=1}^k \det(z_l-H)^n \det(\bar z_l - H^\dagger)^n \right\rangle,
\end{align}
where the angular brackets represent the ensemble averaging with respect to 
the probability measure in Eq.~\eqref{eq:AIdag-gaussian},
$\langle \cdots \rangle=\int dH P(H) \cdots$.
Following the replica trick, the $k$-point correlation functions of complex eigenvalues of $H$ can be determined using 
$Z^{(k)}_n(z_1, \bar{z}_1,\cdots, 
z_k, \bar{z}_k)$ 
(see Refs.~\cite{Nishigaki-02, CKKR-24} for details). For simplicity, we will first derive the replica space matrix integral for $k=1$.  
A few modifications will give the auxiliary  
field matrix integral for general $k$. 

The determinants are re-expressed as Grassmann integrals,
\begin{align}
    Z^{(1)}_n(z,\bar z) 
    &= \int dH P(H) \int d\psi d\bar\psi
    \exp[-\bar\psi^i_a \left( z\delta^{ij} \delta^{ab} - H^{ij}\delta^{ab} \right) \psi^j_a] \nonumber \\
    & \qquad 
    \times \int d\chi d\bar\chi\exp[-\bar\chi^i_a \left( \bar z \delta^{ij} \delta^{ab} - (H^\dagger)^{ij}\delta^{ab} \right) \chi^j_a ], 
    \label{eq 4}
\end{align}
where 
$\psi^i_a, \bar{\psi}^i_a, \chi^i_a, \bar{\chi}^i_a$
are independent Grassmann variables.
They satisfy the anti-commutation relation
$\{\eta,\xi \}= 0$,
 where $\eta,\xi $ are any of $ \psi^i_a, \bar{\psi}^i_a, {\chi}^i_a, 
 \bar{\chi}^i_a$.
In Eq.\ \eqref{eq 4}
and from now on,
we use the convention 
for implicitly summing repeated indices. For Grassmann integrals, we use conventions from Ref.~\cite{Efetov-textbook}. 
The indices 
$i,j = 1, \cdots, N$ are color/matrix indices, and $a,b = 1, \cdots, n$ are flavor/replica indices.
Since $H$ is symmetric, it is only coupled with the symmetric part of the fermion bilinears $\left\lfloor \psi_a \bar\psi^T_a \right\rfloor 
    = \frac 12\left(\psi_a \bar\psi^T_a - \bar\psi_a \psi^T_a \right)$.
Specifically, we have
\begin{align}
    \bar\psi^i_a H^{ij} \psi^j_a
    = - \tr(H \left\lfloor \psi_a \bar\psi^T_a \right\rfloor).
\end{align}
The Gaussian integral over $H$ gives a quartic action for the fermions,
\begin{align}
    Z^{(1)}_n(z,\bar z) &= \int d\psi d\bar\psi d\chi d\bar\chi
    \ e^{S[\psi, \chi]}, \\
    S[\psi,\chi] &= \bar z \tr(\left\lfloor \chi_a \bar\chi^T_a \right\rfloor)
    + z\tr(\left\lfloor \psi_a \bar\psi^T_a \right\rfloor)  
    + g\tr(\left\lfloor \chi_a \bar\chi^T_a \right\rfloor \left\lfloor \psi_b \bar\psi^T_b \right\rfloor). 
\end{align}
The quartic term above is expressed in terms of color-space matrices. We rearrange it as follows to express it in terms of flavor/replica space matrices,
\begin{equation}
    \label{eqn:quartic-term}
    \tr(\left\lfloor \chi_a \bar\chi^T_a \right\rfloor \left\lfloor \psi_b \bar\psi^T_b \right\rfloor)
    = \frac 12 \left((\psi_b^T \chi_a) (\bar\chi_a^T \bar\psi_b) -(\bar\psi_b^T \chi_a) (\bar\chi_a^T \psi_b)  \right).
\end{equation}
Now, we introduce replica space auxiliary fields to decouple these quartic terms into quadratic terms using the Hubbard-Stratonovich transformations. 
We use ``$\Tr$" for the trace in flavor space. We introduce flavor space matrices $Q$ and $ R \in \mathbb C^{n\times n}$ to decouple, respectively, the first and second terms of Eq.~\eqref{eqn:quartic-term},
\begin{equation}
    \exp[-\frac 12 g \Tr((\bar\psi^T \chi) (\bar\chi^T \psi))]
    \propto \int dQ \exp[
    -\frac 12 g^{-1} \Tr(QQ^\dagger) - \frac 12\Tr(\bar\psi^T\chi Q^\dagger) + \frac 12\Tr(Q\bar\chi^T\psi) ].
\end{equation}
Similarly, we have
\begin{equation}
    \exp[\frac 12 g \Tr((\psi^T \chi) (\bar\chi^T \bar\psi) )]
    \propto \int dR \exp[
    -\frac 12 g^{-1} \Tr(RR^\dagger) - \frac 12\Tr(R^\dagger \psi^T\chi) - \frac 12\Tr(\bar\chi^T\bar\psi R) ].
\end{equation}
If we collect all Grassmann variables  into one vector 
\begin{align}
    \Psi^{i} = \begin{pmatrix}
     \psi^i_1, \cdots, \psi^i_n,
     \chi^i_1, \cdots, \chi^i_n,
     \bar\psi^i_1, \cdots, \bar\psi^i_n,
     \bar\chi^i_1, \cdots, \bar\chi^i_n
    \end{pmatrix}^T,
\end{align}
and define the $4n\times 4n$ matrix
\begin{align}
    \mathbf M = \frac 12
    \begin{pmatrix} 
        0 & \bar R & -2z\mathbb I_{n} & Q \\
        -R^\dagger & 0 & -Q^\dagger & -2\bar z\mathbb I_{n} \\
        2z\mathbb I_{n} & \bar Q & 0 & -R \\
        -Q^T & 2\bar z\mathbb I_{n} & R^T & 0
    \end{pmatrix},
\end{align}
then the resulting action has the quadratic form,
$S=-\frac 12 \Psi^{iT} \mathbf{M} \Psi^i
$.
The matrix ${\bf M}$ is an antisymmetric matrix. 
The integral over the Grassmann fields can now be performed, which gives the Pfaffian of ${\bf M}$:
\begin{align}
    Z^{(1)}_n(z,\bar z) 
    &=  \int_{\mathbb C^{n\times n}} dQ \kk{\int_{\mathbb C^{n\times n}}} dR \exp[-\frac 12 g^{-1} \tr(Q^\dagger Q + R^\dagger R)] \text{det}^{\frac N2} \mathbf M.
\end{align}
By a basis change,
$\text{det}\, \mathbf M$
can be rewritten as the determinant of the following 
$2n\times 2n$ matrix:
\begin{align}
    \frac 12
    \begin{pmatrix} 
        2\bar z\mathbb I_{n} & 0 & R^T & -Q^T \\
        0 & 2\bar z\mathbb I_{n} & Q^\dagger & R^\dagger  \\
        -\bar R & -Q &  2z\mathbb I_{n} & 0\\
        \bar Q & -R & 0 & 2z\mathbb I_{n}
    \end{pmatrix}
    =
   \begin{pmatrix} 
   \bar{z}\mathbb{I}_n & {\cal Q}\\
   -{\cal Q}^{\dag} & z \mathbb{I}_n
   \end{pmatrix},
   \quad
       \mathcal Q = \frac 12 \begin{pmatrix} R^T & -Q^T \\ Q^\dagger & R^\dagger \end{pmatrix}.
\end{align}
The matrix ${\cal Q}$ satisfies
\begin{align}
    \Sigma^y_n 
    \mathcal Q
    = \bar{\mathcal Q} \Sigma^y_n,
    \quad
    \Sigma^y_n =
    \sigma^y \otimes \mathbb{I}_n
    =\begin{pmatrix} 
    0 & -i \mathbb{I}_n 
    \\ 
    i \mathbb{I}_n & 0 \end{pmatrix}.
\end{align}
By noticing 
$\tr(\mathcal Q^\dagger \mathcal Q) 
    = (1/2)   \tr( QQ^\dagger + RR^\dagger )
$,
the characteristic polynomial is expressed as
\begin{align}
    Z^{(1)}_n(z,\bar z)
    &= \int_{\mathcal M} d\mathcal Q\ e^{- g^{-1}\tr\mathcal Q^\dagger\mathcal Q }\ \text{det}^{N/2} \begin{pmatrix}
        \bar z \mathbb{I}_n & \mathcal Q \\ -\mathcal Q^\dagger & z \mathbb{I}_n
    \end{pmatrix},
    \label{eq:AI-dag-Zn-k1}
\end{align}
with
\begin{align}
    \mathcal M = \{ \mathcal Q \in \mathbb C^{2n \times 2n} \ | \  \Sigma^y_n \mathcal Q = \bar{\mathcal Q} \Sigma^y_n \}.
\end{align}

Now, we generalize this expression to the $k$-point characteristic polynomial. A straightforward extension of the above procedure gives the same expressions as above, with $z\mathbb{I}_n$ and $\Sigma^y_n$ replaced by matrices $Z$ and $\Sigma^y_{nk}$, defined momentarily. 
Using a suitable (orthogonal) similarity transformation on $Q, Z, $ and $\Sigma^y_{nk}$, we can choose 
$Z = \text{diag}(z_1, \cdots , z_k)\otimes 
\mathbb{I}_{2n}$ and $\Sigma^y_{nk} = \sigma^y \otimes 
\mathbb{I}_{nk}$ without loss of generality. 
The characteristic polynomial for the $k$-point function is then given as 
\begin{align}
    Z^{(k)}_n(z_1, \bar{z}_1,\cdots,z_k,\bar {z}_k)
    &= \int_{\mathcal M} d\mathcal Q\ e^{- g^{-1}\tr\mathcal Q^\dagger\mathcal Q }\ \text{det}^{N/2} \begin{pmatrix}
        \bar Z & \mathcal Q \\ -\mathcal Q^\dagger & Z
    \end{pmatrix},
    \label{eq:AI-dag-Zn-general-k}
\end{align}
with
\begin{align}
    \mathcal M = \{ \mathcal Q \in \mathbb C^{2nk \times 2nk} \ | \  \Sigma^y_{nk} \mathcal Q = \bar{\mathcal Q} \Sigma^y_{nk} \}.
        \label{eq: AIdag-M}
\end{align}
In the subsequent sections, we will compute this integral for $k=1$ and $k=2$ under suitable limits.

\subsection{One-point characteristic polynomial} 
    \label{sec: AIdag - DoS}

The integral expression in Eq.~\eqref{eq:AI-dag-Zn-k1}, 
playing a role of the partition function, 
can be further evaluated in various ways. 
Let us first discuss the behavior in the large-$N$ limit.  
In the limit $N\rightarrow \infty$, we can use the saddle-point approximation to calculate $Z^{(1)}_n(z,\bar z)$. 
The saddle-point equation for the action in Eq.~\eqref{eq:AI-dag-Zn-k1} is
\begin{align}
    \mathcal Q^\dagger \mathcal Q = 
    \left(
    \frac{gN}{2} - |z|^2
    \right)\mathbb{I}_{2n}.
\end{align}
For $|z|^2<gN/2$, the solution is simply 
\begin{equation}
\mathcal Q = \sqrt{\frac{gN}{2} - |z|^2}\,U,
\end{equation}
where $U$ is a unitary matrix that satisfies $\Sigma^y_n U = \bar U \Sigma^y_n$ (or equivalently, $U \Sigma^y_n U^T = \Sigma^y_n$), i.e., $U\in \text{Sp}(n)$. 
Now, we substitute this solution into Eq.~\eqref{eq:AI-dag-Zn-k1} and obtain the dominant contribution at large $N$, leading to
\begin{align}
\label{large N, AId}
    Z^{(1)}_n(z,\bar z)
    &\simeq \int_{\text{Sp}(n)} d\,U \ e^{\left(g^{-1}|z|^2 - \frac N2 \right) \tr \mathbb I_{2n}}\ \text{det}^{N/2} \left(g \frac N2 \mathbb I_{2n} \right) \nonumber \\
    &= e^{2n g^{-1}|z|^2 } \left[ e^{-nN} \left(\frac{gN}{2}\right)^{nN}\ \text{Vol}\,(\text{Sp}(n)) \right].
\end{align}
Note that the factors inside the square bracket drop out in the replica limit $n\rightarrow 0$. 
In the next subsection, we will use this expression to discuss the density of states.

Now, let us go back to Eq.~\eqref{eq:AI-dag-Zn-k1} and evaluate it without the large-$N$ approximation. 
We consider the singular-value decomposition of $\mathcal Q$, 
\begin{equation}
    \mathcal Q = U \Lambda V;  \qquad V \in \text{Sp}(n), \ U \in \text{Sp}(n)/\text{Sp}(1)^{\oplus n}, \qquad
    \Lambda = \text{diag}(\lambda_1^{\frac 12}, \cdots, \lambda_n^{\frac 12} , \lambda_1^{\frac 12}, \cdots, \lambda_n^{\frac 12}) \quad \left( \lambda_a \geq 0 \right).
\end{equation}
To find the measure in terms of $U, \Lambda$, and $V$, we must calculate the determinant of the Jacobian $\frac{\partial \mathcal Q}{\partial[U, \Lambda, V]}$ for this transformation. See Appendix~\ref{asec:SVD}
and Ref.~\cite{Nishigaki:2016random} for more details. 
The measure is given as
\begin{align}
    d\mathcal{Q} = dU dV |\Delta(\lambda)|^4 \prod_{a=1}^{n} d\lambda_a \, \lambda_a, \quad \Delta(\lambda) = \prod_{a>b}^n (\lambda_a-\lambda_b).
\end{align}
Here, $dU$ and $dV$ represent the Haar measure on Sp$(n)$, the integral over which is just an overall constant and is irrelevant in the replica limit.
The only relevant degrees of freedom in the replica space are $\lambda_a$, and hence we have
\begin{align}
    \label{eq:AI-dag-1pt-microscopic}
    Z^{(1)}_n(z,\bar z)
    &\simeq \int_0^\infty \prod_{a=1}^n d\lambda_a \ e^{- 2g^{-1} \lambda_a}\, \left( |z|^2 + \lambda_a \right)^N \lambda_a |\Delta(\lambda)|^4.
\end{align}
We highlight that this is an exact expression for all moments of the one-point characteristic polynomial at arbitrary $N$ and $z$, up to an overall constant. 
This can also be thought of as a partition function for replica space degrees of freedom $\lambda_a$. 
We note that
there exist other results on 
this quantity in 
Refs.\ \cite{Akemann-24, Forrester-24}. 
We also note that
there exists another work 
\cite{Liu:2024xew}
which studies similar objects.

\begin{figure}
    \centering
    \subfloat{\includegraphics[width = 0.4\textwidth]{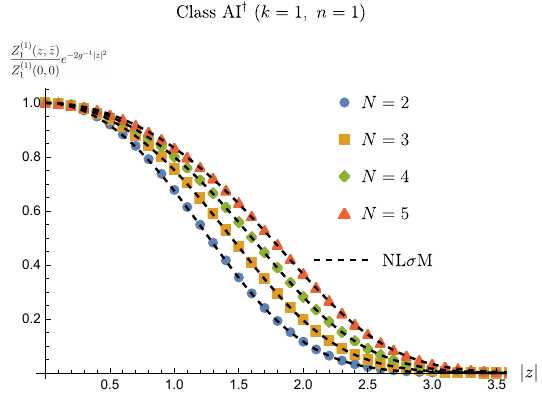}}
    \subfloat{\includegraphics[width = 0.4 \textwidth]{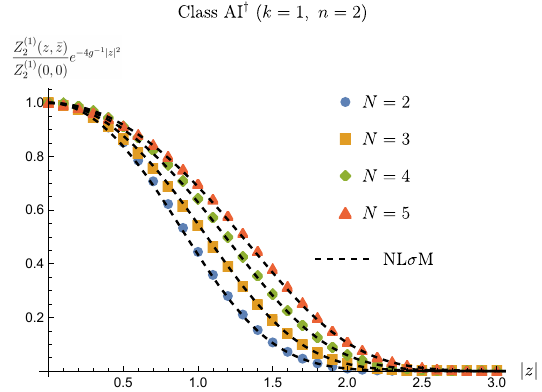}}
    \caption{Plot of 
    $\frac{Z^{(1)}_n(z, \bar{z})}{Z^{(1)}_n(0,0)}e^{-2ng^{-1}|z|^2}$ as a function of $\left| z \right|$ for class AI$^\dagger$. For each value of $N$, we sample $10^5$ non-Hermitian random matrices according to Eq.~\eqref{eq:AIdag-gaussian} with $g=2$. The solid markers show the ensemble-averaged characteristic polynomials. The black dashed curves show the same quantity calculated by the NL$\sigma$M in Eq.~\eqref{eq:AI-dag-1pt-microscopic}.
    }
    \label{fig:charpol-AIdag-smallN}
\end{figure}

In Fig.\ \ref{fig:charpol-AIdag-smallN},
we plot the analytical expression 
in Eq.~\eqref{eq:AI-dag-1pt-microscopic}
of the characteristic polynomial 
for various values of $N$ and $n$
as a function of $|z|$.
We also compare the analytical result
in Eq.~\eqref{fig:charpol-AIdag-smallN}
with numerical results 
where we numerically generate 
non-Hermitian random matrices from the Gaussian ensemble in Eq.~\eqref{eq:AIdag-gaussian}
and calculate
the characteristic polynomial.
For all values of $N$ and $n$ we studied, 
we have a good agreement between 
the analytical and numerical results.
While Fig.\ \ref{fig:charpol-AIdag-smallN} shows a good agreement of the derived formula in Eq.\ \eqref{eq:AI-dag-1pt-microscopic}  and the numerics, 
it should be noted that,
in the interest of the universal statistics of non-Hermitian random matrix theory,
it is known that the small $N$
behaviour is not a good approximation for the large $N$ limit.
This fact was pointed out for class A in Ref.\ \ \cite{Grobe-88}
and was also further studied for 
the $N=2$ behaviour of
classes AI$^{\dag}$ and AII$^{\dag}$
in Ref.\ \cite{Akemann_2022}.

\subsubsection{Density of states}
\label{AId DOS}

The density of states, $R_1(z,\bar{z})$, is determined from the first-order characteristic polynomial by~\cite{Nishigaki-02, CKKR-24}
\begin{align}
    \label{eq:DoS-from-Zn}
    \pi R_1(z, \bar{z}) = \lim_{n\rightarrow 0} \frac 1n \partial_z \partial_{\bar z} Z^{(1)}_n(z,\bar z).
\end{align}
To begin with, we extract the dominant contribution to 
$R_1(z,\bar{z})$ at large $N$.
We expect this contribution to be a uniform distribution on a disk centered at the spectral origin, which is verified from the NL$\sigma$M as follows. 
In the large $N$ limit,
using
Eqs.~\eqref{large N, AId}
and \eqref{eq:DoS-from-Zn},
we obtain
\begin{align}
    \pi 
    R_1(z, \bar{z}) = \lim_{n\rightarrow 0} \frac 1n \partial_{\bar z}\partial_z Z^{(1)}_n(z,\bar z) = \frac{2}{g} \quad \text{for} \quad |z|^2 < \frac{gN}{2}.
\end{align}
Indeed, at large $N$, the density of states is a uniform distribution on a disc of radius $\sqrt{gN/2}$ centered at the origin, which is consistent with Girko's circular law~\cite{Girko-85}.

To discuss  
the density of states 
near the edge
of the spectrum, 
we use 
Eq.~\eqref{eq:AI-dag-1pt-microscopic} and
analyze its behavior 
at large $N$
and in the limit $n\to 0$.
For large $N$, we can further perform a saddle-point approximation. 
We set $g=2$ for simplicity
and consider the factor $I(\lambda) = e^{-\lambda_a}\, \left( |z|^2 + \lambda_a \right)^N$ in the integrand. 
For large $N$, $I(\lambda)$ approaches an un-normalized Gaussian. 
The saddle point obtained by solving $\partial_\lambda \ln(I(\lambda)) = 0$ is $\lambda_{\text{sp}} = N-|z|^2$. 
We Taylor-expand the action around this saddle point, 
$\ln I(\lambda) 
    \sim  -\lambda_{\text{sp}} + {N\ln N} - \frac{(\lambda - \lambda_{\text{sp}})^2}{2N}$,
and substitute this back into 
Eq.\ \eqref{eq:AI-dag-1pt-microscopic}.
Rescaling $\lambda \rightarrow \lambda_{\text{sp}} \lambda$ and 
ignoring irrelevant overall factors,
we obtain
\begin{align}
    Z^{(1)}_n(z,\bar z)
    \simeq&\ e^{-n \lambda_{\text{sp}}} \lambda_{\text{sp}}^{2n^2} \int_0^\infty \prod_{a=1}^n d\lambda_a \ \exp(- \frac{\lambda_{\text{sp}}^2}{2N} \left(\lambda_a  - 1 \right)^2) |\lambda_a| |\Delta(\lambda)|^4.
\end{align}
Now, we rewrite the contour of integration as a sum over two contours: $\int_0^\infty d\lambda \rightarrow \int_{-\infty}^\infty d\lambda  - \int_{-\infty}^0 d\lambda $ for each $\lambda$. 
This gives us
\begin{align}
    \label{eq:AI-dag-1pt-partition-largeN}
    Z^{(1)}_n(z,\bar z)
    &\simeq e^{-n \lambda_{\text{sp}}} \lambda_{\text{sp}}^{2n^2} \sum_{p} \left( -1 \right)^p{n \choose p} \int_{-\infty}^0 \prod_{a=1}^p dx_a \exp(- \frac{\lambda_{\text{sp}}^2}{2N} \left(x_a  - 1 \right)^2) |x_a| |\Delta(x)|^4 \nonumber \\
    &\qquad\qquad \times \int_{-\infty}^\infty \prod_{a=1}^{n-p} dy_a \exp(- \frac{\lambda_{\text{sp}}^2}{2N} \left(y_a  - 1 \right)^2) |y_a| |\Delta(y)|^4.
\end{align}
We have considered the regime where $\lambda_{\text{sp}}^2/2N$ is large enough so that $x$ and $y$ accumulate near the maxima of the weight function, i.e., $0$ and $1$. 
As such, we can approximate $\Delta(\lambda) \simeq \Delta(x)\Delta(y)$, 
and the $x$ and $y$ variables become decoupled. 
In the $y$-integral, since the Gaussian is narrowly peaked at 1, we replace the factor of $y_a$ in the integrand with 1. The remaining integral is then a Selberg integral that is evaluated exactly,
\begin{align}
    \int_{-\infty}^\infty\prod_{a=1}^{n-p} dy_a \exp(- \frac{\lambda_{\text{sp}}^2}{2N} \left(y_a  - 1 \right)^2) |\Delta(y)|^4
    &\simeq \left(\frac{\lambda_{\text{sp}}}{\sqrt{N}}\right)^{- 2(n-p)^2 + (n-p) } \left(\frac 2 \pi \right)^{p/2} \prod_{a=1}^{n-p} \Gamma(1+2a).
\end{align}
Before proceeding to the $x$-integral, we should look at the coefficient of the $p$th term in this expansion and identify which terms survive in the replica limit $n\rightarrow 0$. 
The coefficient is ${n \choose p} \prod_{a=1}^{n-p} \Gamma(1+2a)$. At order $n$, the coefficient is $1$ and $n$ for $p=0$ and $1$, respectively, and vanishes for all $p\geq 2$. 
We then have
\begin{align}
    \int_{-\infty}^0 dx_1 \exp(- \frac{\lambda_{\text{sp}}^2}{2N} \left(x_1  - 1 \right)^2) |x_1| \simeq \frac{N^2}{\lambda_{\text{sp}}^4}\exp(-\frac{\lambda_{\text{sp}}^2}{2N}) \quad \text{for}\quad \frac{\lambda_{\text{sp}}^2}{2N} \gg 1.
\end{align}
Putting it all together, we have
\begin{align}
    Z^{(1)}_n(z,\bar z)
    \simeq &\ 1 - n \lambda_{\text{sp}} + n \ln \lambda_{\text{sp}} - n \left(\frac 2 \pi \right)^{1/2}\left(\frac{\lambda_{\text{sp}}}{\sqrt{N}}\right)^{-7}\exp(-\frac{\lambda_{\text{sp}}^2}{2N})  + \mathcal O(n^2).
\end{align}
Finally,
plugging this into Eq.~\eqref{eq:DoS-from-Zn} and keeping the leading-order terms
lead to
\begin{align}
\label{AId DOS analytical}
    \pi 
    R_1(z,\bar{z})
    = 1 - \frac{1}{4 u^2} - \sqrt{\frac{2}{\pi N}} \frac{e^{-2 u^2} }{ 16 u^4}  - \frac{e^{-2 u^2}}{16 \sqrt{2 \pi } u^5},
\end{align}
where 
we introduce 
$u = \sqrt N - |z|$ as the distance of $z$ from the edge of the spectrum, 
and use 
 $\lambda_{\text{sp}} \simeq 2 u \sqrt N$. 
We expect that the fermionic replica method can only reproduce the tail of the density of states for $|z|<\sqrt{gN/2}$ \cite{Nishigaki-02}.
Furthermore, note that we assumed $|z|\ll\sqrt{gN/2}$ in our approximations. 
In the above analysis and in Eq.\ \eqref{AId DOS analytical},
the $g$ dependence can be added back using dimensional analysis.

\begin{figure}
    \centering
    \includegraphics[width = 0.7\textwidth]{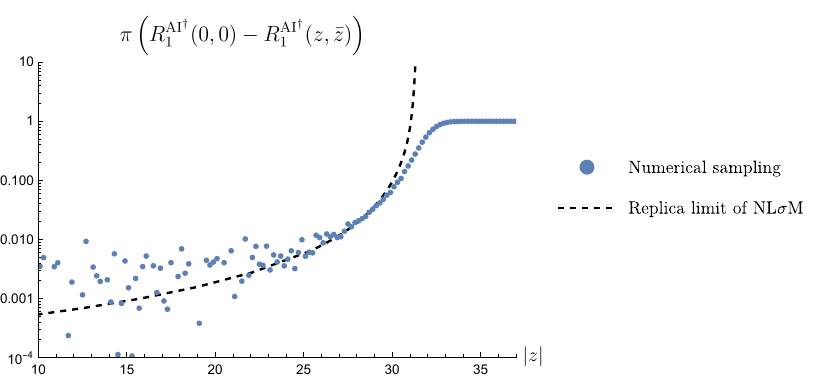}
    \caption{
    Comparison between
    the analytical result in Eq.~\eqref{AId DOS analytical}
    and the density of states 
    obtained from numerical calculations.
    The numerical results are obtained using $2 \times 10^4$ realizations of $10^3 \times 10^3$ non-Hermitian random matrices in class AI$^\dagger$ sampled according to Eq.~\eqref{eq:AIdag-gaussian} with $g=2$.
    Note that this is a logarithmic plot. 
    This might cause the spread of the numerics in the plot.}
    \label{fig:DoS-AIdag}
\end{figure}


In Fig.\ \ref{fig:DoS-AIdag}, we compare the analytical result 
from Eq.\ \eqref{AId DOS analytical} with the density of states obtained from numerical calculations
for $g=2$ and $N=10^3$. Note that Fig.\ \ref{fig:DoS-AIdag} is a logarithmic plot
for the difference $\pi (R_1(0,0) - R_1(z, \bar{z}))$, which may visually amplify the spread of the numerical data. 
Additionally, Eq.\ \eqref{AId DOS analytical} is strictly valid only for $|z| \ll \sqrt{gN}/2 \approx 31.6$ given the choice of parameters in Fig.\ \ref{fig:DoS-AIdag}.
Within these limitations, the NL$\sigma$M result and the numerical calculations appear consistent for 
$22 \lesssim |z| \lesssim 30$. 
For 
$|z| \gtrsim 30$, 
Eq.\ \eqref{AId DOS analytical} 
begins to deviate from the numerical results.
For 
$|z| \lesssim 20$, 
the analytical result and the numerical data 
are consistent 
within
$\sim 10^{-2}$,
although 
the numerical data scatter 
due to the logarithmic plot.
A more precise comparison with large-scale numerics, along with an analysis that includes corrections beyond the saddle-point approximation, 
is left for future study.
Furthermore, 
while our approach of taking the replica limit
is inspired by 
Refs.\ \cite{Kamenev_1999a, Kamenev_1999b, Yurkevich_Lerner_1999}, 
there remains room for improvement on these calculations.
For example, 
in the standard case of 
the Gaussian unitary ensemble in Hermitian RMT, 
the replica limit can be taken 
more systematically
once 
we know the recursion relation satisfied by the replica partition functions
with different replica indices,
and also 
the partition functions with
negative replica indices,
which can be obtained 
from 
bosonic replica 
NL$\sigma$M~\cite{Verbaarschot:1985qx, 
Kamenev_1999a,
Kamenev_1999b,
Yurkevich_Lerner_1999,
zirnbauer1999critiquereplicatrick,
PhysRevLett.89.250201,
Splittorff:2002eb}.
In the context of 
\kk{non-Hermitian}
random matrices,
these issues were discussed in detail for 
the Ginibre unitary ensemble
in Ref.\
\cite{ kanzieper2005exactreplicatreatmentnonhermitean}.
Additionally, 
it should be useful to develop 
a supersymmetric NL$\sigma$M approach.


\subsection{Two-point characteristic polynomial}
    \label{sec: AIdag - 2nd}
 
We now discuss the second-order characteristic polynomial, 
$Z^{(2)}_n$,
starting from 
Eq.~\eqref{eq:AI-dag-Zn-general-k}.
For large $N$, the integral over $\mathcal Q$ can be approximated by an integral over the saddle-point manifold. 
We introduce the radius of the spectrum, $r_s = \sqrt{gN/2}$, and focus on the region in the bulk of the spectrum, $|z_{1,2}| \ll r_s$, where the saddle-point equation is simplified to
\begin{align}
    \mathcal Q^\dagger \mathcal Q = r_s^2\, 
    \mathbb{I}_{4n}.
\end{align}
The solution is $\mathcal Q = r_s U$, with a unitary matrix $U$. 
Moreover, owing to $\mathcal Q \in \mathcal M$ [see Eq.~(\ref{eq: AIdag-M})], we have
\begin{align}
    U\Sigma^y_{nk} U^T = \Sigma^y_{nk} \bar U U^T = \Sigma^y_{nk}.
\end{align}
Thus, $U$ is symplectic and unitary, and thus belongs to the symplectic group $\text{Sp}(2n)$. 
We substitute the saddle-point solution back into the expression for $Z^{(2)}_n(z_1, \bar z_1, z_2, \bar z_2)$, leading to
\begin{align}
    Z^{(2)}_n(z_1, \bar z_1, z_2, \bar z_2) 
    \simeq & \int_{\text{Sp}(2n)} dU\ e^{-g^{-1} r_s^2 4n } \text{det}^{N/2}
    \begin{pmatrix}
        \bar Z & r_s U \\ - r_s U^\dagger & Z
    \end{pmatrix} \nonumber \\
    \simeq& \int_{\text{Sp}(2n)} dU\ \exp[g^{-1}\tr(U^\dagger \bar Z U Z)].
\end{align}
When we introduce $z = (z_1 + z_2)/2$, $\omega=z_1 - z_2$, and $s=\text{diag}(1, - 1)\otimes \mathbb I_{2n}$, we have $Z = z \mathbb I_{4n} + \frac 12 \omega s$ and hence
\begin{align}
    \label{saddle-point-integral}
    Z^{(2)}_n(z_1, \bar z_1, z_2, \bar z_2) 
    &\simeq e^{g^{-1}4n|z|^2}  Y_{\text{Sp}(n)}(|\omega|^2)
    \quad \text{where}\quad 
    Y_{\text{Sp}(n)}(|\omega|^2) = \int_{\text{Sp}(2n)} dU\ \exp[\frac{|\omega|^2}{4g} \tr( U^\dagger s U s )].
\end{align}
Observe that for $U \in \text{Sp}(2n)$, $U^\dagger s U$ lies on the symplectic Grassmannian $\text{Sp}(2n)/[\text{Sp}(n)\times \text{Sp}(n)]$. 
Thus, the target space of this NL$\sigma$M is the symplectic Grassmannian. 
In the following, we calculate this integral in the regime $|\omega|^2\gg g$.

\subsubsection{Cosine-sine decomposition}
    \label{sec: CS decomposition}

Analogous to the singular-value decomposition used in Sec.~\ref{sec: AIdag - DoS}, we find that a cosine-sine (CS) decomposition for the symplectic group is effective in calculating the second-order characteristic polynomial $Z_n^{(2)}$. 
The CS decomposition of a symplectic unitary matrix $U \in \text{Sp}(2n)$ is described below~\cite{Edelman-23}. 
We start with the following parametrization, 
\begin{align}
    U = 
    \begin{pmatrix} u_1 & 0 \\ 0 & u_2 \end{pmatrix}
    \begin{pmatrix} \cos \Theta & \sin\Theta \\ -\sin \Theta & \cos \Theta \end{pmatrix}
    \begin{pmatrix} v_1 & 0 \\ 0 & v_2 \end{pmatrix},
\end{align}
with 
\begin{equation}
\Theta = \text{diag}(\theta_1, \theta_1, \cdots ,\theta_{n}, \theta_{n}), \text{ where } \theta_i \in [0,\pi) , \ u_{1,2}, v_{1,2} \in \text{Sp} (n).
\end{equation}
Notice that each angle is repeated twice. 
One can regard each $\theta_i$ as the commuting part of the $2\times 2$ matrix representation of a quaternion.
While $u_{1,2}$ and $v_{1,2}$ are generic matrices in $\text{Sp} (n)$, they must be compensated by an Sp$(1)^{\otimes n}$ phase factor inside the CS matrix. 
A parametrization for Sp$(1) \cong S^3$ is $e^{i\phi (\mathbf m \cdot \mathbf \sigma)}$ with $\phi \in [0, \pi]$ and $\mathbf m \in \mathbb R^3$ satisfying $|\mathbf m|^2 = 1$ (i.e., $\mathbf{m} \in S^2$). 
We thus obtain a new CS decomposition without redundancy by
\begin{align}
    U = 
    \begin{pmatrix} u_1 & 0 \\ 0 & u_2 \end{pmatrix}
    \begin{pmatrix} \cos \Theta \exp(i\Phi \mathbf M \cdot \mathbf\Sigma) & \sin\Theta \\ -\sin \Theta & \cos \Theta \exp(-i\Phi \mathbf M \cdot \mathbf\Sigma) \end{pmatrix}
    \begin{pmatrix} v_1 & 0 \\ 0 & v_2 \end{pmatrix},
\end{align}
with 
\begin{align}
u_{1,2} \in \text{Sp} (n)/ \text{Sp}(1)^{\otimes n}, & \quad v_{1,2} \in \text{Sp} (n); \\
\mathbf M \cdot \mathbf \Sigma = \text{diag}(\mathbf m_1 \cdot \mathbf \sigma, \cdots, \mathbf m_n \cdot \mathbf \sigma), & \quad \Phi = \text{diag}(\phi_1, \phi_1, \cdots, \phi_n, \phi_n)~~~\left[ \mathbf m_i \in S^2, \phi_i \in [0,\pi] \right].
\end{align}

Now, we proceed to derive the Haar measure in terms of these new variables. 
Let $\Lambda$ be the CS matrix. 
By using the invariance properties of the Haar measure, it is sufficient to consider a neighborhood of $u_{1,2} = v_{1,2} = \mathbb I_{2n}$, leading to 
\begin{align}
    U^\dagger dU
    =& 
    \begin{pmatrix}
        dv_1 & 0 \\ 0 & dv_2
    \end{pmatrix}
    + \Lambda^\dagger
    \begin{pmatrix}
        du_1 & 0 \\ 0 & du_2
    \end{pmatrix}
    \Lambda
    + \Lambda^\dagger d\Lambda.
\end{align}
The calculations follow quite similarly to the unitary case from here on. 
The contribution from $\Lambda^\dagger d\Lambda$ is simply an extension of the measure for Sp$(2)$ matrices given in Ref.~\cite{Liu_Kudler-Flam_Kawabata_2023}. 
The second term is the non-trivial part, which by comparison to the unitary case contributes $\Delta(\cos^2\Theta)^4$. 
Finally, the Haar measure is obtained as
\begin{align}
    dU^{\text{Haar}} \propto du_1^{\text{Haar}} du_2^{\text{Haar}} dv_1^{\text{Haar}} dv_2^{\text{Haar}} |\Delta(\cos^2\Theta)|^4 
    \prod_{i=1}^n \sin^3(2\theta_i)\sin^2(\phi_i) d\theta_i d\phi_i d^2\mathbf m_i.
\end{align}
Now, we recall the saddle-point integral in Eq.~\eqref{saddle-point-integral}.
The integrand in terms of the new variables is
\begin{align}
    \exp(\frac{|\omega|^2}{4g}\tr(U s U^\dagger s))
    = \exp(\frac{|\omega|^2}{2g}\tr( \cos 2\Theta)).
\end{align}
Thus, ignoring irrelevant overall constants, we have
\begin{align}
    Y_{\text{Sp}(n)}(|\omega|^2)
    &\simeq \int_0^\pi \prod_{i=1}^n d\theta_i \sin^3(2\theta_i) \exp(\frac{|\omega|^2}{2g}\tr \cos 2\Theta ) |\Delta(\cos^2\Theta)|^4 \nonumber \\
    &=  \int_{-1}^1 
    \prod_{i=1}^n d\lambda_i\, (1-\lambda_i^2) \, e^{|\omega|^2 \sum_i\lambda_i } |\Delta(\lambda)|^4,
\end{align}
where we set $g=1$ for simplicity and introduce $\lambda_i =\cos2\theta_i$, satisfying $\tr \cos 2\Theta = 2\sum_i \lambda_i$.
To make a comparison with Selberg integrals, we make the shift $\lambda \rightarrow 1 - \lambda$ and then rescale $\lambda \rightarrow 2\lambda$, resulting in
\begin{align}
    \label{eq:AI-dag-2pt-microscopic}
    &
   Z^{(2)}_n(z_1, \bar{z}_1, z_2, \bar{z}_2)
   \simeq
   e^{4n |z|^2}
   Y_{\text{Sp}(n)}(|\omega|^2),
    \\
    &
   Y_{\mathrm{Sp}(n)}(|\omega|^2)
    \simeq
    e^{n{|\omega|^2}} \int_{0}^1 
    \prod_{i=1}^n d\lambda_i\, \lambda_i(1 - \lambda_i) \, e^{-2|\omega|^2\sum_i\lambda_i} |\Delta(\lambda)|^4.
\end{align}
\begin{figure}
    \centering
    \subfloat{\includegraphics[width = 0.4 \textwidth]{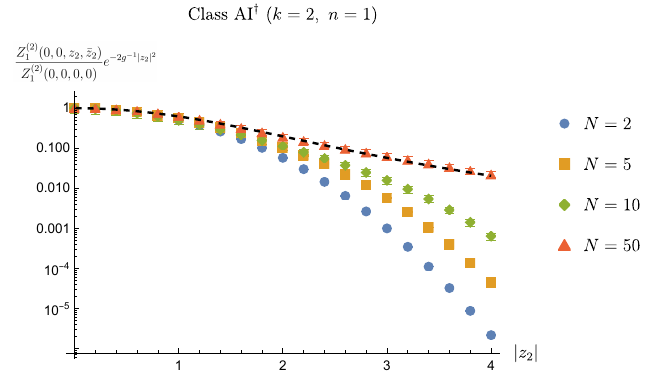}}
    \subfloat{\includegraphics[width = 0.4 \textwidth]{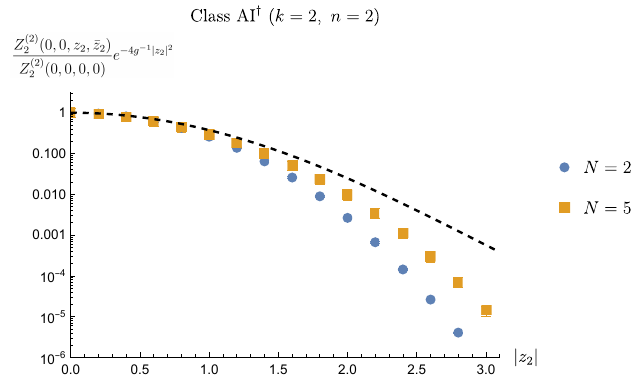}}
    \caption{Logarithmic plot of the two-point characteristic polynomial in class AI$^\dagger$. We plot the expression 
    $\frac{Z^{(2)}_n(0,0,z_2, \bar{z}_2)}{Z^{(2)}_n(0,0,0,0)}e^{-2ng^{-1}|z_2|^2}$ instead of $Z^{(2)}_n(0,0,z_2, \bar{z}_2)$ for better visualization. For each value of $N=2$ and $N=5$, we collect $10^6$ realizations of non-Hermitian random matrices, while for $N=10$ and $N=50$ we collect $5\times 10^6$ realizations. In all cases, we use $g=2$. 
    The solid markers show the numerically computed ensemble-averaged moments of characteristic polynomials. The black dashed curves show the same quantity calculated with the NL$\sigma$M for large $N$ in Eq.~\eqref{eq:AI-dag-2pt-microscopic}.
    }
    \label{fig:charpol-2pt-AIdag}
\end{figure}

In Fig.\ \ref{fig:charpol-2pt-AIdag},
we plot 
the analytical 
expression in Eq.~\eqref{eq:AI-dag-2pt-microscopic}
of 
the second-order characteristic 
polynomial 
for various values of $N$ and $n$
as a function of $|z_2-z_1|$.
We also compare the analytical result
in Eq.~\eqref{eq:AI-dag-2pt-microscopic}
with numerical results 
where we numerically generate 
non-Hermitian random matrices from the Gaussian ensemble in Eq.~\eqref{eq:AIdag-gaussian}
and calculate
the characteristic polynomial.
Here, it should be noted that 
Eq.~\eqref{eq:AI-dag-2pt-microscopic}
is valid only for large $N$
although it allows us to access 
the small $|\omega|$ regime,
which is relevant for level repulsion.
While the numerics for small $N$
deviates from 
Eq.~\eqref{eq:AI-dag-2pt-microscopic}
for large $z_2$,
it is consistent with 
Eq.~\eqref{eq:AI-dag-2pt-microscopic}
for large enough $N$ and for small enough $z_2$.

\subsubsection{Two-point correlation function}
\label{Two-point correlation function}

We now discuss the two-point correlation function.
The two-point correlation function is obtained from the second-order characteristic polynomial $Z^{(2)}_n$ by~\cite{Nishigaki-02, CKKR-24}
\begin{align}
    \label{eqn:two-point-from-Zn}
    \pi^2 R_2(z_1, \bar{z}_1,z_2, \bar{z}_2) = \lim_{n\rightarrow 0} \frac{1}{n^2} \partial_{z_1} \partial_{\bar z_1} \partial_{z_2} \partial_{\bar z_2} Z^{(2)}_n(z_1, \bar z_1, z_2, \bar z_2),
\end{align}
where $Z^{(2)}_n$ is given by Eq.~\eqref{eq:AI-dag-Zn-general-k} with $k=2$. 
Using the fact that $Z_n^{(2)}$ factorizes into a $z$-dependent and $\omega$-dependent part, we can simplify this as follows,
\begin{align}
\label{R2andYa}
    \pi^2 R_2(z_1,\bar{z}_1, z_2, \bar{z}_2)
    &= \lim_{n\rightarrow 0} \frac{1}{n^2} ( \partial^2_\omega - \partial^2_{2z} ) ( \partial^2_{\bar\omega} - \partial^2_{2\bar z} ) e^{4n|z|^2} Y_{\text{Sp}(n)}(|\omega|^2) \nonumber\\
    &= 2 + \lim_{n\rightarrow 0} \frac{1}{n^2} \partial^2_\omega\partial^2_{\bar\omega} Y_{\text{Sp}(n)}(|\omega|^2).
\end{align}
To evaluate $R_2$, 
we go back to Eq.~\eqref{eq:AI-dag-2pt-microscopic}
and analyze its behavior
in the limit $n\to 0$.
It is difficult to evaluate this integral exactly on the domain $[0,1]^n$. 
However, we aim to evaluate it only up to order $n^2$. 
For this purpose, we can expand the domain to $[0,\infty]^n$ and then systematically add or subtract corrections at increasing orders in $n$. 
Only finitely many terms in this series survive at order $n^2$. 
To perform this procedure, 
we rewrite the contour of integration in Eq.~\eqref{eq:AI-dag-2pt-microscopic} as a sum over two contours, $\int_0^1 d\lambda \rightarrow \int_{0}^\infty d\lambda  - \int_{1}^\infty d\lambda$. This gives
\begin{align}
    &Y_{\text{Sp}(n)}(|\omega|^2)
    \simeq e^{n\frac t2} \int_{0}^\infty 
    \prod_{i=1}^n d\lambda_i\, |\lambda_i(1 - \lambda_i) | (1-\Theta(\lambda_i -1)) \, e^{-t\sum_i\lambda_i} |\Delta(\lambda)|^4 \nonumber \\
    &\qquad \simeq e^{4n|z|^2}  e^{n\frac t2} \sum_{p=0}^n {n \choose p} 
    (-1)^p
    \int_{1}^\infty \prod_{i=1}^{p} dx_i\, | x_i(1 - x_i) | \, e^{-t\sum_ix_i} \int_{0}^\infty \prod_{i=1}^{n-p} dy_i\, | y_i(1 - y_i)|  \, e^{-t\sum_iy_i} |\Delta(x,y)|^4
\end{align}
with $t=2|\omega|^2$.
For large $t$, the $x$ and $y$ variables accumulate near the peak of the weight function $|\lambda(1-\lambda)|e^{-t\lambda}$ within their respective domain. 
To the leading order in $1/t$, these are simply located at $x=1$ and $y=0$, respectively, and hence we can make the approximation $\Delta(x,y) \simeq \Delta(x)\Delta(y)$. 
Let us focus on the $y$-integral.
Since the $y$ variables accumulate close to the origin, we can approximate $|y_i(1 - y_i)| \simeq y_i$.
Thus, we have
\begin{align}
    \int_{0}^\infty \prod_{i=1}^{n-p} dy_i\, y_i \, e^{-t\sum_iy_i} |\Delta(y)|^4
    &= \left(\frac{1}{t}\right)^{2(n-p)^2}\Gamma(3)^{p-n} \prod_{k=1}^{n-p} \Gamma(2k) \Gamma(1+2k) \nonumber \\
    &\simeq \left(\frac{1}{t}\right)^{2(n-p)^2}\Gamma(3)^{p} \prod_{k=n-p+1}^n \left(\Gamma(2k) \Gamma(1+2k)\right)^{-1}.
\end{align}
After taking the ${n \choose p}$ factor into account, only $p=0,1$ terms survive at order $n^2$. 
On the other hand, the $x$-integral for $p=0$ is trivially 1. 
For $p=1$, it is
\begin{align}
    \int_{1}^\infty dx_1\, x_1(x_1 - 1)  \, e^{-tx_1} = \frac{e^{-t} (t+2)}{t^3}.
\end{align}
Thus, we have
\begin{align}
    Y_{\text{Sp}(n)}(|\omega|^2)
    &\simeq  e^{n\frac t2} 
    \left(  \frac{1}{t^{2n^2}}
    - \frac{4n^2 e^{-t} \left( t+2 \right)}{t^5} \right) 
    \nonumber \\
    &\simeq
    \left(1 + \frac{nt}{2} + \frac{n^2 t^2}{8} - 2n^2 \ln t - \frac{4n^2e^{-t}}{t^4} \right).
\end{align}
We rewrite Eq.~\eqref{eqn:two-point-from-Zn} in terms of $\partial_\omega, \partial_z, \partial_{\bar\omega}$ and $ \partial_{\bar z}$ to obtain
\begin{align}
    \pi^2 R_2(z_1,\bar{z}_1, z_2, \bar{z}_2)
    &= \lim_{n\rightarrow 0} \frac{1}{n^2} ( \partial^2_\omega - \partial^2_{2z} ) ( \partial^2_{\bar\omega} - \partial^2_{2\bar z} ) e^{4n|z|^2} \left(1 + \frac{nt}{2} + \frac{n^2 t^2}{8} - 2n^2 \ln t - \frac{4n^2e^{-t}}{t^4} \right) \nonumber\\
    &= 2 + \lim_{n\rightarrow 0} \frac{1}{n^2} \partial^2_\omega\partial^2_{\bar\omega} \left(1 + {n|\omega|^2} + \frac{n^2 |\omega|^4}{2} - 2n^2 \ln (|\omega|^2) - \frac{n^2e^{-2|\omega|^2}}{4|\omega|^8} \right).
\end{align}
Then, the two-point correlation function is obtained from Eq.~\eqref{eqn:two-point-from-Zn}, to leading order in $1/|\omega|$, as
\begin{align}
\label{two-pt AIdag}
    \pi^2 R_2^{\text{AI}^\dagger}(\omega) \simeq 4 \left( 1 -  \frac{e^{-2|\omega|^2}}{|\omega|^4} \right).
\end{align}
A similar saddle-point integral was also calculated in Ref.~\cite{Yurkevich_Lerner_1999} (see Appendix~\ref{asec: Yurkevich-Lerner} for details).
However, the relevant parameter $\mu$ is assumed to be real in Ref.~\cite{Yurkevich_Lerner_1999}, while it should be imaginary in our calculations.
For comparison, we also analytically present the two-point correlation function $R_2 \left( z_1, \bar{z}_1, z_2, \bar{z}_2 \right)$ for $2\times 2$ non-Hermitian random matrices in class AI$^{\dag}$ in Appendix~\ref{asec: small}.

In Fig.\ \ref{fig:level-corr-AIdag},
we compare 
the result in Eq.~\eqref{two-pt AIdag}
with numerical calculations. 
We recall that the regime of validity for this result is $\sqrt g \ll |z_1-z_2|$ and $|z_1|,|z_2| \ll \sqrt{gN/2}$,
although in this regime the effect of level repulsion is very small and difficult to detect numerically.
A more interesting regime is $|z_1-z_2|\simeq \sqrt{g}$.
In this regime,
our result for the two-point correlation function appears to deviate from the numerics substantially. 
In contrast, 
for the case of 
the one-point function and the density of states, 
in Sec.\ \ref{AId DOS},
the analytical result
in Eq.~\eqref{AId DOS analytical}
appear to do a better job even 
outside of the regime of its validity.
Taking Eq.~\eqref{eq:AI-dag-2pt-microscopic} as a starting point, future work may 
investigate other contributions and 
provide important insights into the correlations of nearby eigenvalues in class AI$^\dagger$. 
Moreover, as we discussed previously  
at the end of Sec.\ \ref{AId DOS},
we should explore  
more systematic approaches to taking the replica limit.

\begin{figure}
    \centering    \includegraphics[width = 0.7\textwidth]{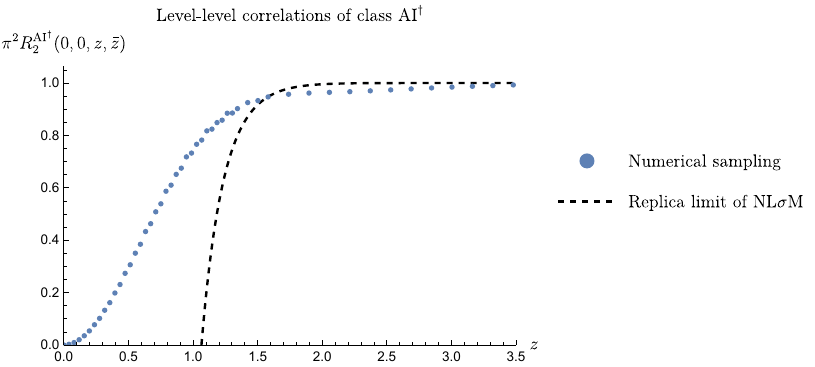}
    \caption{
    Comparison between
    the analytical result in Eq.~\eqref{two-pt AIdag}
    and the two-point correlation function obtained from numerical calculations. The numerical results are obtained by $2 \times 10^4$ realizations of $10^3 \times 10^3$ non-Hermitian random matrices in class AI$^\dagger$ sampled 
    according to Eq.~\eqref{eq:AIdag-gaussian} with $g=2$.}
    \label{fig:level-corr-AIdag}
\end{figure}

\section{NL$\sigma$M for class AII$^\dagger$}
\label{sec:AIIdagger-fermion}

\subsection{Replica space matrix integral for characteristic polynomials}

Non-Hermitian matrices in class AII$^\dagger$ respect TRS$^\dagger$ with sign $\mathcal T^2 = -1$. 
We choose the symmetry operator to be $\mathcal T = \Sigma^y \mathcal K$, with complex conjugation $\mathcal K$ and $\Sigma^y = \sigma^y\otimes \mathbb I_{N}$. 
Thus, the symmetry condition reads
\begin{equation}
H^\dagger = \mathcal T H \mathcal T^{-1},\quad \mathrm{i.e.}, \quad \Sigma^y H^T \Sigma^y = H. 
\end{equation}
Here, $H$ is a $2N\times 2N$ complex matrix.
The symmetry condition may equivalently be expressed by representing $H$ as follows,
\begin{align}
    H = \begin{pmatrix} a & b \\ c & a^T \end{pmatrix}
    \ \text{where} \
    a, b, c \in \mathbb C^{N\times N} \ \text{and}\ b^T = -b,\  c^T = -c.
\end{align}
The Gaussian probability measure on $H$ is given as follows,
\begin{align}
\label{eq:AIIdag-gaussian}
P(H) dH &= dH \exp[-g^{-1}\text{tr } H ^\dagger H] \nonumber
\\
&=\mathcal N_
{\text{AII}^\dagger}  \prod_{i,j =1}^Nda_{ij}da^*_{ij} \prod_{1\leq i<j \leq N}db_{ij}db^*_{ij}dc_{ij}dc^*_{ij}\exp[-2g^{-1}\left(\sum_{i, j = 1}^N |a_{ij}|^2  + \sum_{1\leq i<j \leq N} (|b_{ij}|^2 + |c_{ij}|^2)\right)],
\end{align}
where $g$ parameterizes the width of 
the Gaussian and the normalization $\mathcal N_{\text{AII}^\dagger}$ is defined by setting $\int dHP(H) = 1$.
As before, the general characteristic polynomial for $k$-point functions is defined as in Eq.~\eqref{eq:def-characteristic-polynomial}. 
We first focus on the $k=1$ case and then generalize to $k>1$. 
After rewriting the determinants in terms of Grassmann variables, the replica characteristic polynomial for the one-point function is given by
\begin{align} 
    Z^{(1)}_n(z,\bar z) = \int dH P(H) 
    \int d\psi d\bar\psi d\chi d\bar\chi
    \exp[-\bar\psi^i_a \left( z\delta^{ij} \delta^{ab} - H^{ij}\delta^{ab} \right) \psi^j_a   -\bar\chi^i_a \left( \bar z \delta^{ij} \delta^{ab} - (H^\dagger)^{ij}\delta^{ab} \right) \chi^j_a ],
\end{align}
where the indices $i,j$ now run from $1$ to $2N$, and $a,b$ run from $1$ to $n$. 
Now, we account for the symmetry of $H$ and distill the relevant fermionic degrees of freedom by noting
\begin{align}
    \bar\psi^i_a H^{ij} \psi^j_a
    =& -\tr(H \psi_a \bar\psi^T_a)
    = +\tr(H \Sigma^y \bar\psi_a \psi^T_a \Sigma^y)
    = -\tr(H \lceil \psi_a \bar\psi^T_a \rceil),
\end{align}
where we define the modified symmetric part of the fermion bilinear as
\begin{align}
    \lceil \psi_a \bar\psi^T_a \rceil
    = \frac 12 \big( \psi_a \bar\psi^T_a + (\Sigma^y \psi_a \bar\psi^T_a \Sigma^y )^T \big)
    = \frac 12 \big( \psi_a \bar\psi^T_a - \Sigma^y \bar\psi_a \psi^T_a \Sigma^y \big).
\end{align}
Indeed, the modified symmetric part respects TRS$^{\dag}$ in class AII$^\dagger$: 
$\Sigma^y \lceil \psi_a \bar\psi^T_a \rceil^T \Sigma^y  = \lceil \psi_a \bar\psi^T_a \rceil$. 
In perfect analogy to class AI$^\dagger$ (see Sec.~\ref{sec: AIdag - replica}), we integrate out the $H$ variables and are left with the following integral,
\begin{equation}
    Z^{(1)}_n(z,\bar z) = \int d\psi d\bar\psi d\chi d\bar\chi\ \exp[g\tr(\lceil \chi_a \bar\chi^T_a \rceil \lceil \psi_b \bar\psi^T_b \rceil)
    + \bar z \tr(\lceil \chi_a \bar\chi^T_a \rceil)
    + z\tr(\lceil \psi_a \bar\psi^T_a \rceil) 
    ].
\end{equation}
We expand the four-fermion term as 
\begin{equation}
\label{eq:quartic-term-AII-dagger}
    g\tr( \lceil  \chi_a \bar\chi^T_a \rceil   \lceil  \psi_b \bar\psi^T_b \rceil )
    = \frac 12 g \left[ -(\bar\psi_b^T \chi_a) (\bar\chi_a^T \psi_b) + (\psi_b^T \Sigma^y \chi_a) (\bar\chi_a^T \Sigma^y \bar\psi_b) \right].
\end{equation}
We introduce flavor-space matrices $Q$ and $ R \in \mathbb C^{n\times n}$ to decouple, respectively, the first and second terms in Eq.~\eqref{eq:quartic-term-AII-dagger},
\begin{equation}
    \exp[-\frac 12 g \Tr((\bar\psi^T \chi) (\bar\chi^T \psi))]
    \propto \int dQ \exp[
    -\frac 12 g^{-1} \Tr(QQ^\dagger) - \frac 12\Tr(\bar\psi^T\chi Q^\dagger) + \frac 12\Tr(Q\bar\chi^T\psi) ].
\end{equation}
Similarly, we have
\begin{equation}
    \exp[\frac 12 g \Tr((\psi^T \Sigma^y \chi) (\bar\chi^T \Sigma^y \bar\psi) )]
    \propto \int dR \exp[
    -\frac 12 g^{-1} \Tr(RR^\dagger) - \frac 12\Tr(R^\dagger \psi^T \Sigma^y \chi) - \frac 12\Tr(\bar\chi^T \Sigma^y \bar\psi R) ].
\end{equation}
Following similar steps as before, we express the characteristic polynomial as
\begin{align}
    \label{eq:AII-dag-Zn-k1}
    Z^{(1)}_n(z,\bar z)
    \simeq\ \int_{\mathbb R^{2n\times 2n}} d\mathcal Q\ e^{- g^{-1} \tr\mathcal Q^T\mathcal Q } \ \text{det}^{N}
    \begin{pmatrix}
         -z & \mathcal Q \\ \mathcal Q^T & \bar z
    \end{pmatrix}.
\end{align}
Now, we generalize this expression to the $k$-point characteristic polynomial. 
The procedure is largely identical to the one above, resulting in
\begin{equation}
    Z^{(k)}_n(z_1,\bar{z}_1, \cdots, z_k ,\bar z_k)
    = \int_{\mathbb R^{2nk\times 2nk}} d\mathcal Q\ e^{- g^{-1}\tr\mathcal Q^T \mathcal Q }\ \text{det}^{N} \begin{pmatrix}
        - Z & \mathcal Q \\ \mathcal Q^T & \bar Z
    \end{pmatrix},
    \label{eq:AII-dag-Zn-general-k}
\end{equation}
with
\begin{equation}
    Z = \text{diag}(z_1, \cdots , z_k)\otimes \mathbb{I}_{2n}.
\end{equation}
In the subsequent sections, we will compute this integral for $k=1$ and $k=2$ under suitable limits.

\subsection{One-point characteristic polynomial} 
    \label{sec: AIIdag - DoS}

We study the first-order characteristic polynomial $Z_n^{(1)}(z,\bar z)$ in more detail and use it to derive the density of states for class AII$^\dagger$. 
First, we extract only the dominant contribution for large $N$. 
In this limit, the integral in Eq.~\eqref{eq:AII-dag-Zn-k1} can be approximated by integrating over the saddle point. 
The saddle-point equation is
\begin{align}
    \mathcal Q^T \mathcal Q = 
    \left(
    gN - |z|^2\right)\mathbb{I}_{2n}.
\end{align}
For $|z|<\sqrt{gN}$, it is solved by $\mathcal Q = \sqrt{gN - |z|^2}\, O$, with $O \in \text O(2n)$. 
Dropping the overall factors irrelevant in the replica limit, we get $Z^{(1)}_n(z,\bar z) \simeq e^{2ng^{-1}|z|^2}$, which gives $\pi R_1(z, \bar{z}) = 2g^{-1}$ from Eq.~\eqref{eq:DoS-from-Zn}. 
This is consistent with numerical calculations and Girko's circular law~\cite{Girko-85}.

Now, let us study the characteristic polynomial $Z^{(1)}_n(z, \bar z)$ in more detail without any approximations. 
Below, we assume $g=1$ for simplicity. 
Notice that the integrand in Eq.~\eqref{eq:AII-dag-Zn-k1} remains invariant if we multiply $\mathcal Q$ by an orthogonal matrix on the left or on the right. 
Thus, we can reduce the matrix integral to an integral over singular values of $Q$. 
Let the singular-value decomposition of $\mathcal Q\in \mathbb R^{2n\times 2n}$ be
\begin{equation}
    \mathcal Q = U \Lambda V; \qquad U,V \in \text O(2n), \qquad
    \Lambda = \text{diag}(\lambda_1^{\frac 12 }, \cdots, \lambda_{2n}^{\frac 12 }) \quad \left( \lambda_a \geq 0 \right).
\end{equation}
The integration measure on $\mathcal Q$ is transformed into the following measure on $U$, $V$, and $\Lambda$,
\begin{align}
    dQ = dU dV |\Delta(\lambda)| \prod_{a=1}^{2n} d\lambda_a\, \lambda_a^{-\frac 12 }, \quad \Delta(\lambda) = \prod_{a>b}^n (\lambda_a-\lambda_b).
\end{align}
Since the integrand is independent of $U$ and $V$, the integrals over $U$ and $V$ simply give the volume of $\text O(2n)$. 
This is an irrelevant overall factor in the replica limit and hence ignored. 
The remaining integral is
\begin{align}
    \label{eq:AII-dag-1pt-microscopic}
    Z^{(1)}_n(z,\bar z)
    \simeq&\ \int_0^\infty \prod_{a=1}^{2n} d\lambda_a I(\lambda_a) \lambda_a^{-\frac 12 } |\Delta(\lambda)|, \qquad I(\lambda) = e^{-\lambda }\ \left( \lambda + |z|^2 \right)^N.
\end{align}
We refer the reader to Refs.\ \cite{Akemann-24, Forrester-24} 
for other results on this quantity.

\begin{figure}
    \centering
    \subfloat{\includegraphics[width = 0.4\textwidth]{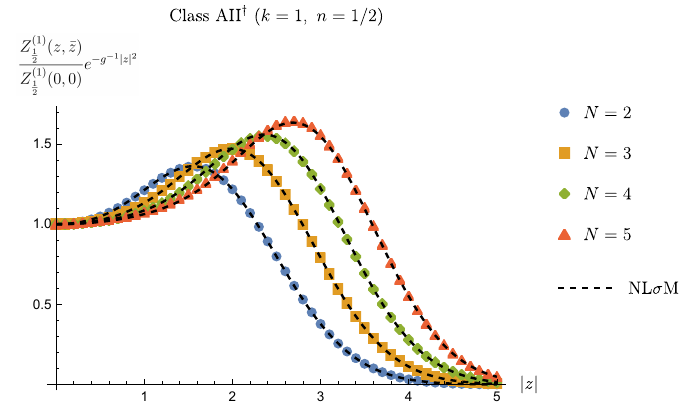}}
    \subfloat{\includegraphics[width = 0.4 \textwidth]{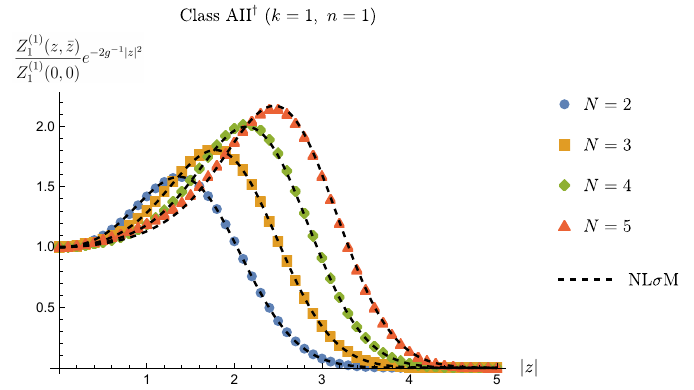}}
    \caption{Plot of $\frac{Z^{(1)}_n(z, \bar{z})}{Z^{(1)}_n(0,0)}e^{-2ng^{-1}|z|^2}$ as a function of $\left| z \right|$ for class AII$^\dagger$. 
    For each value of $N$, we sample $10^5$ realizations of 
    non-Hermitian random matrices according to Eq.~\eqref{eq:AIIdag-gaussian} with $g=2$. 
    The solid markers show the ensemble-averaged characteristic polynomials. 
    The black dashed curves show the same quantity calculated by Eq.~\eqref{eq:AII-dag-1pt-microscopic} obtained from the NL$\sigma$M.}
    \label{fig:charpol-AIIdag-smallN}
\end{figure}

In Fig.\ \ref{fig:charpol-AIIdag-smallN},
we plot 
the analytical 
expression in Eq.~\eqref{eq:AII-dag-1pt-microscopic}
of 
the first-order characteristic 
polynomial 
for various values of $N$ and $n$
as a function of $|z|$.
We also compare the analytical result
in Eq.~\eqref{eq:AII-dag-1pt-microscopic}
with numerical results 
where we numerically generate 
non-Hermitian random matrices from the Gaussian ensemble in Eq.~\eqref{eq:AIIdag-gaussian}
and calculate the characteristic polynomial.
For all values of $N$ and $n$ we studied,
the analytical and numerical results are consistent.

\subsubsection{Density of states}
\label{Density of states AIId}

As in the case of class AI$^\dagger$, we consider the large-$N$ limit and approximate $I(\lambda)$ with an un-normalized Gaussian: 
$I(\lambda) \simeq N^N e^{-\lambda_{\text{sp}}} \exp(-\frac{(\lambda - \lambda_{\text{sp}})^2}{2N}  )$. 
Here, $\lambda_{\text{sp}} = N - |z|^2$ is the solution to the saddle-point equation $\partial_\lambda \ln I(\lambda) = 0$. 
Rescaling $\lambda \rightarrow \lambda_{\text{sp}}\lambda$, we obtain
\begin{align}
    Z^{(1)}_n(z,\bar z)
    \simeq&\ e^{-2n \lambda_{\text{sp}}} \lambda_{\text{sp}}^{2n^2} \int_0^\infty \prod_{a=1}^{2n} \frac{d\lambda_a}{\sqrt{|\lambda_a|}} \ \exp(- \frac{\lambda_{\text{sp}}^2}{2N} \left(\lambda_a  - 1 \right)^2) |\Delta(\lambda)|.
\end{align}
We rewrite the range of integration as $\int_0^\infty d\lambda = \int_{-\infty}^\infty d\lambda - \int_{-\infty}^0 d\lambda$. 
We consider the regime where $\frac{\lambda_{\text{sp}}^2}{2N}$ is large enough so that $x$ and $y$ accumulate near the maxima of the weight function, i.e., $0$ and $1$, respectively.
As such, we can approximate $\Delta(\lambda) \simeq \Delta(x)\Delta(y)$, and hence the $x$ and $y$ variables become decoupled, as follows:
\begin{align}
    Z^{(1)}_n(z,\bar z)
    &\simeq e^{-2n \lambda_{\text{sp}}} \lambda_{\text{sp}}^{2n^2} \sum_{p} \left( -1 \right)^p {2n \choose p} \int_{-\infty}^0 \prod_{a=1}^p \frac{dx_a}{\sqrt{|x_a|}} \exp(- \frac{\lambda_{\text{sp}}^2}{2N} \left(x_a  - 1 \right)^2)|\Delta(x)| \nonumber \\
    &\qquad\qquad \times \int_{-\infty}^\infty \prod_{a=1}^{2n-p} \frac{dy_a}{\sqrt{|y_a|}} \exp(- \frac{\lambda_{\text{sp}}^2}{2N} \left(y_a  - 1 \right)^2) |\Delta(y)|.
\end{align} 
Furthermore, in the $y$-integral, since the Gaussian is narrowly peaked at $1$, we replace the factor of $|y_a|^{-\frac 12}$ in the integrand with $1$. The remaining integral is a Selberg integral, evaluated exactly as
\begin{align}
    \int_{-\infty}^\infty\prod_{a=1}^{2n-p} dy_a\exp(- \frac{\lambda_{\text{sp}}^2}{2N} \left(y_a  - 1 \right)^2) |\Delta(y)|
    &\simeq \left(\frac{\lambda_{\text{sp}}}{\sqrt{N}}\right)^{- \frac{(2n-p+1)(2n-p)}{2} } 2^{\frac 32 (2n-p)} \prod_{a=1}^{2n-p} \Gamma\left( 1 + \frac 12 a \right).
\end{align}
We should check the coefficient of the $p$th term in this expansion and identify which terms survive in the limit $n\rightarrow 0$. 
The coefficient is ${2n \choose p} \prod_{a=1}^{2n-p} \Gamma\left( 1 + \frac 12 a \right)$. 
At order $n$, the coefficients are $1$, $2n$, and $-n/\sqrt{\pi}$ for $p=0$, $1$, and $2$, respectively, and zero for all $p\geq 3$. 
On the other hand, there is no $x$-integral for $p=0$. 
For $p=1$, it is
\begin{equation}
    \int_{-\infty}^0 \frac{dx_1}{\sqrt{|x_1|}} \exp(- \frac{\lambda_{\text{sp}}^2}{2N} \left(x_1  - 1 \right)^2) \simeq \frac{\sqrt{\pi N}}{\lambda_{\text{sp}}}\exp(-\frac{\lambda_{\text{sp}}^2}{2N}) \quad \text{for}\quad \frac{\lambda_{\text{sp}}^2}{2N} \gg 1.
\end{equation}
For $p=2$, it is
\begin{equation}
    \int_{-\infty}^0 \frac{dx_1}{\sqrt{|x_1|}} \int_{-\infty}^0 \frac{dx_2}{\sqrt{|x_2|}} \exp(- \frac{\lambda_{\text{sp}}^2}{2N} \left[\left(x_1  - 1 \right)^2 + \left(x_2  - 1 \right)^2\right]) |x_1 - x_2| \simeq \frac{2N^2}{\lambda_{\text{sp}}^4}\exp(-\frac{\lambda_{\text{sp}}^2}{N}) \quad \text{for}\quad \frac{\lambda_{\text{sp}}^2}{2N} \gg 1.
\end{equation}
Putting it all together, we have
\begin{align}
    Z^{(1)}_n(z,\bar z)
    &\simeq  1 - 2 n \lambda_{\text{sp}} - n \ln \lambda_{\text{sp}} - \frac{n}{\lambda_{\text{sp}}} \sqrt{\frac{\pi N}{2}} \exp(-\frac{\lambda_{\text{sp}}^2}{2N}) -  \frac{n}{4\lambda_{\text{sp}}^5} \sqrt{\frac{N^5}{\pi}} \exp(-\frac{\lambda_{\text{sp}}^2}{N}) + \mathcal O(n^2).
\end{align}
Plugging this into Eq.~\eqref{eq:DoS-from-Zn} and keeping the leading-order terms, we obtain
\begin{align}
\label{DOS AIIdag}
    \pi 
    R_1(z, \bar{z}) = 
    2 +\frac{1}{4 u^2} - \sqrt{2\pi} e^{-2 u^2} \left(u +\frac{1 }{4u} + \frac{1}{8 u^3}\right)-\frac{e^{-4 u^2}}{8 \sqrt{\pi } u^3} + \mathcal O\left(\frac{1}{u^5}\right),
\end{align}
where 
we introduce $ u = \sqrt N - |z|$ as the distance of $z$ from the edge of the spectrum and use
$\lambda_{\text{sp}} \simeq 2 u \sqrt N$.
Similar to 
Eq.\ \eqref{AId DOS analytical},
we expect that the fermionic replica method can only reproduce the tail of the density of states for $|z|<\sqrt{gN}$ \cite{Nishigaki-02}.

\begin{figure}
    \centering    \includegraphics[width = 0.7\textwidth]{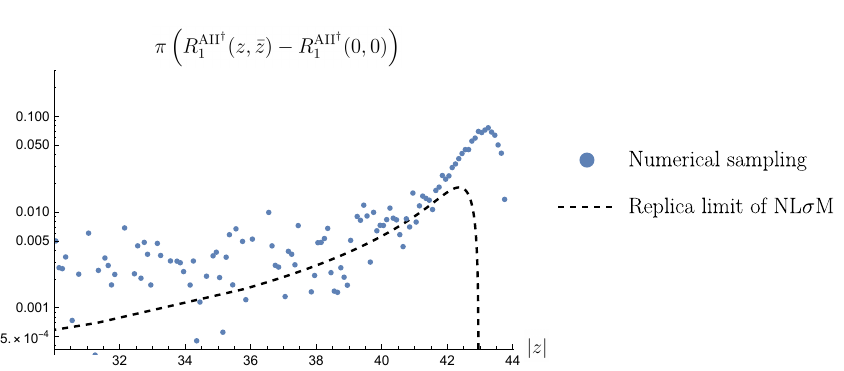}
    \caption{
    Comparison between
    the analytical result in Eq.~\eqref{DOS AIIdag}
    and the density of states 
    obtained from numerical calculations. 
    The numerical results are obtained by $10^4$ realizations of $\left( 2\times 10^3 \right) \times \left( 2\times 10^3 \right)$ non-Hermitian random matrices in class AII$^\dagger$ sampled according to Eq.~\eqref{eq:AIIdag-gaussian} with $g=2$.
    Note that this is a logarithmic plot. 
    This might cause the spread of the numerics in the plot.}
    \label{fig:DoS-AIIdag}
\end{figure}



In Fig.\ \ref{fig:DoS-AIIdag},
we compare the analytical result from 
Eq.\ \eqref{DOS AIIdag},
with the density of states obtained from numerical calculations. Notably, the positive term $+1/4u^2$ 
in Eq.\ \eqref{DOS AIIdag}
is unique to class AII$^{\dag}$ and does not appear in class A or AI$^{\dag}$. 
As seen in Fig.\
\ref{fig:DoS-AIIdag},
this term may explain the observed trend in the numerical data—specifically, the increase in the density of states near the edge of the spectrum.
Similar to the discussion in 
Sec.\ \ref{AId DOS}, 
two caveats should be noted: (i) the validity of Eq.\ \eqref{DOS AIIdag}, which holds only for $|z| \ll \sqrt{gN} \approx 44.72$, and (ii) the spread of numerical data due to the logarithmic scale of the plot. A more precise comparison with large-scale numerics, including corrections beyond the saddle-point approximation, as well as a more systematic approach to taking the replica limit, are left for future study. 
 
\subsection{Two-point characteristic polynomial}
    \label{sec: AIIdag - 2nd}

We recall the replica space matrix integral for the second-order characteristic polynomial
in Eq.\ \eqref{eq:AII-dag-Zn-general-k}.
For large $N$, we can approximate the $\mathcal Q$-integral by an integral over the saddle-point manifold of $\mathcal Q$. 
Unlike our discussion in Sec.~\ref{sec: AIIdag - DoS} for the density of states, $Z$ and $\mathcal Q$ do not commute, which makes the saddle-point equation more involved. 
However, in the regime $z_{1,2} \ll \sqrt{gN}$, it is simplified to
\begin{align}
    \mathcal Q^T \mathcal Q = gN\, 
    \mathbb{I}_{4n},
\end{align}
which is solved as $\mathcal Q = \sqrt{gN} O$ with an orthogonal matrix $O\in \text{O}(4n)$.
The integral over the saddle-point manifold is
\begin{align}
    Z^{(2)}_n(z_1, \bar z_1, z_2, \bar z_2) 
    &= \int_{\text{O}(4n)} dO\ e^{-4nN} \text{det}^{N}
    \begin{pmatrix}
        - Z &  \sqrt{gN} O \\ \sqrt{gN} O^T & \bar{Z} 
    \end{pmatrix} \nonumber \\
    &\simeq \int_{\text{O}(4n)} dO \exp[ g^{-1} \tr(O^T Z O  \bar{ Z} )].
\end{align}
Here, we introduce $z = (z_1 + z_2)/2$, $\omega = (z_1 - z_2)$, and $s=\text{diag}(1, - 1)\otimes \mathbb I_{2n}$, and then have $Z = z \mathbb I_{4n} + \frac 12 \omega s$, leading to
\begin{align}
    Z^{(2)}_n(z_1, \bar z_1, z_2, \bar z_2) 
    &\simeq e^{g^{-1} 4n |z|^2}\int_{\text{O}(4n)} dO \exp[ g^{-1} \tr(\frac{|\omega|^2}{4} O^T s O s )].
\end{align}
When we define $W=O^T s O$, this change of variables maps the integration manifold to the orthogonal Grassmannian, 
\begin{align}
    Z^{(2)}_n(z_1, \bar z_1, z_2, \bar z_2) 
    &\simeq e^{g^{-1} 4n |z|^2}\int_{\text{O}(4n)/[\text{O}(2n)\times \text{O} (2n)]} dW \exp[ g^{-1} \tr(\frac{|\omega|^2}{4} W s )].
\end{align}

\subsubsection{Cosine-sine decomposition}
\begin{figure}
    \centering
    \subfloat{\includegraphics[width = 0.4 \textwidth]{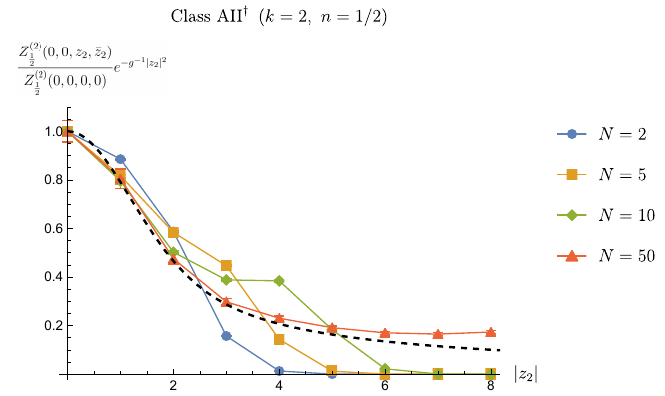}}
    \subfloat{\includegraphics[width = 0.4 \textwidth]{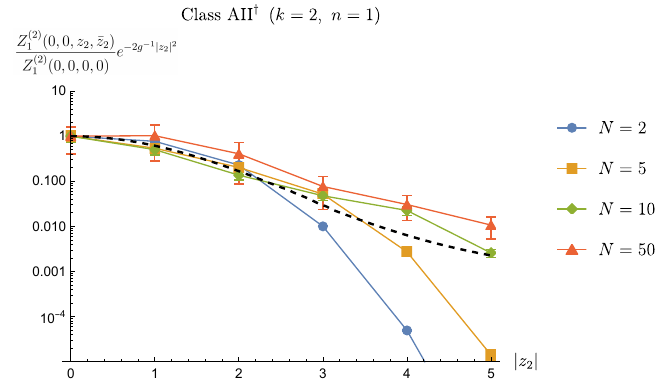}}
    \caption{Plots of the two-point characteristic polynomial in class AII$^\dagger$. We plot the expression 
    $\frac{Z^{(2)}_n(0,0,z_2, \bar{z}_2)}{Z^{(2)}_n(0,0,0,0)}
    e^{-2ng^{-1}|z_2|^2}$ instead of $Z^{(2)}_n(0,0,z_2, \bar{z}_2)$ for better visualization. For each value of $N$, we sample $2\times 10^6$ realizations of non-Hermitian random matrices in class AII$^\dagger$. In all cases, we use $g=2$. 
    The solid markers show the ensemble-averaged characteristic polynomials. The black dashed curves show the same quantity calculated by the NL$\sigma$M for large $N$ in Eq.~\eqref{eq:AII-dag-2pt-microscopic}.}
    \label{fig:charpol-2pt-AIIdag}
\end{figure}

We further simplify the above integrals by using the CS decomposition of the orthogonal group O$(4n)$,
\begin{equation}
    O = 
    \begin{pmatrix} u_1 & 0 \\ 0 & u_2 \end{pmatrix}
    \begin{pmatrix} \cos \Theta & \sin\Theta \\ -\sin \Theta & \cos \Theta \end{pmatrix}
    \begin{pmatrix} v_1 & 0 \\ 0 & v_2 \end{pmatrix},
\end{equation}
with 
\begin{equation}
    \Theta = \text{diag}(\theta_1, \cdots, \theta_{2n}), \qquad \theta_i \in [0,\pi), \qquad u_{1,2}, v_{1,2} \in \text O (2n).
\end{equation}
The integrand is only a function of $\Theta$, not dependent on $u$ and $v$,
\begin{align}
    \exp(\frac{|\omega|^2}{4}\tr(O s O^T s)) 
    = \exp(\frac{|\omega|^2}{2}\cos 2\Theta ),
\end{align}
where we have set $g=1$ to lighten the notation. Now, we express the Haar measure on $\text O(4n)$ in terms of $u,v$, and $\Theta$. Due to the invariance properties of the Haar measure, we can assume that $u_{1,2}$ and $v_{1,2}$ are in the neighborhood of identity. 
Thus, the Haar measure in terms of $u$, $v$, and $\Theta$ is given as 
\begin{align}
    \bigwedge_{4n\geq i>j \geq 1} (O^T dO)_{ij} 
    = \bigwedge_{k>l}(d v_1)_{kl} (d v_2)_{kl} \bigwedge_m d\theta_m \bigwedge_{i>j} (d u_1)_{ij} \wedge (d u_2)_{ij} \Delta(\cos 2\Theta).
\end{align}
For $u$ and $v$, the measure is just the Haar measure on $\text O(2n)$, which only contributes an irrelevant overall factor. 
The relevant part is the measure on $\Theta$.
Thus, we finally have
\begin{equation}
    Z^{(2)}_n(z_1, \bar z_1, z_2 \bar z_2) 
    \simeq e^{4n |z|^2} \int_0^\pi d\Theta\ |\Delta(\cos 2\Theta)| \exp(\frac{|\omega|^2}{2}\sum_i \cos 2\theta_i).
\end{equation}
Introducing $\lambda_i = \sin(\theta_i)^2$ and $t=|\omega|^2$, we further have
\begin{align}
    \label{eq:AII-dag-2pt-microscopic}
    Z^{(2)}_n(z_1, \bar z_1, z_2, \bar z_2) 
    &\simeq e^{ 4n |z|^2} Y_{\text O (2n)}(|\omega|^2), \\
    Y_{\text O (m)}(t) &=  e^{mt/2} \int_{0}^1 \prod_{i=1}^{m} d\lambda_i \ (\lambda_i(1-\lambda_i))^{-\frac 12} e^{-t\sum_i \lambda_i}  |\Delta(\lambda)|.
\end{align}
In Fig. \ref{fig:charpol-2pt-AIIdag}, we plot the above analytical result for the second-order characteristic polynomial as a function of $|z_2 - z_1|, N$, and $n$ and compare this with numerically generated data. For the nuremcis, we sample non-Hermitian random matrices in class AII$^\dagger$ from the Gaussian ensemble in Eq.\ \eqref{eq:AIIdag-gaussian} and calculate
the characteristic polynomial by exact diagonalization. Note that while Eq.\ \eqref{eq:AII-dag-2pt-microscopic} is valid only for large $N$, it allows us to access the small $|z_1 - z_2|$ regime, which is relevant for level repulsion. Indeed, in Fig. \ref{fig:charpol-2pt-AIIdag} we see that while the numerics for small $N$ deviates from
Eq.\ \eqref{eq:AII-dag-2pt-microscopic} for large $z_2$, it is consistent with Eq.\ \eqref{eq:AII-dag-2pt-microscopic} for large enough $N$ and for small enough $z_2$.

\subsubsection{Two-point correlation function}

\label{Two-point correlation function AIId}

As before, we extend the range of integration from $[0,1]$ to $[0,\infty)$ and deform the integration contour for each $\lambda_i$ as $\int_0^1 d\lambda \rightarrow \int_{0}^{\infty} d\lambda - \int_{1}^{\infty} d\lambda$.
Now, we make several approximations. 
We observe that in each branch of the contour, the dominant contribution comes from the endpoints $0$ and $1$.
At these endpoints, $(\lambda_i(1-\lambda_i))^{-\frac 12}$ can be approximated as $\lambda^{-\frac 12}$ and $(1-\lambda)^{-\frac 12}$. 
We can also factorize the Vandermonde determinant such that variables on different contours are decoupled.
Introducing $t=|\omega|^2$, we thus have
\begin{align}
    Y_{\text O (m)}(t)
    =& e^{mt/2}\sum_{p=0}^{m} {m \choose p}(-1)^p \int_{0}^{\infty} \prod_{i=1}^{m-p} dy_i \ |y_i|^{-\frac 12} e^{-t\sum_i y_i}  |\Delta(y)|  \times
    e^{-pt}\int_{0}^{\infty} \prod_{i=1}^{p} dx_i \ |x_i|^{-\frac 12} e^{-t\sum_i x_i}  |\Delta(x)|.
\end{align}
The $y$-integral is a Selberg integral evaluated as
\begin{align}
    \int_{0}^{\infty} \prod_{i=1}^{m-p} dy_i \ |y_i|^{-\frac 12} e^{-t\sum_i y_i}  |\Delta(y)|
    &= t^{-(m-p)^2/2} \frac{1}{\Gamma(\frac 32)^{m-p}} \prod_{k=1}^{m-p} \Gamma\left( \frac k2\right) \Gamma\left(1 + \frac k2\right).
\end{align}
The combinatorial factor to be expanded in small $n$ is then
\begin{align}
    F^p_n = \frac{\Gamma(m+1)}{\Gamma(p+1)\Gamma(m-p+1)}
\prod_{k=m-p+1}^{m} \Gamma\left( \frac k2\right)^{-1} \Gamma\left(1 + \frac k2\right)^{-1}.
\end{align}
At order $n^2$, the non-zero terms are $p=0,1,2$. The relevant $x$-integrals for $p=1$ and $2$ are respectively
\begin{align}
&
    \int_{0}^{\infty} dx_1 \ x_1^{-\frac 12} e^{-t x_1}
    = \sqrt{\frac \pi t},
    \\
    &
    \int_{0}^{\infty} dx_1 dx_2 \ (x_1 x_2)^{-\frac 12} e^{-t(x_1+x_2)}  |x_1-x_2|
    = \frac{2}{t^2}.
\end{align}
Thus, to leading order in $t$, we have
\begin{equation}
    Y_{\text O (m)}(t)
    \simeq
    1 + \frac{mt}{2} + \frac{m^2 t^2}{4} - \frac{m^2}{2} \ln t
    - \frac{m^2 \pi}{4} \frac{ e^{-t} }{t}
    + \frac{m^2}{16} \frac{e^{-2t}}{t^4} + \mathcal O(m^3).
\end{equation}
Now we use Eq.~\eqref{eqn:two-point-from-Zn} to calculate the two-point function. Notice once again that the factorization of $Z^{(2)}_n$ into separate $z$-dependent and $\omega$-dependent parts yields
\begin{align}
\label{R2andY_AIIdag}
    \pi^2 R_2(z_1,\bar{z}_1, z_2, \bar{z}_2)
    &= \lim_{n\rightarrow 0} \frac{1}{n^2} ( \partial^2_\omega - \partial^2_{2z} ) ( \partial^2_{\bar\omega} - \partial^2_{2\bar z} ) e^{4n|z|^2} Y_{\text{O}(2n)}(|\omega|^2) \nonumber\\
    &= 2 + \lim_{n\rightarrow 0} \frac{1}{n^2} \partial^2_\omega\partial^2_{\bar\omega} Y_{\text{O}(2n)}(|\omega|^2).
\end{align}
Combining this with the approximation for $Y_{\text{O}(2n)}$ we obtain the following expression for the two-point function of class AII$^\dagger$
\begin{align}
\label{two-pt AIIdag}
    \pi^2 R_2(\omega) \simeq 4 - \pi |\omega|^2 e^{-|\omega|^2}.
\end{align}
This is compatible with the saddle-point integral in Ref.~\cite{Yurkevich_Lerner_1999} (see Appendix~\ref{asec: Yurkevich-Lerner} for details).
For comparison, we also analytically present the two-point correlation function $R_2 \left( z_1, \bar{z}_1,z_2, \bar{z}_2 \right)$ for $4\times 4$ non-Hermitian random matrices in class AII$^{\dag}$ in Appendix~\ref{asec: small}.
In Fig.\ \ref{fig:level-corr-AIIdag},
we compare the analytical result 
in Eq.~\eqref{two-pt AIIdag} and 
the numerically generated 
two-point correlation function. 

Similar remarks that we made for Fig.\ \ref{fig:level-corr-AIdag} apply here.
Namely, the analytical result 
in Eq.~\eqref{two-pt AIIdag}
deviates substantially from the numerics in the regime 
$|z_1-z_2|\lesssim \sqrt{g}$.
However, in future works, Eq.\ \eqref{eq:AII-dag-2pt-microscopic}
can be used as a starting point to study this regime.
Finally, as we discussed in Secs.~\ref{AId DOS},
\ref{Two-point correlation function},
and
\ref{Density of states AIId},
we should keep in mind that there may be more systematic ways of taking the replica limit.

%
\begin{figure}
    \centering    \includegraphics[width = 0.7\textwidth]{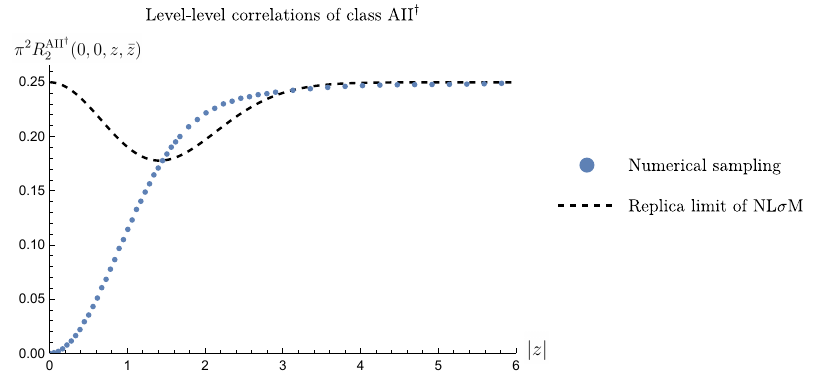}
    \caption{
     Comparison between
    the analytical result in Eq.~\eqref{two-pt AIIdag}
    and the two-point correlation function obtained from numerical calculations. The numerical results are obtained by $10^4$ realizations of $\left( 2 \times 10^3 \right) \times \left( 2 \times 10^3 \right)$ non-Hermitian random matrices in class AII$^\dagger$ sampled according to Eq.~\eqref{eq:AIIdag-gaussian} with $g=2$.
    }
    \label{fig:level-corr-AIIdag}
\end{figure}

\section{Conclusion}
    \label{sec: discussions}

In this work,
we investigated the spectral properties of Gaussian non-Hermitian 
random matrices in 
symmetry classes AI$^{\dagger}$
and AII$^{\dagger}$.
Using the fermionic replica NL$\sigma$Ms,
we computed the one-point and two-point 
characteristic polynomials. 
We compared our results with 
finite-$N$ numerical calculations and found good agreements.
The method developed in this work can be applied to other symmetry classes of non-Hermitian RMT~\cite{CKKR-24}.
Taking the replica limit, 
we also calculated the density of states and the two-point correlation functions. 
As discussed in Sec.\ \ref{AId DOS},
there remains room for improvement on these calculations.
Another important challenge is to establish the universality of level statistics. 
Our derivation of the NL$\sigma$Ms
yields different target spaces for different symmetry classes which provides support to the universality conjecture. However, our treatment is currently limited to Gaussian ensembles. Extending our treatment to non-Gaussian ensembles will be necessary to rigorously establish universality.
Further studies will be required to address this limitation.
Lastly, it is interesting to explore applications of our results to physical systems, 
such as Lindblad superoperators.

\bigskip
{\it Note added}.---A part of this work was presented in Ref.~\cite{Kulkarni-SCGP}.
When we were finalizing the draft, we learned about a related work~\cite{Akemann-24}.
After the initial submission of this work, we also learned about another related work~\cite{Forrester-24}.

\begin{acknowledgments}
We thank 
Ze Chen,
Giorgio Cipolloni,
and Jacobus Verbaarschot
for helpful discussion.
K.K. is supported by MEXT KAKENHI Grant-in-Aid for Transformative Research Areas A ``Extreme Universe" No.~24H00945.
S.R. is supported by Simons Investigator Grant from the Simons Foundation (Grant No.~566116). 
We gratefully acknowledge
support from the Simons Center for Geometry and Physics, Stony Brook University at which some of the research for this work was performed.
\end{acknowledgments}

\appendix

\section{Saddle-point integral}
    \label{asec: Yurkevich-Lerner}

\subsection{Class AI$^\dagger$}

We calculate the saddle-point integral using the results from Ref.~\cite{Yurkevich_Lerner_1999}, which computed the following sigma model matrix integral:
\begin{align}
    Y_{\text{S}(n)}(-i\mu) = \int_{\text{S}(2n)/[\text{S}(n) \times \text{S}(n)]} \mathcal D W \exp[-\frac{i\mu\alpha}{4}\Tr(s W)].
\end{align}
The group S$(n)$ can be O$(n)$, U$(n)$, or Sp$(n)$, corresponding to $\alpha = 2$, $\alpha = 2$, or $\alpha = 1$, respectively. 
When we define $W=U^\dagger sU$ and substitute $\mu = 2i g^{-1} |\omega|^2$, this becomes almost the same matrix integral in Eq.~\eqref{saddle-point-integral}. 
We note, however, that $\mu$ is assumed to be real in Ref.~\cite{Yurkevich_Lerner_1999}, while it should be imaginary in our calculations.
Assuming that the measure is identical, we have the following expression for 
$Z_n^{(2)}$ up to $z$-independent factors that reduce to $1$ in the replica limit,
\begin{align}
    Z_n^{(2)}
    (z_1, \bar z_1, z_2, \bar z_2)
    = e^{4g^{-1}n|z|^2} Y_n(g^{-1} |\omega|^2).
\end{align}
We rewrite $R_2$ in terms of $Y_n$ and $\partial_\omega, \partial_z$, and so on, and assume $g=1$ for simplicity, leading to
\begin{align}
\label{R2andY}
    \pi^2 R_2(z_1,\bar{z}_1, z_2, \bar{z}_2)
    &= \lim_{n\rightarrow 0} \frac{1}{n^2} ( \partial^2_\omega - \partial^2_{2z} ) ( \partial^2_{\bar\omega} - \partial^2_{2\bar z} ) e^{4n|z|^2} Y_{n}(|\omega|^2) \nonumber\\
    &= 2 + \lim_{n\rightarrow 0} \frac{1}{n^2} \partial^2_\omega\partial^2_{\bar\omega} Y_{n}(|\omega|^2).
\end{align}
Hence, we only need to keep terms of order $n^2$ in $Y_n$. 
For the symplectic case, at large $\mu$, we use the $\beta=1$ result from Ref.~\cite{Yurkevich_Lerner_1999}:
\begin{align}
    Y_{n\rightarrow 0}(-i\mu) = 1 - i\mu n + n^2 \left( -\frac{\mu^2}{2} - 2 \ln \mu + \frac{e^{2i\mu}}{4\mu^4} \right).
\end{align}
In our calculations, we instead need $\mu = i\nu,\ \nu\in \mathbb R_{>0}$. The sign of the $e^{2i\mu}/{\mu^4}$ term is ambiguous for $\mu = i\nu$. 
We here choose the sign based on a physical expectation and replace $\ln \mu \rightarrow \ln|\mu|$. 
Then, we have
\begin{align}
    Y_{n\rightarrow 0}(\nu) = 1 + \nu n + n^2 \left( \frac{\nu^2}{2} - 2 \ln \nu -  \frac{e^{-2\nu}}{4\nu^4}\right)
\end{align}
for $\nu \in \mathbb R_{>0}$.
By using the above expression for $Y_n$, the term $z^2 \partial_\omega^2  + \bar z^2 \partial_{\bar\omega}^2$
should vanish in the $n\rightarrow 0$ limit. 
We are thus left with 
\begin{align}
    \pi^2 R_2(z_1,\bar{z}_1,z_2,\bar{z}_2)
    &=\ 2 + \partial_\omega^2 \partial_{\bar\omega}^2 \left( \frac{\nu^2}{2} - 2 \ln \nu -  \frac{e^{-2\nu}}{4\nu^4} \right) \nonumber \\
    &=\
     4 - \frac{4 e^{- 2|\omega|^2}}{|\omega|^4} + \mathcal{O}\left(\frac{e^{-|\omega|^2}}{|\omega|^2}\right) + \mathcal{O}\left(\frac{e^{-2|\omega|^2}}{|\omega|^4}\right),
\end{align}
where we introduce $\omega=z_1-z_2$ and $\nu = g^{-1}\omega\bar\omega$.
While we here assume $g=1$ for simplicity, 
the $g$ dependence is recovered as
\begin{align}
    \pi^2 R_2(z_1,\bar{z}_1,z_2, \bar{z}_2) 
    = 4 g^{-2}\left( 1 
    - \frac{e^{- 2g^{-1}|\omega|^2}}{g^{-2}|\omega|^4} + \mathcal{O}\left(\frac{e^{-g^{-1}|\omega|^2}}{g^{-1}|\omega|^2}\right) + \mathcal{O}\left(\frac{e^{-2g^{-1}|\omega|^2}}{g^{-2}|\omega|^4}\right) \right).
\end{align}

\subsection{Class AII$^\dagger$}.

For the orthogonal case, at large $\mu$, we use the $\beta=4$ result from Ref.~\cite{Yurkevich_Lerner_1999}:
\begin{align}
    Y_{\text{O}(2n)}(-i\mu) = 1 - 2i\mu n + 4n^2 \left( -\frac{\mu^2}{2} -\frac 12 \ln \mu + \Gamma^2(3/2)\frac{e^{2i\mu}}{2\mu}  + \frac{e^{4i\mu}}{2^8\mu^4} \right) + \mathcal O(n^2)
\end{align}
for $\mu \in \mathbb R$.
As Eq.\ \eqref{R2andY},
we rewrite the two-point correlation function in terms of $z$ and $\omega$ derivatives and assume $g=1$ for simplicity, leading to
\begin{align}
    \pi^2 R_2(z_1,\bar{z}_1,z_2, \bar{z}_2)
    &= 2 + \lim_{n\rightarrow 0} \frac{1}{n^2} \partial^2_\omega\partial^2_{\bar\omega} Y_{\text O(2n)}(|\omega|^2/2).
\end{align}
Thus, we have
\begin{align}
    \pi^2 R_2(z_1,\bar{z}_1,z_2, \bar{z}_2) 
    &=\ 2 + 4\partial_\omega^2 \partial_{\bar\omega}^2 \left( \frac{|\omega|^4}{8} -\frac 12 \ln \left( \frac{i|\omega|^2}{2} \right) 
     - 
     \frac \pi 4 \frac{e^{-|\omega|^2}}{|\omega|^2}  + \frac{e^{-2|\omega|^2}}{2^4|\omega|^8}  \right) \nonumber \\
    &=\ 4 - \pi |\omega|^2 e^{-|\omega|^2} + e^{-2 |\omega|^2} \mathcal O\left(\frac{1}{|\omega|^6}\right).
\end{align}
Putting the $g$ dependence back by dimensional analysis, we have
\begin{equation}
    \pi^2 R_2(z_1,\bar{z}_1,z_2, \bar{z}_2) 
    =\ g^{-2}\left(4 - \pi g^{-1}|\omega|^2 e^{-g^{-1}|\omega|^2} + e^{-2 g^{-1} |\omega|^2} \mathcal O\left(\frac{1}{g^{-3}|\omega|^6}\right) \right).
\end{equation}
\section{Jacobian of quaternion SVD}
\label{asec:SVD}

In this appendix, we derive the Jacobian of the singular value decomposition (SVD) of a quaternion matrix. Let $Q$ be an $n\times n$ quaternion matrix. The SVD is given by
\begin{align}
    Q = U \Lambda V, \quad \Lambda = \text{diag}(\sqrt \lambda_1, \cdots, \sqrt \lambda_n),\ \lambda_i\in \mathbb R_{\geq 0}, \  V \in \text{Sp}(n), \ U\in \text{Sp}(n)/\text{Sp}(1)^{\oplus n}.
\end{align}
The Euclidean measure on $Q$ written as a differential form is $\bigwedge_{c}\bigwedge_{i,j} dQ^{(c)}_{i,j}$ where the index $c = 0, \cdots ,3$ represents the various components of the quaternion numbers. This measure is invariant under multiplying $Q$ on the left or right with a quaternion unitary matrix. We use this freedom to choose $U,V$ in the neighborhood of identity. We then have
\begin{align}
    dQ = dU \Lambda + \Lambda dV + d\Lambda.
\end{align}
Due to unitarity, we also have $dU = - dU^\dagger$ and $dV = -dV^\dagger$. In particular, this means that the diagonal of $dU$ is zero and the real part of the diagonal of $dV$ is zero. The Euclidean measure (up to an overall sign) is therefore given as
\begin{align}
    \bigwedge_{c=0}^3\bigwedge_{i,j} dQ^{(c)}_{i,j} &=
    \left(\bigwedge_{i} d\sqrt\lambda_i 
    \right) \left(\bigwedge_{c=1}^3\bigwedge_{i} dV^{(c)}_{ii} \sqrt\lambda_i
    \right) \left(\bigwedge_{c=0}^3\bigwedge_{i<j} (dU^{(c)}_{ij}\sqrt\lambda_j + \sqrt\lambda_idV^{(c)}_{ij})\wedge (dU^{(c)}_{ij}\sqrt\lambda_i + \sqrt\lambda_jdV^{(c)}_{ij})\right)
    \nonumber \\
    &= \left(\bigwedge_{i} \sqrt \lambda_i ^3 d\sqrt\lambda_i
    \right)
    \left(\bigwedge_{c=1}^3\bigwedge_{i} dV^{(c)}_{ii}
    \right)
    \left(\bigwedge_{c=0}^3\bigwedge_{i<j} dU^{(c)}_{ij} \wedge dV^{(c)}_{ij} (\lambda_i-\lambda_j)\right)
    \nonumber \\
     &= \ dU_{\text{Haar}} dV_{\text{Haar}}2^{-n}\bigwedge_{i}d\lambda_i \lambda_i
    \Delta(\lambda)^4.
\end{align}
\section{Two-point correlation functions of small non-Hermitian random matrices}
    \label{asec: small}

\begin{figure}[t]
\centering
\includegraphics[width=0.3\linewidth]{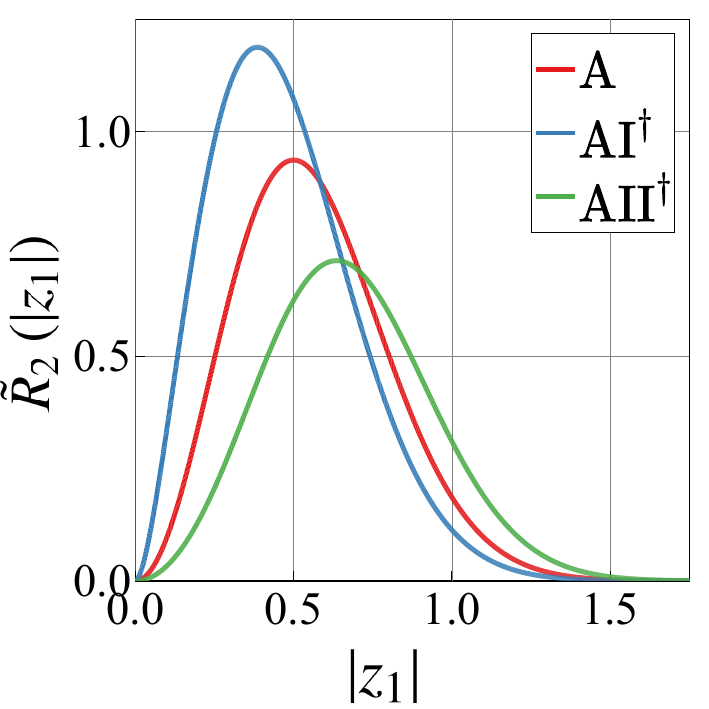} 
\caption{Two-point correlation functions $R_2 \left( z_1, z_2 \right)$ of $2\times 2$ ($4 \times 4$) non-Hermitian random matrices in classes A and AI$^{\dag}$ (AII$^{\dag}$).
The vertical axis $\tilde{R}_2 \left( \left| z_1 \right| \right)$ is defined by $R_2 \left( z_1, z_2 \right) \eqqcolon \delta \left( z_1 + z_2 \right) \tilde{R}_2 \left( \left| z_1 \right| \right)$.
The normalization constants are chosen as $C_2 = C_3 = C_5 = 1$.}
	\label{fig: NH-RMT-correlation-Wigner-surmise}
\end{figure}


In this appendix, 
we calculate the two-point correlation function of $2\times 2$ ($4 \times 4$) non-Hermitian random matrices in classes A and AI$^{\dag}$ (class AII$^{\dag}$).
Similar to 
the Wigner surmise, 
this gives a simple
analytical expression for the 
two-point function.
We should however 
keep in mind that 
the small-$N$ result 
may not be a good approximation 
to the large-$N$
limit.
See similar comments around Eq.\  
\eqref{eq:AI-dag-1pt-microscopic}.

In these cases,
non-Hermitian random matrices host only an opposite-sign pair of complex eigenvalues.
The two-point correlation function reads
\begin{align}
    R_2 \left( z_1, z_2 \right) = \delta \left( z_1 + z_2 \right) \ev{\mathrm{tr}\,\delta \left( z_1 - H \right)}.
\end{align}
Here, $\ev{\mathrm{tr}\,\delta \left( z_1 - z \right)}$ depends only on $\left|z_1\right|$ and is given as $2\ev{\delta \left( z_1 - z \right)}$ with an eigenvalue $z \in \mathbb{C}$ of $H$.
Hence, we have
\begin{align}
    R_2 \left( z_1, z_2 \right) 
    = \frac{\delta \left( z_1 + z_2 \right) \ev{\delta \left( \left| z_1 \right| - \left| z \right| \right)}}{\pi \left| z_1 \right|}.
\end{align}
On the other hand, the level-spacing distributions are defined as
\begin{align}
    p_{\rm s} \left( s \right) \coloneqq \ev{\delta \left( s - 2\left| z\right| \right)}. 
\end{align}
Consequently, we have (Fig.~\ref{fig: NH-RMT-correlation-Wigner-surmise})
\onecolumngrid
\begin{align}
    R_2 \left( z_1, z_2 \right) 
    = \frac{2 \delta \left( z_1 + z_2 \right) p_{\rm s} \left( 2 \left| z_1 \right| \right)}{\pi \left| z_1 \right|} 
    = \frac{2}{\pi} \delta \left( z_1 + z_2 \right) \begin{cases}
       16 C_2^4 \left| z_1 \right|^2 K_0\,( 4 C_2^2 \left| z_1 \right|^2 ) & ( \text{class AI}^{\dag} ); \\
       16 C_3^4 \left| z_1 \right|^2 e^{-4 C_3^2 \left| z_1 \right|^2} & ( \text{class A} ); \\
       ( 16 C_5^4 \left| z_1\right|^2/3 )\,( 1 + 4 C_5^2 \left| z_1 \right|^2 )\,e^{-4 C_5^2 \left| z_1 \right|^2}  & ( \text{class AII}^{\dag} ), \\
    \end{cases}
\end{align}
with the modified Bessel function of the second kind, $K_0 \left( x \right)$.
Here, we use the analytical results of $p_{\rm s} \left( s \right)$ in Ref.~\cite{Hamazaki-20} with arbitrary positive constants $C_2$, $C_3$, $C_5 \geq 0$.
For $\left|z_1 \right|$, $\left|z_2 \right| \ll 1$, we have
\begin{align}
    R_2 \left( z_1, z_2 \right) \simeq \frac{2}{\pi} \delta \left( z_1 + z_2 \right) \begin{cases}
       - 16 C_2^4 \left| z_1 \right|^2 \ln\,( 4 C_2^2 \left| z_1 \right|^2 ) & ( \text{class AI}^{\dag} ); \\
       16 C_3^4 \left| z_1 \right|^2 & ( \text{class A} ); \\
       16 C_5^4 \left| z_1\right|^2/3  & ( \text{class AII}^{\dag} ). \\
    \end{cases}
\end{align}

\bibliography{NH_NLSM_RMT.bib}

\begin{thebibliography}{89}%
\makeatletter
\providecommand \@ifxundefined [1]{%
 \@ifx{#1\undefined}
}%
\providecommand \@ifnum [1]{%
 \ifnum #1\expandafter \@firstoftwo
 \else \expandafter \@secondoftwo
 \fi
}%
\providecommand \@ifx [1]{%
 \ifx #1\expandafter \@firstoftwo
 \else \expandafter \@secondoftwo
 \fi
}%
\providecommand \natexlab [1]{#1}%
\providecommand \enquote  [1]{``#1''}%
\providecommand \bibnamefont  [1]{#1}%
\providecommand \bibfnamefont [1]{#1}%
\providecommand \citenamefont [1]{#1}%
\providecommand \href@noop [0]{\@secondoftwo}%
\providecommand \href [0]{\begingroup \@sanitize@url \@href}%
\providecommand \@href[1]{\@@startlink{#1}\@@href}%
\providecommand \@@href[1]{\endgroup#1\@@endlink}%
\providecommand \@sanitize@url [0]{\catcode `\\12\catcode `\$12\catcode
  `\&12\catcode `\#12\catcode `\^12\catcode `\_12\catcode `\%12\relax}%
\providecommand \@@startlink[1]{}%
\providecommand \@@endlink[0]{}%
\providecommand \url  [0]{\begingroup\@sanitize@url \@url }%
\providecommand \@url [1]{\endgroup\@href {#1}{\urlprefix }}%
\providecommand \urlprefix  [0]{URL }%
\providecommand \Eprint [0]{\href }%
\providecommand \doibase [0]{https://doi.org/}%
\providecommand \selectlanguage [0]{\@gobble}%
\providecommand \bibinfo  [0]{\@secondoftwo}%
\providecommand \bibfield  [0]{\@secondoftwo}%
\providecommand \translation [1]{[#1]}%
\providecommand \BibitemOpen [0]{}%
\providecommand \bibitemStop [0]{}%
\providecommand \bibitemNoStop [0]{.\EOS\space}%
\providecommand \EOS [0]{\spacefactor3000\relax}%
\providecommand \BibitemShut  [1]{\csname bibitem#1\endcsname}%
\let\auto@bib@innerbib\@empty
\bibitem [{\citenamefont {Wigner}(1951)}]{Wigner-51}%
  \BibitemOpen
  \bibfield  {author} {\bibinfo {author} {\bibfnamefont {E.~P.}\ \bibnamefont
  {Wigner}},\ }\bibfield  {title} {\bibinfo {title} {{On the statistical
  distribution of the widths and spacings of nuclear resonance levels}},\
  }\href {https://doi.org/10.1017/S0305004100027237} {\bibfield  {journal}
  {\bibinfo  {journal} {Math. Proc. Cambridge Philos. Soc.}\ }\textbf {\bibinfo
  {volume} {47}},\ \bibinfo {pages} {790} (\bibinfo {year} {1951})}\BibitemShut
  {NoStop}%
\bibitem [{\citenamefont {Wigner}(1958)}]{Wigner-58}%
  \BibitemOpen
  \bibfield  {author} {\bibinfo {author} {\bibfnamefont {E.~P.}\ \bibnamefont
  {Wigner}},\ }\bibfield  {title} {\bibinfo {title} {{On the Distribution of
  the Roots of Certain Symmetric Matrices}},\ }\href
  {https://doi.org/10.2307/1970008} {\bibfield  {journal} {\bibinfo  {journal}
  {Ann. Math.}\ }\textbf {\bibinfo {volume} {67}},\ \bibinfo {pages} {325}
  (\bibinfo {year} {1958})}\BibitemShut {NoStop}%
\bibitem [{\citenamefont {Akemann}\ \emph {et~al.}(2015)\citenamefont
  {Akemann}, \citenamefont {Baik},\ and\ \citenamefont
  {Di~Francesco}}]{Akemann:2011csh}%
  \BibitemOpen
  \bibinfo {editor} {\bibfnamefont {G.}~\bibnamefont {Akemann}}, \bibinfo
  {editor} {\bibfnamefont {J.}~\bibnamefont {Baik}},\ and\ \bibinfo {editor}
  {\bibfnamefont {P.}~\bibnamefont {Di~Francesco}},\ eds.,\ \href
  {https://doi.org/https://doi.org/10.1093/oxfordhb/9780198744191.001.0001}
  {\emph {\bibinfo {title} {{The Oxford Handbook of Random Matrix Theory}}}},\
  Oxford Handbooks in Mathematics\ (\bibinfo  {publisher} {Oxford University
  Press},\ \bibinfo {address} {Oxford},\ \bibinfo {year} {2015})\BibitemShut
  {NoStop}%
\bibitem [{\citenamefont {Byun}\ and\ \citenamefont
  {Forrester}(2024)}]{Byun-Forrester-book}%
  \BibitemOpen
  \bibfield  {author} {\bibinfo {author} {\bibfnamefont {S.-S.}\ \bibnamefont
  {Byun}}\ and\ \bibinfo {author} {\bibfnamefont {P.~J.}\ \bibnamefont
  {Forrester}},\ }\href
  {https://doi.org/https://doi.org/10.1007/978-981-97-5173-0} {\emph {\bibinfo
  {title} {Progress on the Study of the Ginibre Ensembles}}}\ (\bibinfo
  {publisher} {Springer},\ \bibinfo {year} {2024})\BibitemShut {NoStop}%
\bibitem [{\citenamefont {Konotop}\ \emph {et~al.}(2016)\citenamefont
  {Konotop}, \citenamefont {Yang},\ and\ \citenamefont
  {Zezyulin}}]{Konotop-review}%
  \BibitemOpen
  \bibfield  {author} {\bibinfo {author} {\bibfnamefont {V.~V.}\ \bibnamefont
  {Konotop}}, \bibinfo {author} {\bibfnamefont {J.}~\bibnamefont {Yang}},\ and\
  \bibinfo {author} {\bibfnamefont {D.~A.}\ \bibnamefont {Zezyulin}},\
  }\bibfield  {title} {\bibinfo {title} {{Nonlinear waves in
  $\mathcal{PT}$-symmetric systems}},\ }\href
  {https://doi.org/10.1103/RevModPhys.88.035002} {\bibfield  {journal}
  {\bibinfo  {journal} {Rev. Mod. Phys.}\ }\textbf {\bibinfo {volume} {88}},\
  \bibinfo {pages} {035002} (\bibinfo {year} {2016})}\BibitemShut {NoStop}%
\bibitem [{\citenamefont {El-Ganainy}\ \emph {et~al.}(2018)\citenamefont
  {El-Ganainy}, \citenamefont {Makris}, \citenamefont {Khajavikhan},
  \citenamefont {Musslimani}, \citenamefont {Rotter},\ and\ \citenamefont
  {Christodoulides}}]{Christodoulides-review}%
  \BibitemOpen
  \bibfield  {author} {\bibinfo {author} {\bibfnamefont {R.}~\bibnamefont
  {El-Ganainy}}, \bibinfo {author} {\bibfnamefont {K.~G.}\ \bibnamefont
  {Makris}}, \bibinfo {author} {\bibfnamefont {M.}~\bibnamefont {Khajavikhan}},
  \bibinfo {author} {\bibfnamefont {Z.~H.}\ \bibnamefont {Musslimani}},
  \bibinfo {author} {\bibfnamefont {S.}~\bibnamefont {Rotter}},\ and\ \bibinfo
  {author} {\bibfnamefont {D.~N.}\ \bibnamefont {Christodoulides}},\ }\bibfield
   {title} {\bibinfo {title} {{Non-Hermitian physics and PT symmetry}},\ }\href
  {https://doi.org/10.1038/nphys4323} {\bibfield  {journal} {\bibinfo
  {journal} {Nat. Phys.}\ }\textbf {\bibinfo {volume} {14}},\ \bibinfo {pages}
  {11} (\bibinfo {year} {2018})}\BibitemShut {NoStop}%
\bibitem [{\citenamefont {Ginibre}(1965)}]{Ginibre-65}%
  \BibitemOpen
  \bibfield  {author} {\bibinfo {author} {\bibfnamefont {J.}~\bibnamefont
  {Ginibre}},\ }\bibfield  {title} {\bibinfo {title} {{Statistical Ensembles of
  Complex, Quaternion, and Real Matrices}},\ }\href
  {https://doi.org/10.1063/1.1704292} {\bibfield  {journal} {\bibinfo
  {journal} {J. Math. Phys.}\ }\textbf {\bibinfo {volume} {6}},\ \bibinfo
  {pages} {440} (\bibinfo {year} {1965})}\BibitemShut {NoStop}%
\bibitem [{\citenamefont {Girko}(1985)}]{Girko-85}%
  \BibitemOpen
  \bibfield  {author} {\bibinfo {author} {\bibfnamefont {V.~L.}\ \bibnamefont
  {Girko}},\ }\bibfield  {title} {\bibinfo {title} {{Circular Law}},\ }\href
  {https://doi.org/10.1137/1129095} {\bibfield  {journal} {\bibinfo  {journal}
  {Theory Probab. Appl.}\ }\textbf {\bibinfo {volume} {29}},\ \bibinfo {pages}
  {694} (\bibinfo {year} {1985})}\BibitemShut {NoStop}%
\bibitem [{\citenamefont {Verbaarschot}\ \emph {et~al.}(1985)\citenamefont
  {Verbaarschot}, \citenamefont {Weidenm\"uller},\ and\ \citenamefont
  {Zirnbauer}}]{Verbaarschot:1985jn}%
  \BibitemOpen
  \bibfield  {author} {\bibinfo {author} {\bibfnamefont {J.~J.~M.}\
  \bibnamefont {Verbaarschot}}, \bibinfo {author} {\bibfnamefont {H.~A.}\
  \bibnamefont {Weidenm\"uller}},\ and\ \bibinfo {author} {\bibfnamefont
  {M.~R.}\ \bibnamefont {Zirnbauer}},\ }\bibfield  {title} {\bibinfo {title}
  {{Grassmann integration in stochastic quantum physics: The case of
  compound-nucleus scattering}},\ }\href
  {https://doi.org/10.1016/0370-1573(85)90070-5} {\bibfield  {journal}
  {\bibinfo  {journal} {Phys. Rep.}\ }\textbf {\bibinfo {volume} {129}},\
  \bibinfo {pages} {367} (\bibinfo {year} {1985})}\BibitemShut {NoStop}%
\bibitem [{\citenamefont {Sokolov}\ and\ \citenamefont
  {Zelevinsky}(1988)}]{Sokolov:1988ata}%
  \BibitemOpen
  \bibfield  {author} {\bibinfo {author} {\bibfnamefont {V.~V.}\ \bibnamefont
  {Sokolov}}\ and\ \bibinfo {author} {\bibfnamefont {V.~G.}\ \bibnamefont
  {Zelevinsky}},\ }\bibfield  {title} {\bibinfo {title} {{On a statistical
  theory of overlapping resonances}},\ }\href
  {https://doi.org/10.1016/0370-2693(88)90844-1} {\bibfield  {journal}
  {\bibinfo  {journal} {Phys. Lett. B}\ }\textbf {\bibinfo {volume} {202}},\
  \bibinfo {pages} {10} (\bibinfo {year} {1988})}\BibitemShut {NoStop}%
\bibitem [{\citenamefont {Sokolov}\ and\ \citenamefont
  {Zelevinsky}(1989)}]{Sokolov:1988df}%
  \BibitemOpen
  \bibfield  {author} {\bibinfo {author} {\bibfnamefont {V.~V.}\ \bibnamefont
  {Sokolov}}\ and\ \bibinfo {author} {\bibfnamefont {V.~G.}\ \bibnamefont
  {Zelevinsky}},\ }\bibfield  {title} {\bibinfo {title} {{Dynamics and
  statistics of unstable quantum states}},\ }\href
  {https://doi.org/10.1016/0375-9474(89)90558-7} {\bibfield  {journal}
  {\bibinfo  {journal} {Nucl. Phys. A}\ }\textbf {\bibinfo {volume} {504}},\
  \bibinfo {pages} {562} (\bibinfo {year} {1989})}\BibitemShut {NoStop}%
\bibitem [{\citenamefont {Haake}\ \emph {et~al.}(1992)\citenamefont {Haake},
  \citenamefont {Izrailev}, \citenamefont {Lehmann}, \citenamefont {Saher},\
  and\ \citenamefont {Sommers}}]{Haake-92}%
  \BibitemOpen
  \bibfield  {author} {\bibinfo {author} {\bibfnamefont {F.}~\bibnamefont
  {Haake}}, \bibinfo {author} {\bibfnamefont {F.}~\bibnamefont {Izrailev}},
  \bibinfo {author} {\bibfnamefont {N.}~\bibnamefont {Lehmann}}, \bibinfo
  {author} {\bibfnamefont {D.}~\bibnamefont {Saher}},\ and\ \bibinfo {author}
  {\bibfnamefont {H.-J.}\ \bibnamefont {Sommers}},\ }\bibfield  {title}
  {\bibinfo {title} {{Statistics of complex levels of random matrices for
  decaying systems}},\ }\href
  {https://doi.org/https://doi.org/10.1007/BF01470925} {\bibfield  {journal}
  {\bibinfo  {journal} {Z. Physik B}\ }\textbf {\bibinfo {volume} {88}},\
  \bibinfo {pages} {359} (\bibinfo {year} {1992})}\BibitemShut {NoStop}%
\bibitem [{\citenamefont {Stephanov}(1996)}]{Stephanov_1996}%
  \BibitemOpen
  \bibfield  {author} {\bibinfo {author} {\bibfnamefont {M.~A.}\ \bibnamefont
  {Stephanov}},\ }\bibfield  {title} {\bibinfo {title} {{Random Matrix Model of
  QCD at Finite Density and the Nature of the Quenched Limit}},\ }\href
  {https://doi.org/10.1103/PhysRevLett.76.4472} {\bibfield  {journal} {\bibinfo
   {journal} {Phys. Rev. Lett.}\ }\textbf {\bibinfo {volume} {76}},\ \bibinfo
  {pages} {4472} (\bibinfo {year} {1996})}\BibitemShut {NoStop}%
\bibitem [{\citenamefont {Fyodorov}\ \emph {et~al.}(1997)\citenamefont
  {Fyodorov}, \citenamefont {Khoruzhenko},\ and\ \citenamefont
  {Sommers}}]{Fyodorov_1997}%
  \BibitemOpen
  \bibfield  {author} {\bibinfo {author} {\bibfnamefont {Y.~V.}\ \bibnamefont
  {Fyodorov}}, \bibinfo {author} {\bibfnamefont {B.~A.}\ \bibnamefont
  {Khoruzhenko}},\ and\ \bibinfo {author} {\bibfnamefont {H.-J.}\ \bibnamefont
  {Sommers}},\ }\bibfield  {title} {\bibinfo {title} {{Almost-Hermitian random
  matrices: eigenvalue density in the complex plane}},\ }\href
  {https://doi.org/10.1016/s0375-9601(96)00904-8} {\bibfield  {journal}
  {\bibinfo  {journal} {Phys. Lett. A}\ }\textbf {\bibinfo {volume} {226}},\
  \bibinfo {pages} {46} (\bibinfo {year} {1997})}\BibitemShut {NoStop}%
\bibitem [{\citenamefont {Kogut}\ \emph {et~al.}(2000)\citenamefont {Kogut},
  \citenamefont {Stephanov}, \citenamefont {Toublan}, \citenamefont
  {Verbaarschot},\ and\ \citenamefont {Zhitnitsky}}]{Kogut:2000ek}%
  \BibitemOpen
  \bibfield  {author} {\bibinfo {author} {\bibfnamefont {J.~B.}\ \bibnamefont
  {Kogut}}, \bibinfo {author} {\bibfnamefont {M.~A.}\ \bibnamefont
  {Stephanov}}, \bibinfo {author} {\bibfnamefont {D.}~\bibnamefont {Toublan}},
  \bibinfo {author} {\bibfnamefont {J.~J.~M.}\ \bibnamefont {Verbaarschot}},\
  and\ \bibinfo {author} {\bibfnamefont {A.}~\bibnamefont {Zhitnitsky}},\
  }\bibfield  {title} {\bibinfo {title} {{QCD-like theories at finite baryon
  density}},\ }\href {https://doi.org/10.1016/S0550-3213(00)00242-X} {\bibfield
   {journal} {\bibinfo  {journal} {Nucl. Phys. B}\ }\textbf {\bibinfo {volume}
  {582}},\ \bibinfo {pages} {477} (\bibinfo {year} {2000})}\BibitemShut
  {NoStop}%
\bibitem [{\citenamefont {Hatano}\ and\ \citenamefont
  {Nelson}(1996)}]{Hatano-Nelson-96}%
  \BibitemOpen
  \bibfield  {author} {\bibinfo {author} {\bibfnamefont {N.}~\bibnamefont
  {Hatano}}\ and\ \bibinfo {author} {\bibfnamefont {D.~R.}\ \bibnamefont
  {Nelson}},\ }\bibfield  {title} {\bibinfo {title} {{Localization Transitions
  in Non-Hermitian Quantum Mechanics}},\ }\href
  {https://doi.org/10.1103/PhysRevLett.77.570} {\bibfield  {journal} {\bibinfo
  {journal} {Phys. Rev. Lett.}\ }\textbf {\bibinfo {volume} {77}},\ \bibinfo
  {pages} {570} (\bibinfo {year} {1996})}\BibitemShut {NoStop}%
\bibitem [{\citenamefont {Efetov}(1997{\natexlab{a}})}]{Efetov-97}%
  \BibitemOpen
  \bibfield  {author} {\bibinfo {author} {\bibfnamefont {K.~B.}\ \bibnamefont
  {Efetov}},\ }\bibfield  {title} {\bibinfo {title} {{Directed Quantum
  Chaos}},\ }\href {https://doi.org/10.1103/PhysRevLett.79.491} {\bibfield
  {journal} {\bibinfo  {journal} {Phys. Rev. Lett.}\ }\textbf {\bibinfo
  {volume} {79}},\ \bibinfo {pages} {491} (\bibinfo {year}
  {1997}{\natexlab{a}})}\BibitemShut {NoStop}%
\bibitem [{\citenamefont {Feinberg}\ and\ \citenamefont
  {Zee}(1997)}]{Feinberg-97}%
  \BibitemOpen
  \bibfield  {author} {\bibinfo {author} {\bibfnamefont {J.}~\bibnamefont
  {Feinberg}}\ and\ \bibinfo {author} {\bibfnamefont {A.}~\bibnamefont {Zee}},\
  }\bibfield  {title} {\bibinfo {title} {{Non-hermitian random matrix theory:
  Method of hermitian reduction}},\ }\href
  {https://doi.org/10.1016/S0550-3213(97)00502-6} {\bibfield  {journal}
  {\bibinfo  {journal} {Nucl. Phys. B}\ }\textbf {\bibinfo {volume} {504}},\
  \bibinfo {pages} {579} (\bibinfo {year} {1997})}\BibitemShut {NoStop}%
\bibitem [{\citenamefont {Efetov}(1997{\natexlab{b}})}]{Efetov-97B}%
  \BibitemOpen
  \bibfield  {author} {\bibinfo {author} {\bibfnamefont {K.~B.}\ \bibnamefont
  {Efetov}},\ }\bibfield  {title} {\bibinfo {title} {{Quantum disordered
  systems with a direction}},\ }\href
  {https://doi.org/10.1103/PhysRevB.56.9630} {\bibfield  {journal} {\bibinfo
  {journal} {Phys. Rev. B}\ }\textbf {\bibinfo {volume} {56}},\ \bibinfo
  {pages} {9630} (\bibinfo {year} {1997}{\natexlab{b}})}\BibitemShut {NoStop}%
\bibitem [{\citenamefont {Hatano}\ and\ \citenamefont
  {Nelson}(1997)}]{Hatano-Nelson-97}%
  \BibitemOpen
  \bibfield  {author} {\bibinfo {author} {\bibfnamefont {N.}~\bibnamefont
  {Hatano}}\ and\ \bibinfo {author} {\bibfnamefont {D.~R.}\ \bibnamefont
  {Nelson}},\ }\bibfield  {title} {\bibinfo {title} {{Vortex pinning and
  non-Hermitian quantum mechanics}},\ }\href
  {https://doi.org/10.1103/PhysRevB.56.8651} {\bibfield  {journal} {\bibinfo
  {journal} {Phys. Rev. B}\ }\textbf {\bibinfo {volume} {56}},\ \bibinfo
  {pages} {8651} (\bibinfo {year} {1997})}\BibitemShut {NoStop}%
\bibitem [{\citenamefont {Brouwer}\ \emph {et~al.}(1997)\citenamefont
  {Brouwer}, \citenamefont {Silvestrov},\ and\ \citenamefont
  {Beenakker}}]{Brouwer-97}%
  \BibitemOpen
  \bibfield  {author} {\bibinfo {author} {\bibfnamefont {P.~W.}\ \bibnamefont
  {Brouwer}}, \bibinfo {author} {\bibfnamefont {P.~G.}\ \bibnamefont
  {Silvestrov}},\ and\ \bibinfo {author} {\bibfnamefont {C.~W.~J.}\
  \bibnamefont {Beenakker}},\ }\bibfield  {title} {\bibinfo {title} {{Theory of
  directed localization in one dimension}},\ }\href
  {https://doi.org/10.1103/PhysRevB.56.R4333} {\bibfield  {journal} {\bibinfo
  {journal} {Phys. Rev. B}\ }\textbf {\bibinfo {volume} {56}},\ \bibinfo
  {pages} {R4333(R)} (\bibinfo {year} {1997})}\BibitemShut {NoStop}%
\bibitem [{\citenamefont {Feinberg}\ and\ \citenamefont
  {Zee}(1999)}]{Feinberg-99}%
  \BibitemOpen
  \bibfield  {author} {\bibinfo {author} {\bibfnamefont {J.}~\bibnamefont
  {Feinberg}}\ and\ \bibinfo {author} {\bibfnamefont {A.}~\bibnamefont {Zee}},\
  }\bibfield  {title} {\bibinfo {title} {{Non-Hermitian localization and
  delocalization}},\ }\href {https://doi.org/10.1103/PhysRevE.59.6433}
  {\bibfield  {journal} {\bibinfo  {journal} {Phys. Rev. E}\ }\textbf {\bibinfo
  {volume} {59}},\ \bibinfo {pages} {6433} (\bibinfo {year}
  {1999})}\BibitemShut {NoStop}%
\bibitem [{\citenamefont {Longhi}(2019)}]{Longhi-19}%
  \BibitemOpen
  \bibfield  {author} {\bibinfo {author} {\bibfnamefont {S.}~\bibnamefont
  {Longhi}},\ }\bibfield  {title} {\bibinfo {title} {{Topological Phase
  Transition in non-Hermitian Quasicrystals}},\ }\href
  {https://doi.org/10.1103/PhysRevLett.122.237601} {\bibfield  {journal}
  {\bibinfo  {journal} {Phys. Rev. Lett.}\ }\textbf {\bibinfo {volume} {122}},\
  \bibinfo {pages} {237601} (\bibinfo {year} {2019})}\BibitemShut {NoStop}%
\bibitem [{\citenamefont {Zeng}\ \emph {et~al.}(2020)\citenamefont {Zeng},
  \citenamefont {Yang},\ and\ \citenamefont {Xu}}]{Zeng-20}%
  \BibitemOpen
  \bibfield  {author} {\bibinfo {author} {\bibfnamefont {Q.-B.}\ \bibnamefont
  {Zeng}}, \bibinfo {author} {\bibfnamefont {Y.-B.}\ \bibnamefont {Yang}},\
  and\ \bibinfo {author} {\bibfnamefont {Y.}~\bibnamefont {Xu}},\ }\bibfield
  {title} {\bibinfo {title} {{Topological phases in non-Hermitian
  Aubry-Andr\'e-Harper models}},\ }\href
  {https://doi.org/10.1103/PhysRevB.101.020201} {\bibfield  {journal} {\bibinfo
   {journal} {Phys. Rev. B}\ }\textbf {\bibinfo {volume} {101}},\ \bibinfo
  {pages} {020201} (\bibinfo {year} {2020})}\BibitemShut {NoStop}%
\bibitem [{\citenamefont {Tzortzakakis}\ \emph {et~al.}(2020)\citenamefont
  {Tzortzakakis}, \citenamefont {Makris},\ and\ \citenamefont
  {Economou}}]{Tzortzakakis-20}%
  \BibitemOpen
  \bibfield  {author} {\bibinfo {author} {\bibfnamefont {A.~F.}\ \bibnamefont
  {Tzortzakakis}}, \bibinfo {author} {\bibfnamefont {K.~G.}\ \bibnamefont
  {Makris}},\ and\ \bibinfo {author} {\bibfnamefont {E.~N.}\ \bibnamefont
  {Economou}},\ }\bibfield  {title} {\bibinfo {title} {{Non-Hermitian disorder
  in two-dimensional optical lattices}},\ }\href
  {https://doi.org/10.1103/PhysRevB.101.014202} {\bibfield  {journal} {\bibinfo
   {journal} {Phys. Rev. B}\ }\textbf {\bibinfo {volume} {101}},\ \bibinfo
  {pages} {014202} (\bibinfo {year} {2020})}\BibitemShut {NoStop}%
\bibitem [{\citenamefont {Huang}\ and\ \citenamefont
  {Shklovskii}(2020)}]{Huang-20}%
  \BibitemOpen
  \bibfield  {author} {\bibinfo {author} {\bibfnamefont {Y.}~\bibnamefont
  {Huang}}\ and\ \bibinfo {author} {\bibfnamefont {B.~I.}\ \bibnamefont
  {Shklovskii}},\ }\bibfield  {title} {\bibinfo {title} {{Anderson transition
  in three-dimensional systems with non-Hermitian disorder}},\ }\href
  {https://doi.org/10.1103/PhysRevB.101.014204} {\bibfield  {journal} {\bibinfo
   {journal} {Phys. Rev. B}\ }\textbf {\bibinfo {volume} {101}},\ \bibinfo
  {pages} {014204} (\bibinfo {year} {2020})}\BibitemShut {NoStop}%
\bibitem [{\citenamefont {Kawabata}\ and\ \citenamefont {Ryu}(2021)}]{KR-21}%
  \BibitemOpen
  \bibfield  {author} {\bibinfo {author} {\bibfnamefont {K.}~\bibnamefont
  {Kawabata}}\ and\ \bibinfo {author} {\bibfnamefont {S.}~\bibnamefont {Ryu}},\
  }\bibfield  {title} {\bibinfo {title} {{Nonunitary Scaling Theory of
  Non-Hermitian Localization}},\ }\href
  {https://doi.org/10.1103/PhysRevLett.126.166801} {\bibfield  {journal}
  {\bibinfo  {journal} {Phys. Rev. Lett.}\ }\textbf {\bibinfo {volume} {126}},\
  \bibinfo {pages} {166801} (\bibinfo {year} {2021})}\BibitemShut {NoStop}%
\bibitem [{\citenamefont {Claes}\ and\ \citenamefont
  {Hughes}(2021)}]{Claes-21}%
  \BibitemOpen
  \bibfield  {author} {\bibinfo {author} {\bibfnamefont {J.}~\bibnamefont
  {Claes}}\ and\ \bibinfo {author} {\bibfnamefont {T.~L.}\ \bibnamefont
  {Hughes}},\ }\bibfield  {title} {\bibinfo {title} {{Skin effect and winding
  number in disordered non-Hermitian systems}},\ }\href
  {https://doi.org/10.1103/PhysRevB.103.L140201} {\bibfield  {journal}
  {\bibinfo  {journal} {Phys. Rev. B}\ }\textbf {\bibinfo {volume} {103}},\
  \bibinfo {pages} {L140201} (\bibinfo {year} {2021})}\BibitemShut {NoStop}%
\bibitem [{\citenamefont {Luo}\ \emph {et~al.}(2021{\natexlab{a}})\citenamefont
  {Luo}, \citenamefont {Ohtsuki},\ and\ \citenamefont {Shindou}}]{Luo-21L}%
  \BibitemOpen
  \bibfield  {author} {\bibinfo {author} {\bibfnamefont {X.}~\bibnamefont
  {Luo}}, \bibinfo {author} {\bibfnamefont {T.}~\bibnamefont {Ohtsuki}},\ and\
  \bibinfo {author} {\bibfnamefont {R.}~\bibnamefont {Shindou}},\ }\bibfield
  {title} {\bibinfo {title} {{Universality Classes of the Anderson Transitions
  Driven by Non-Hermitian Disorder}},\ }\href
  {https://doi.org/10.1103/PhysRevLett.126.090402} {\bibfield  {journal}
  {\bibinfo  {journal} {Phys. Rev. Lett.}\ }\textbf {\bibinfo {volume} {126}},\
  \bibinfo {pages} {090402} (\bibinfo {year} {2021}{\natexlab{a}})}\BibitemShut
  {NoStop}%
\bibitem [{\citenamefont {Luo}\ \emph {et~al.}(2021{\natexlab{b}})\citenamefont
  {Luo}, \citenamefont {Ohtsuki},\ and\ \citenamefont {Shindou}}]{Luo-21B}%
  \BibitemOpen
  \bibfield  {author} {\bibinfo {author} {\bibfnamefont {X.}~\bibnamefont
  {Luo}}, \bibinfo {author} {\bibfnamefont {T.}~\bibnamefont {Ohtsuki}},\ and\
  \bibinfo {author} {\bibfnamefont {R.}~\bibnamefont {Shindou}},\ }\bibfield
  {title} {\bibinfo {title} {{Transfer matrix study of the Anderson transition
  in non-Hermitian systems}},\ }\href
  {https://doi.org/10.1103/PhysRevB.104.104203} {\bibfield  {journal} {\bibinfo
   {journal} {Phys. Rev. B}\ }\textbf {\bibinfo {volume} {104}},\ \bibinfo
  {pages} {104203} (\bibinfo {year} {2021}{\natexlab{b}})}\BibitemShut
  {NoStop}%
\bibitem [{\citenamefont {Luo}\ \emph {et~al.}(2022)\citenamefont {Luo},
  \citenamefont {Xiao}, \citenamefont {Kawabata}, \citenamefont {Ohtsuki},\
  and\ \citenamefont {Shindou}}]{Luo-22R}%
  \BibitemOpen
  \bibfield  {author} {\bibinfo {author} {\bibfnamefont {X.}~\bibnamefont
  {Luo}}, \bibinfo {author} {\bibfnamefont {Z.}~\bibnamefont {Xiao}}, \bibinfo
  {author} {\bibfnamefont {K.}~\bibnamefont {Kawabata}}, \bibinfo {author}
  {\bibfnamefont {T.}~\bibnamefont {Ohtsuki}},\ and\ \bibinfo {author}
  {\bibfnamefont {R.}~\bibnamefont {Shindou}},\ }\bibfield  {title} {\bibinfo
  {title} {{Unifying the Anderson transitions in Hermitian and non-Hermitian
  systems}},\ }\href {https://doi.org/10.1103/PhysRevResearch.4.L022035}
  {\bibfield  {journal} {\bibinfo  {journal} {Phys. Rev. Research}\ }\textbf
  {\bibinfo {volume} {4}},\ \bibinfo {pages} {L022035} (\bibinfo {year}
  {2022})}\BibitemShut {NoStop}%
\bibitem [{\citenamefont {Liu}\ \emph {et~al.}(2021)\citenamefont {Liu},
  \citenamefont {You}, \citenamefont {Ryu},\ and\ \citenamefont
  {Fulga}}]{Liu-Fulga-21}%
  \BibitemOpen
  \bibfield  {author} {\bibinfo {author} {\bibfnamefont {H.}~\bibnamefont
  {Liu}}, \bibinfo {author} {\bibfnamefont {J.-S.}\ \bibnamefont {You}},
  \bibinfo {author} {\bibfnamefont {S.}~\bibnamefont {Ryu}},\ and\ \bibinfo
  {author} {\bibfnamefont {I.~C.}\ \bibnamefont {Fulga}},\ }\bibfield  {title}
  {\bibinfo {title} {{Supermetal-insulator transition in a non-Hermitian
  network model}},\ }\href {https://doi.org/10.1103/PhysRevB.104.155412}
  {\bibfield  {journal} {\bibinfo  {journal} {Phys. Rev. B}\ }\textbf {\bibinfo
  {volume} {104}},\ \bibinfo {pages} {155412} (\bibinfo {year}
  {2021})}\BibitemShut {NoStop}%
\bibitem [{\citenamefont {Bergholtz}\ \emph {et~al.}(2021)\citenamefont
  {Bergholtz}, \citenamefont {Budich},\ and\ \citenamefont
  {Kunst}}]{BBK-review}%
  \BibitemOpen
  \bibfield  {author} {\bibinfo {author} {\bibfnamefont {E.~J.}\ \bibnamefont
  {Bergholtz}}, \bibinfo {author} {\bibfnamefont {J.~C.}\ \bibnamefont
  {Budich}},\ and\ \bibinfo {author} {\bibfnamefont {F.~K.}\ \bibnamefont
  {Kunst}},\ }\bibfield  {title} {\bibinfo {title} {{Exceptional topology of
  non-Hermitian systems}},\ }\href
  {https://doi.org/10.1103/RevModPhys.93.015005} {\bibfield  {journal}
  {\bibinfo  {journal} {Rev. Mod. Phys.}\ }\textbf {\bibinfo {volume} {93}},\
  \bibinfo {pages} {015005} (\bibinfo {year} {2021})}\BibitemShut {NoStop}%
\bibitem [{\citenamefont {Okuma}\ and\ \citenamefont
  {Sato}(2023)}]{Okuma-Sato-review}%
  \BibitemOpen
  \bibfield  {author} {\bibinfo {author} {\bibfnamefont {N.}~\bibnamefont
  {Okuma}}\ and\ \bibinfo {author} {\bibfnamefont {M.}~\bibnamefont {Sato}},\
  }\bibfield  {title} {\bibinfo {title} {{Non-Hermitian Topological Phenomena:
  A Review}},\ }\href
  {https://doi.org/10.1146/annurev-conmatphys-040521-033133} {\bibfield
  {journal} {\bibinfo  {journal} {Annu. Rev. Condens. Matter Phys.}\ }\textbf
  {\bibinfo {volume} {14}},\ \bibinfo {pages} {83} (\bibinfo {year}
  {2023})}\BibitemShut {NoStop}%
\bibitem [{\citenamefont {Altland}\ and\ \citenamefont
  {Zirnbauer}(1997)}]{AZ-97}%
  \BibitemOpen
  \bibfield  {author} {\bibinfo {author} {\bibfnamefont {A.}~\bibnamefont
  {Altland}}\ and\ \bibinfo {author} {\bibfnamefont {M.~R.}\ \bibnamefont
  {Zirnbauer}},\ }\bibfield  {title} {\bibinfo {title} {{Nonstandard symmetry
  classes in mesoscopic normal-superconducting hybrid structures}},\ }\href
  {https://doi.org/10.1103/PhysRevB.55.1142} {\bibfield  {journal} {\bibinfo
  {journal} {Phys. Rev. B}\ }\textbf {\bibinfo {volume} {55}},\ \bibinfo
  {pages} {1142} (\bibinfo {year} {1997})}\BibitemShut {NoStop}%
\bibitem [{\citenamefont {Evers}\ and\ \citenamefont
  {Mirlin}(2008)}]{Evers-review}%
  \BibitemOpen
  \bibfield  {author} {\bibinfo {author} {\bibfnamefont {F.}~\bibnamefont
  {Evers}}\ and\ \bibinfo {author} {\bibfnamefont {A.~D.}\ \bibnamefont
  {Mirlin}},\ }\bibfield  {title} {\bibinfo {title} {{Anderson transitions}},\
  }\href {https://doi.org/10.1103/RevModPhys.80.1355} {\bibfield  {journal}
  {\bibinfo  {journal} {Rev. Mod. Phys.}\ }\textbf {\bibinfo {volume} {80}},\
  \bibinfo {pages} {1355} (\bibinfo {year} {2008})}\BibitemShut {NoStop}%
\bibitem [{\citenamefont {Chiu}\ \emph {et~al.}(2016)\citenamefont {Chiu},
  \citenamefont {Teo}, \citenamefont {Schnyder},\ and\ \citenamefont
  {Ryu}}]{CTSR-review}%
  \BibitemOpen
  \bibfield  {author} {\bibinfo {author} {\bibfnamefont {C.-K.}\ \bibnamefont
  {Chiu}}, \bibinfo {author} {\bibfnamefont {J.~C.~Y.}\ \bibnamefont {Teo}},
  \bibinfo {author} {\bibfnamefont {A.~P.}\ \bibnamefont {Schnyder}},\ and\
  \bibinfo {author} {\bibfnamefont {S.}~\bibnamefont {Ryu}},\ }\bibfield
  {title} {\bibinfo {title} {{Classification of topological quantum matter with
  symmetries}},\ }\href {https://doi.org/10.1103/RevModPhys.88.035005}
  {\bibfield  {journal} {\bibinfo  {journal} {Rev. Mod. Phys.}\ }\textbf
  {\bibinfo {volume} {88}},\ \bibinfo {pages} {035005} (\bibinfo {year}
  {2016})}\BibitemShut {NoStop}%
\bibitem [{\citenamefont {Haake}\ \emph {et~al.}(2018)\citenamefont {Haake},
  \citenamefont {Gnutzmann},\ and\ \citenamefont {Ku\'{s}}}]{Haake-textbook}%
  \BibitemOpen
  \bibfield  {author} {\bibinfo {author} {\bibfnamefont {F.}~\bibnamefont
  {Haake}}, \bibinfo {author} {\bibfnamefont {S.}~\bibnamefont {Gnutzmann}},\
  and\ \bibinfo {author} {\bibfnamefont {M.}~\bibnamefont {Ku\'{s}}},\ }\href
  {https://doi.org/10.1007/978-3-319-97580-1} {\emph {\bibinfo {title}
  {{Quantum Signatures of Chaos}}}}\ (\bibinfo  {publisher} {Springer},\
  \bibinfo {address} {Cham},\ \bibinfo {year} {2018})\BibitemShut {NoStop}%
\bibitem [{\citenamefont {Bernard}\ and\ \citenamefont
  {LeClair}(2002)}]{Bernard-LeClair-02}%
  \BibitemOpen
  \bibfield  {author} {\bibinfo {author} {\bibfnamefont {D.}~\bibnamefont
  {Bernard}}\ and\ \bibinfo {author} {\bibfnamefont {A.}~\bibnamefont
  {LeClair}},\ }\bibinfo {title} {{A Classification of Non-Hermitian Random
  Matrices}},\ in\ \href {https://doi.org/10.1007/978-94-010-0514-2_19} {\emph
  {\bibinfo {booktitle} {Statistical Field Theories}}},\ \bibinfo {editor}
  {edited by\ \bibinfo {editor} {\bibfnamefont {A.}~\bibnamefont {Cappelli}}\
  and\ \bibinfo {editor} {\bibfnamefont {G.}~\bibnamefont {Mussardo}}}\
  (\bibinfo  {publisher} {Springer},\ \bibinfo {address} {Dordrecht},\ \bibinfo
  {year} {2002})\ pp.\ \bibinfo {pages} {207--214}\BibitemShut {NoStop}%
\bibitem [{\citenamefont {Kawabata}\ \emph {et~al.}(2019)\citenamefont
  {Kawabata}, \citenamefont {Shiozaki}, \citenamefont {Ueda},\ and\
  \citenamefont {Sato}}]{KSUS-19}%
  \BibitemOpen
  \bibfield  {author} {\bibinfo {author} {\bibfnamefont {K.}~\bibnamefont
  {Kawabata}}, \bibinfo {author} {\bibfnamefont {K.}~\bibnamefont {Shiozaki}},
  \bibinfo {author} {\bibfnamefont {M.}~\bibnamefont {Ueda}},\ and\ \bibinfo
  {author} {\bibfnamefont {M.}~\bibnamefont {Sato}},\ }\bibfield  {title}
  {\bibinfo {title} {{Symmetry and Topology in Non-Hermitian Physics}},\ }\href
  {https://doi.org/10.1103/PhysRevX.9.041015} {\bibfield  {journal} {\bibinfo
  {journal} {Phys. Rev. X}\ }\textbf {\bibinfo {volume} {9}},\ \bibinfo {pages}
  {041015} (\bibinfo {year} {2019})}\BibitemShut {NoStop}%
\bibitem [{\citenamefont {Grobe}\ \emph {et~al.}(1988)\citenamefont {Grobe},
  \citenamefont {Haake},\ and\ \citenamefont {Sommers}}]{Grobe-88}%
  \BibitemOpen
  \bibfield  {author} {\bibinfo {author} {\bibfnamefont {R.}~\bibnamefont
  {Grobe}}, \bibinfo {author} {\bibfnamefont {F.}~\bibnamefont {Haake}},\ and\
  \bibinfo {author} {\bibfnamefont {H.-J.}\ \bibnamefont {Sommers}},\
  }\bibfield  {title} {\bibinfo {title} {{Quantum Distinction of Regular and
  Chaotic Dissipative Motion}},\ }\href
  {https://doi.org/10.1103/PhysRevLett.61.1899} {\bibfield  {journal} {\bibinfo
   {journal} {Phys. Rev. Lett.}\ }\textbf {\bibinfo {volume} {61}},\ \bibinfo
  {pages} {1899} (\bibinfo {year} {1988})}\BibitemShut {NoStop}%
\bibitem [{\citenamefont {Grobe}\ and\ \citenamefont {Haake}(1989)}]{Grobe-89}%
  \BibitemOpen
  \bibfield  {author} {\bibinfo {author} {\bibfnamefont {R.}~\bibnamefont
  {Grobe}}\ and\ \bibinfo {author} {\bibfnamefont {F.}~\bibnamefont {Haake}},\
  }\bibfield  {title} {\bibinfo {title} {{Universality of cubic-level repulsion
  for dissipative quantum chaos}},\ }\href
  {https://doi.org/10.1103/PhysRevLett.62.2893} {\bibfield  {journal} {\bibinfo
   {journal} {Phys. Rev. Lett.}\ }\textbf {\bibinfo {volume} {62}},\ \bibinfo
  {pages} {2893} (\bibinfo {year} {1989})}\BibitemShut {NoStop}%
\bibitem [{\citenamefont {Xu}\ \emph {et~al.}(2019)\citenamefont {Xu},
  \citenamefont {Garc\'{\i}a-Pintos}, \citenamefont {Chenu},\ and\
  \citenamefont {del Campo}}]{Xu-19}%
  \BibitemOpen
  \bibfield  {author} {\bibinfo {author} {\bibfnamefont {Z.}~\bibnamefont
  {Xu}}, \bibinfo {author} {\bibfnamefont {L.~P.}\ \bibnamefont
  {Garc\'{\i}a-Pintos}}, \bibinfo {author} {\bibfnamefont {A.}~\bibnamefont
  {Chenu}},\ and\ \bibinfo {author} {\bibfnamefont {A.}~\bibnamefont {del
  Campo}},\ }\bibfield  {title} {\bibinfo {title} {{Extreme Decoherence and
  Quantum Chaos}},\ }\href {https://doi.org/10.1103/PhysRevLett.122.014103}
  {\bibfield  {journal} {\bibinfo  {journal} {Phys. Rev. Lett.}\ }\textbf
  {\bibinfo {volume} {122}},\ \bibinfo {pages} {014103} (\bibinfo {year}
  {2019})}\BibitemShut {NoStop}%
\bibitem [{\citenamefont {Hamazaki}\ \emph {et~al.}(2019)\citenamefont
  {Hamazaki}, \citenamefont {Kawabata},\ and\ \citenamefont
  {Ueda}}]{Hamazaki-19}%
  \BibitemOpen
  \bibfield  {author} {\bibinfo {author} {\bibfnamefont {R.}~\bibnamefont
  {Hamazaki}}, \bibinfo {author} {\bibfnamefont {K.}~\bibnamefont {Kawabata}},\
  and\ \bibinfo {author} {\bibfnamefont {M.}~\bibnamefont {Ueda}},\ }\bibfield
  {title} {\bibinfo {title} {{Non-Hermitian Many-Body Localization}},\ }\href
  {https://doi.org/10.1103/PhysRevLett.123.090603} {\bibfield  {journal}
  {\bibinfo  {journal} {Phys. Rev. Lett.}\ }\textbf {\bibinfo {volume} {123}},\
  \bibinfo {pages} {090603} (\bibinfo {year} {2019})}\BibitemShut {NoStop}%
\bibitem [{\citenamefont {Denisov}\ \emph {et~al.}(2019)\citenamefont
  {Denisov}, \citenamefont {Laptyeva}, \citenamefont {Tarnowski}, \citenamefont
  {Chru\ifmmode \acute{s}\else \'{s}\fi{}ci\ifmmode~\acute{n}\else
  \'{n}\fi{}ski},\ and\ \citenamefont {\ifmmode~\dot{Z}\else
  \.{Z}\fi{}yczkowski}}]{Denisov-19}%
  \BibitemOpen
  \bibfield  {author} {\bibinfo {author} {\bibfnamefont {S.}~\bibnamefont
  {Denisov}}, \bibinfo {author} {\bibfnamefont {T.}~\bibnamefont {Laptyeva}},
  \bibinfo {author} {\bibfnamefont {W.}~\bibnamefont {Tarnowski}}, \bibinfo
  {author} {\bibfnamefont {D.}~\bibnamefont {Chru\ifmmode \acute{s}\else
  \'{s}\fi{}ci\ifmmode~\acute{n}\else \'{n}\fi{}ski}},\ and\ \bibinfo {author}
  {\bibfnamefont {K.}~\bibnamefont {\ifmmode~\dot{Z}\else
  \.{Z}\fi{}yczkowski}},\ }\bibfield  {title} {\bibinfo {title} {{Universal
  Spectra of Random Lindblad Operators}},\ }\href
  {https://doi.org/10.1103/PhysRevLett.123.140403} {\bibfield  {journal}
  {\bibinfo  {journal} {Phys. Rev. Lett.}\ }\textbf {\bibinfo {volume} {123}},\
  \bibinfo {pages} {140403} (\bibinfo {year} {2019})}\BibitemShut {NoStop}%
\bibitem [{\citenamefont {Can}\ \emph {et~al.}(2019)\citenamefont {Can},
  \citenamefont {Oganesyan}, \citenamefont {Orgad},\ and\ \citenamefont
  {Gopalakrishnan}}]{Can-19PRL}%
  \BibitemOpen
  \bibfield  {author} {\bibinfo {author} {\bibfnamefont {T.}~\bibnamefont
  {Can}}, \bibinfo {author} {\bibfnamefont {V.}~\bibnamefont {Oganesyan}},
  \bibinfo {author} {\bibfnamefont {D.}~\bibnamefont {Orgad}},\ and\ \bibinfo
  {author} {\bibfnamefont {S.}~\bibnamefont {Gopalakrishnan}},\ }\bibfield
  {title} {\bibinfo {title} {{Spectral Gaps and Midgap States in Random Quantum
  Master Equations}},\ }\href {https://doi.org/10.1103/PhysRevLett.123.234103}
  {\bibfield  {journal} {\bibinfo  {journal} {Phys. Rev. Lett.}\ }\textbf
  {\bibinfo {volume} {123}},\ \bibinfo {pages} {234103} (\bibinfo {year}
  {2019})}\BibitemShut {NoStop}%
\bibitem [{\citenamefont {Can}(2019)}]{Can-19JPhysA}%
  \BibitemOpen
  \bibfield  {author} {\bibinfo {author} {\bibfnamefont {T.}~\bibnamefont
  {Can}},\ }\bibfield  {title} {\bibinfo {title} {{Random Lindblad dynamics}},\
  }\href {https://doi.org/10.1088/1751-8121/ab4d26} {\bibfield  {journal}
  {\bibinfo  {journal} {J. Phys. A}\ }\textbf {\bibinfo {volume} {52}},\
  \bibinfo {pages} {485302} (\bibinfo {year} {2019})}\BibitemShut {NoStop}%
\bibitem [{\citenamefont {Hamazaki}\ \emph {et~al.}(2020)\citenamefont
  {Hamazaki}, \citenamefont {Kawabata}, \citenamefont {Kura},\ and\
  \citenamefont {Ueda}}]{Hamazaki-20}%
  \BibitemOpen
  \bibfield  {author} {\bibinfo {author} {\bibfnamefont {R.}~\bibnamefont
  {Hamazaki}}, \bibinfo {author} {\bibfnamefont {K.}~\bibnamefont {Kawabata}},
  \bibinfo {author} {\bibfnamefont {N.}~\bibnamefont {Kura}},\ and\ \bibinfo
  {author} {\bibfnamefont {M.}~\bibnamefont {Ueda}},\ }\bibfield  {title}
  {\bibinfo {title} {{Universality classes of non-Hermitian random matrices}},\
  }\href {https://doi.org/10.1103/PhysRevResearch.2.023286} {\bibfield
  {journal} {\bibinfo  {journal} {Phys. Rev. Research}\ }\textbf {\bibinfo
  {volume} {2}},\ \bibinfo {pages} {023286} (\bibinfo {year}
  {2020})}\BibitemShut {NoStop}%
\bibitem [{\citenamefont {Akemann}\ \emph {et~al.}(2019)\citenamefont
  {Akemann}, \citenamefont {Kieburg}, \citenamefont {Mielke},\ and\
  \citenamefont {Prosen}}]{Akemann-19}%
  \BibitemOpen
  \bibfield  {author} {\bibinfo {author} {\bibfnamefont {G.}~\bibnamefont
  {Akemann}}, \bibinfo {author} {\bibfnamefont {M.}~\bibnamefont {Kieburg}},
  \bibinfo {author} {\bibfnamefont {A.}~\bibnamefont {Mielke}},\ and\ \bibinfo
  {author} {\bibfnamefont {T.}~\bibnamefont {Prosen}},\ }\bibfield  {title}
  {\bibinfo {title} {{Universal Signature from Integrability to Chaos in
  Dissipative Open Quantum Systems}},\ }\href
  {https://doi.org/10.1103/PhysRevLett.123.254101} {\bibfield  {journal}
  {\bibinfo  {journal} {Phys. Rev. Lett.}\ }\textbf {\bibinfo {volume} {123}},\
  \bibinfo {pages} {254101} (\bibinfo {year} {2019})}\BibitemShut {NoStop}%
\bibitem [{\citenamefont {S\'a}\ \emph {et~al.}(2020)\citenamefont {S\'a},
  \citenamefont {Ribeiro},\ and\ \citenamefont {Prosen}}]{Sa-20}%
  \BibitemOpen
  \bibfield  {author} {\bibinfo {author} {\bibfnamefont {L.}~\bibnamefont
  {S\'a}}, \bibinfo {author} {\bibfnamefont {P.}~\bibnamefont {Ribeiro}},\ and\
  \bibinfo {author} {\bibfnamefont {T.}~\bibnamefont {Prosen}},\ }\bibfield
  {title} {\bibinfo {title} {{Complex Spacing Ratios: A Signature of
  Dissipative Quantum Chaos}},\ }\href
  {https://doi.org/10.1103/PhysRevX.10.021019} {\bibfield  {journal} {\bibinfo
  {journal} {Phys. Rev. X}\ }\textbf {\bibinfo {volume} {10}},\ \bibinfo
  {pages} {021019} (\bibinfo {year} {2020})}\BibitemShut {NoStop}%
\bibitem [{\citenamefont {Wang}\ \emph {et~al.}(2020)\citenamefont {Wang},
  \citenamefont {Piazza},\ and\ \citenamefont {Luitz}}]{Wang-20}%
  \BibitemOpen
  \bibfield  {author} {\bibinfo {author} {\bibfnamefont {K.}~\bibnamefont
  {Wang}}, \bibinfo {author} {\bibfnamefont {F.}~\bibnamefont {Piazza}},\ and\
  \bibinfo {author} {\bibfnamefont {D.~J.}\ \bibnamefont {Luitz}},\ }\bibfield
  {title} {\bibinfo {title} {{Hierarchy of Relaxation Timescales in Local
  Random Liouvillians}},\ }\href
  {https://doi.org/10.1103/PhysRevLett.124.100604} {\bibfield  {journal}
  {\bibinfo  {journal} {Phys. Rev. Lett.}\ }\textbf {\bibinfo {volume} {124}},\
  \bibinfo {pages} {100604} (\bibinfo {year} {2020})}\BibitemShut {NoStop}%
\bibitem [{\citenamefont {Xu}\ \emph {et~al.}(2021)\citenamefont {Xu},
  \citenamefont {Chenu}, \citenamefont {Prosen},\ and\ \citenamefont {del
  Campo}}]{Xu-21}%
  \BibitemOpen
  \bibfield  {author} {\bibinfo {author} {\bibfnamefont {Z.}~\bibnamefont
  {Xu}}, \bibinfo {author} {\bibfnamefont {A.}~\bibnamefont {Chenu}}, \bibinfo
  {author} {\bibfnamefont {T.}~\bibnamefont {Prosen}},\ and\ \bibinfo {author}
  {\bibfnamefont {A.}~\bibnamefont {del Campo}},\ }\bibfield  {title} {\bibinfo
  {title} {{Thermofield dynamics: Quantum chaos versus decoherence}},\ }\href
  {https://doi.org/10.1103/PhysRevB.103.064309} {\bibfield  {journal} {\bibinfo
   {journal} {Phys. Rev. B}\ }\textbf {\bibinfo {volume} {103}},\ \bibinfo
  {pages} {064309} (\bibinfo {year} {2021})}\BibitemShut {NoStop}%
\bibitem [{\citenamefont {Garc\'{\i}a-Garc\'{\i}a}\ \emph
  {et~al.}(2022{\natexlab{a}})\citenamefont {Garc\'{\i}a-Garc\'{\i}a},
  \citenamefont {Jia}, \citenamefont {Rosa},\ and\ \citenamefont
  {Verbaarschot}}]{GarciaGarcia-22PRL}%
  \BibitemOpen
  \bibfield  {author} {\bibinfo {author} {\bibfnamefont {A.~M.}\ \bibnamefont
  {Garc\'{\i}a-Garc\'{\i}a}}, \bibinfo {author} {\bibfnamefont
  {Y.}~\bibnamefont {Jia}}, \bibinfo {author} {\bibfnamefont {D.}~\bibnamefont
  {Rosa}},\ and\ \bibinfo {author} {\bibfnamefont {J.~J.~M.}\ \bibnamefont
  {Verbaarschot}},\ }\bibfield  {title} {\bibinfo {title} {{Dominance of
  Replica Off-Diagonal Configurations and Phase Transitions in a $PT$ Symmetric
  Sachdev-Ye-Kitaev Model}},\ }\href
  {https://doi.org/10.1103/PhysRevLett.128.081601} {\bibfield  {journal}
  {\bibinfo  {journal} {Phys. Rev. Lett.}\ }\textbf {\bibinfo {volume} {128}},\
  \bibinfo {pages} {081601} (\bibinfo {year} {2022}{\natexlab{a}})}\BibitemShut
  {NoStop}%
\bibitem [{\citenamefont {Li}\ \emph {et~al.}(2021)\citenamefont {Li},
  \citenamefont {Prosen},\ and\ \citenamefont {Chan}}]{JiachenLi-21}%
  \BibitemOpen
  \bibfield  {author} {\bibinfo {author} {\bibfnamefont {J.}~\bibnamefont
  {Li}}, \bibinfo {author} {\bibfnamefont {T.}~\bibnamefont {Prosen}},\ and\
  \bibinfo {author} {\bibfnamefont {A.}~\bibnamefont {Chan}},\ }\bibfield
  {title} {\bibinfo {title} {{Spectral Statistics of Non-Hermitian Matrices and
  Dissipative Quantum Chaos}},\ }\href
  {https://doi.org/10.1103/PhysRevLett.127.170602} {\bibfield  {journal}
  {\bibinfo  {journal} {Phys. Rev. Lett.}\ }\textbf {\bibinfo {volume} {127}},\
  \bibinfo {pages} {170602} (\bibinfo {year} {2021})}\BibitemShut {NoStop}%
\bibitem [{\citenamefont {Cornelius}\ \emph {et~al.}(2022)\citenamefont
  {Cornelius}, \citenamefont {Xu}, \citenamefont {Saxena}, \citenamefont
  {Chenu},\ and\ \citenamefont {del Campo}}]{Cornelius-22}%
  \BibitemOpen
  \bibfield  {author} {\bibinfo {author} {\bibfnamefont {J.}~\bibnamefont
  {Cornelius}}, \bibinfo {author} {\bibfnamefont {Z.}~\bibnamefont {Xu}},
  \bibinfo {author} {\bibfnamefont {A.}~\bibnamefont {Saxena}}, \bibinfo
  {author} {\bibfnamefont {A.}~\bibnamefont {Chenu}},\ and\ \bibinfo {author}
  {\bibfnamefont {A.}~\bibnamefont {del Campo}},\ }\bibfield  {title} {\bibinfo
  {title} {{Spectral Filtering Induced by Non-Hermitian Evolution with Balanced
  Gain and Loss: Enhancing Quantum Chaos}},\ }\href
  {https://doi.org/10.1103/PhysRevLett.128.190402} {\bibfield  {journal}
  {\bibinfo  {journal} {Phys. Rev. Lett.}\ }\textbf {\bibinfo {volume} {128}},\
  \bibinfo {pages} {190402} (\bibinfo {year} {2022})}\BibitemShut {NoStop}%
\bibitem [{\citenamefont {Garc\'{\i}a-Garc\'{\i}a}\ \emph
  {et~al.}(2022{\natexlab{b}})\citenamefont {Garc\'{\i}a-Garc\'{\i}a},
  \citenamefont {S\'a},\ and\ \citenamefont
  {Verbaarschot}}]{GarciaGarcia-22PRX}%
  \BibitemOpen
  \bibfield  {author} {\bibinfo {author} {\bibfnamefont {A.~M.}\ \bibnamefont
  {Garc\'{\i}a-Garc\'{\i}a}}, \bibinfo {author} {\bibfnamefont
  {L.}~\bibnamefont {S\'a}},\ and\ \bibinfo {author} {\bibfnamefont {J.~J.~M.}\
  \bibnamefont {Verbaarschot}},\ }\bibfield  {title} {\bibinfo {title}
  {{Symmetry Classification and Universality in Non-Hermitian Many-Body Quantum
  Chaos by the Sachdev-Ye-Kitaev Model}},\ }\href
  {https://doi.org/10.1103/PhysRevX.12.021040} {\bibfield  {journal} {\bibinfo
  {journal} {Phys. Rev. X}\ }\textbf {\bibinfo {volume} {12}},\ \bibinfo
  {pages} {021040} (\bibinfo {year} {2022}{\natexlab{b}})}\BibitemShut
  {NoStop}%
\bibitem [{\citenamefont {Prasad}\ \emph {et~al.}(2022)\citenamefont {Prasad},
  \citenamefont {Yadalam}, \citenamefont {Aron},\ and\ \citenamefont
  {Kulkarni}}]{Prasad-22}%
  \BibitemOpen
  \bibfield  {author} {\bibinfo {author} {\bibfnamefont {M.}~\bibnamefont
  {Prasad}}, \bibinfo {author} {\bibfnamefont {H.~K.}\ \bibnamefont {Yadalam}},
  \bibinfo {author} {\bibfnamefont {C.}~\bibnamefont {Aron}},\ and\ \bibinfo
  {author} {\bibfnamefont {M.}~\bibnamefont {Kulkarni}},\ }\bibfield  {title}
  {\bibinfo {title} {{Dissipative quantum dynamics, phase transitions, and
  non-Hermitian random matrices}},\ }\href
  {https://doi.org/10.1103/PhysRevA.105.L050201} {\bibfield  {journal}
  {\bibinfo  {journal} {Phys. Rev. A}\ }\textbf {\bibinfo {volume} {105}},\
  \bibinfo {pages} {L050201} (\bibinfo {year} {2022})}\BibitemShut {NoStop}%
\bibitem [{\citenamefont {S\'a}\ \emph {et~al.}(2022)\citenamefont {S\'a},
  \citenamefont {Ribeiro},\ and\ \citenamefont {Prosen}}]{Sa-22-SYK}%
  \BibitemOpen
  \bibfield  {author} {\bibinfo {author} {\bibfnamefont {L.}~\bibnamefont
  {S\'a}}, \bibinfo {author} {\bibfnamefont {P.}~\bibnamefont {Ribeiro}},\ and\
  \bibinfo {author} {\bibfnamefont {T.}~\bibnamefont {Prosen}},\ }\bibfield
  {title} {\bibinfo {title} {{Lindbladian dissipation of strongly-correlated
  quantum matter}},\ }\href {https://doi.org/10.1103/PhysRevResearch.4.L022068}
  {\bibfield  {journal} {\bibinfo  {journal} {Phys. Rev. Research}\ }\textbf
  {\bibinfo {volume} {4}},\ \bibinfo {pages} {L022068} (\bibinfo {year}
  {2022})}\BibitemShut {NoStop}%
\bibitem [{\citenamefont {Kulkarni}\ \emph {et~al.}(2022)\citenamefont
  {Kulkarni}, \citenamefont {Numasawa},\ and\ \citenamefont
  {Ryu}}]{Kulkarni-22-SYK}%
  \BibitemOpen
  \bibfield  {author} {\bibinfo {author} {\bibfnamefont {A.}~\bibnamefont
  {Kulkarni}}, \bibinfo {author} {\bibfnamefont {T.}~\bibnamefont {Numasawa}},\
  and\ \bibinfo {author} {\bibfnamefont {S.}~\bibnamefont {Ryu}},\ }\bibfield
  {title} {\bibinfo {title} {{Lindbladian dynamics of the Sachdev-Ye-Kitaev
  model}},\ }\href {https://doi.org/10.1103/PhysRevB.106.075138} {\bibfield
  {journal} {\bibinfo  {journal} {Phys. Rev. B}\ }\textbf {\bibinfo {volume}
  {106}},\ \bibinfo {pages} {075138} (\bibinfo {year} {2022})}\BibitemShut
  {NoStop}%
\bibitem [{\citenamefont {Garc\'{\i}a-Garc\'{\i}a}\ \emph
  {et~al.}(2022{\natexlab{c}})\citenamefont {Garc\'{\i}a-Garc\'{\i}a},
  \citenamefont {Jia}, \citenamefont {Rosa},\ and\ \citenamefont
  {Verbaarschot}}]{GarciaGarcia-22PRD}%
  \BibitemOpen
  \bibfield  {author} {\bibinfo {author} {\bibfnamefont {A.~M.}\ \bibnamefont
  {Garc\'{\i}a-Garc\'{\i}a}}, \bibinfo {author} {\bibfnamefont
  {Y.}~\bibnamefont {Jia}}, \bibinfo {author} {\bibfnamefont {D.}~\bibnamefont
  {Rosa}},\ and\ \bibinfo {author} {\bibfnamefont {J.~J.~M.}\ \bibnamefont
  {Verbaarschot}},\ }\bibfield  {title} {\bibinfo {title} {{Replica symmetry
  breaking in random non-Hermitian systems}},\ }\href
  {https://doi.org/10.1103/PhysRevD.105.126027} {\bibfield  {journal} {\bibinfo
   {journal} {Phys. Rev. D}\ }\textbf {\bibinfo {volume} {105}},\ \bibinfo
  {pages} {126027} (\bibinfo {year} {2022}{\natexlab{c}})}\BibitemShut
  {NoStop}%
\bibitem [{\citenamefont {Cipolloni}\ and\ \citenamefont
  {Kudler-Flam}(2023)}]{GJ-23}%
  \BibitemOpen
  \bibfield  {author} {\bibinfo {author} {\bibfnamefont {G.}~\bibnamefont
  {Cipolloni}}\ and\ \bibinfo {author} {\bibfnamefont {J.}~\bibnamefont
  {Kudler-Flam}},\ }\bibfield  {title} {\bibinfo {title} {{Entanglement Entropy
  of Non-Hermitian Eigenstates and the Ginibre Ensemble}},\ }\href
  {https://doi.org/10.1103/PhysRevLett.130.010401} {\bibfield  {journal}
  {\bibinfo  {journal} {Phys. Rev. Lett.}\ }\textbf {\bibinfo {volume} {130}},\
  \bibinfo {pages} {010401} (\bibinfo {year} {2023})}\BibitemShut {NoStop}%
\bibitem [{\citenamefont {Xiao}\ \emph {et~al.}(2022)\citenamefont {Xiao},
  \citenamefont {Kawabata}, \citenamefont {Luo}, \citenamefont {Ohtsuki},\ and\
  \citenamefont {Shindou}}]{Xiao-22}%
  \BibitemOpen
  \bibfield  {author} {\bibinfo {author} {\bibfnamefont {Z.}~\bibnamefont
  {Xiao}}, \bibinfo {author} {\bibfnamefont {K.}~\bibnamefont {Kawabata}},
  \bibinfo {author} {\bibfnamefont {X.}~\bibnamefont {Luo}}, \bibinfo {author}
  {\bibfnamefont {T.}~\bibnamefont {Ohtsuki}},\ and\ \bibinfo {author}
  {\bibfnamefont {R.}~\bibnamefont {Shindou}},\ }\bibfield  {title} {\bibinfo
  {title} {{Level statistics of real eigenvalues in non-Hermitian systems}},\
  }\href {https://doi.org/10.1103/PhysRevResearch.4.043196} {\bibfield
  {journal} {\bibinfo  {journal} {Phys. Rev. Research}\ }\textbf {\bibinfo
  {volume} {4}},\ \bibinfo {pages} {043196} (\bibinfo {year}
  {2022})}\BibitemShut {NoStop}%
\bibitem [{\citenamefont {Shivam}\ \emph {et~al.}(2023)\citenamefont {Shivam},
  \citenamefont {De~Luca}, \citenamefont {Huse},\ and\ \citenamefont
  {Chan}}]{Shivam-22}%
  \BibitemOpen
  \bibfield  {author} {\bibinfo {author} {\bibfnamefont {S.}~\bibnamefont
  {Shivam}}, \bibinfo {author} {\bibfnamefont {A.}~\bibnamefont {De~Luca}},
  \bibinfo {author} {\bibfnamefont {D.~A.}\ \bibnamefont {Huse}},\ and\
  \bibinfo {author} {\bibfnamefont {A.}~\bibnamefont {Chan}},\ }\bibfield
  {title} {\bibinfo {title} {{Many-Body Quantum Chaos and Emergence of Ginibre
  Ensemble}},\ }\href {https://doi.org/10.1103/PhysRevLett.130.140403}
  {\bibfield  {journal} {\bibinfo  {journal} {Phys. Rev. Lett.}\ }\textbf
  {\bibinfo {volume} {130}},\ \bibinfo {pages} {140403} (\bibinfo {year}
  {2023})}\BibitemShut {NoStop}%
\bibitem [{\citenamefont {Ghosh}\ \emph {et~al.}(2022)\citenamefont {Ghosh},
  \citenamefont {Gupta},\ and\ \citenamefont {Kulkarni}}]{Ghosh-22}%
  \BibitemOpen
  \bibfield  {author} {\bibinfo {author} {\bibfnamefont {S.}~\bibnamefont
  {Ghosh}}, \bibinfo {author} {\bibfnamefont {S.}~\bibnamefont {Gupta}},\ and\
  \bibinfo {author} {\bibfnamefont {M.}~\bibnamefont {Kulkarni}},\ }\bibfield
  {title} {\bibinfo {title} {{Spectral properties of disordered interacting
  non-Hermitian systems}},\ }\href
  {https://doi.org/10.1103/PhysRevB.106.134202} {\bibfield  {journal} {\bibinfo
   {journal} {Phys. Rev. B}\ }\textbf {\bibinfo {volume} {106}},\ \bibinfo
  {pages} {134202} (\bibinfo {year} {2022})}\BibitemShut {NoStop}%
\bibitem [{\citenamefont {S\'a}\ \emph {et~al.}(2023)\citenamefont {S\'a},
  \citenamefont {Ribeiro},\ and\ \citenamefont {Prosen}}]{Sa-23}%
  \BibitemOpen
  \bibfield  {author} {\bibinfo {author} {\bibfnamefont {L.}~\bibnamefont
  {S\'a}}, \bibinfo {author} {\bibfnamefont {P.}~\bibnamefont {Ribeiro}},\ and\
  \bibinfo {author} {\bibfnamefont {T.}~\bibnamefont {Prosen}},\ }\bibfield
  {title} {\bibinfo {title} {{Symmetry Classification of Many-Body
  Lindbladians: Tenfold Way and Beyond}},\ }\href
  {https://doi.org/10.1103/PhysRevX.13.031019} {\bibfield  {journal} {\bibinfo
  {journal} {Phys. Rev. X}\ }\textbf {\bibinfo {volume} {13}},\ \bibinfo
  {pages} {031019} (\bibinfo {year} {2023})}\BibitemShut {NoStop}%
\bibitem [{\citenamefont {Kawabata}\ \emph
  {et~al.}(2023{\natexlab{a}})\citenamefont {Kawabata}, \citenamefont
  {Kulkarni}, \citenamefont {Li}, \citenamefont {Numasawa},\ and\ \citenamefont
  {Ryu}}]{Kawabata-23}%
  \BibitemOpen
  \bibfield  {author} {\bibinfo {author} {\bibfnamefont {K.}~\bibnamefont
  {Kawabata}}, \bibinfo {author} {\bibfnamefont {A.}~\bibnamefont {Kulkarni}},
  \bibinfo {author} {\bibfnamefont {J.}~\bibnamefont {Li}}, \bibinfo {author}
  {\bibfnamefont {T.}~\bibnamefont {Numasawa}},\ and\ \bibinfo {author}
  {\bibfnamefont {S.}~\bibnamefont {Ryu}},\ }\bibfield  {title} {\bibinfo
  {title} {{Symmetry of Open Quantum Systems: Classification of Dissipative
  Quantum Chaos}},\ }\href {https://doi.org/10.1103/PRXQuantum.4.030328}
  {\bibfield  {journal} {\bibinfo  {journal} {PRX Quantum}\ }\textbf {\bibinfo
  {volume} {4}},\ \bibinfo {pages} {030328} (\bibinfo {year}
  {2023}{\natexlab{a}})}\BibitemShut {NoStop}%
\bibitem [{\citenamefont {Cipolloni}\ and\ \citenamefont
  {Kudler-Flam}(2024)}]{GJ-24-ETH}%
  \BibitemOpen
  \bibfield  {author} {\bibinfo {author} {\bibfnamefont {G.}~\bibnamefont
  {Cipolloni}}\ and\ \bibinfo {author} {\bibfnamefont {J.}~\bibnamefont
  {Kudler-Flam}},\ }\bibfield  {title} {\bibinfo {title} {{Non-Hermitian
  Hamiltonians violate the eigenstate thermalization hypothesis}},\ }\href
  {https://doi.org/10.1103/PhysRevB.109.L020201} {\bibfield  {journal}
  {\bibinfo  {journal} {Phys. Rev. B}\ }\textbf {\bibinfo {volume} {109}},\
  \bibinfo {pages} {L020201} (\bibinfo {year} {2024})}\BibitemShut {NoStop}%
\bibitem [{\citenamefont {Kawabata}\ \emph
  {et~al.}(2023{\natexlab{b}})\citenamefont {Kawabata}, \citenamefont {Xiao},
  \citenamefont {Ohtsuki},\ and\ \citenamefont {Shindou}}]{Kawabata-23SVD}%
  \BibitemOpen
  \bibfield  {author} {\bibinfo {author} {\bibfnamefont {K.}~\bibnamefont
  {Kawabata}}, \bibinfo {author} {\bibfnamefont {Z.}~\bibnamefont {Xiao}},
  \bibinfo {author} {\bibfnamefont {T.}~\bibnamefont {Ohtsuki}},\ and\ \bibinfo
  {author} {\bibfnamefont {R.}~\bibnamefont {Shindou}},\ }\bibfield  {title}
  {\bibinfo {title} {{Singular-Value Statistics of Non-Hermitian Random
  Matrices and Open Quantum Systems}},\ }\href
  {https://doi.org/10.1103/PRXQuantum.4.040312} {\bibfield  {journal} {\bibinfo
   {journal} {PRX Quantum}\ }\textbf {\bibinfo {volume} {4}},\ \bibinfo {pages}
  {040312} (\bibinfo {year} {2023}{\natexlab{b}})}\BibitemShut {NoStop}%
\bibitem [{\citenamefont {Xiao}\ \emph {et~al.}(2024)\citenamefont {Xiao},
  \citenamefont {Shindou},\ and\ \citenamefont {Kawabata}}]{Xiao-24}%
  \BibitemOpen
  \bibfield  {author} {\bibinfo {author} {\bibfnamefont {Z.}~\bibnamefont
  {Xiao}}, \bibinfo {author} {\bibfnamefont {R.}~\bibnamefont {Shindou}},\ and\
  \bibinfo {author} {\bibfnamefont {K.}~\bibnamefont {Kawabata}},\ }\bibfield
  {title} {\bibinfo {title} {{Universal hard-edge statistics of non-Hermitian
  random matrices}},\ }\href {https://doi.org/10.1103/PhysRevResearch.6.023303}
  {\bibfield  {journal} {\bibinfo  {journal} {Phys. Rev. Research}\ }\textbf
  {\bibinfo {volume} {6}},\ \bibinfo {pages} {023303} (\bibinfo {year}
  {2024})}\BibitemShut {NoStop}%
\bibitem [{\citenamefont {Efetov}(1996)}]{Efetov-textbook}%
  \BibitemOpen
  \bibfield  {author} {\bibinfo {author} {\bibfnamefont {K.}~\bibnamefont
  {Efetov}},\ }\href {https://doi.org/10.1017/CBO9780511573057} {\emph
  {\bibinfo {title} {{Supersymmetry in Disorder and Chaos}}}}\ (\bibinfo
  {publisher} {Cambridge University Press},\ \bibinfo {address} {Cambridge,
  England},\ \bibinfo {year} {1996})\BibitemShut {NoStop}%
\bibitem [{\citenamefont {Nishigaki}\ and\ \citenamefont
  {Kamenev}(2002)}]{Nishigaki-02}%
  \BibitemOpen
  \bibfield  {author} {\bibinfo {author} {\bibfnamefont {S.~M.}\ \bibnamefont
  {Nishigaki}}\ and\ \bibinfo {author} {\bibfnamefont {A.}~\bibnamefont
  {Kamenev}},\ }\bibfield  {title} {\bibinfo {title} {{Replica treatment of
  non-Hermitian disordered Hamiltonians}},\ }\href
  {https://doi.org/10.1088/0305-4470/35/21/307} {\bibfield  {journal} {\bibinfo
   {journal} {J. Phys. A}\ }\textbf {\bibinfo {volume} {35}},\ \bibinfo {pages}
  {4571} (\bibinfo {year} {2002})}\BibitemShut {NoStop}%
\bibitem [{\citenamefont {Kamenev}\ and\ \citenamefont
  {Mézard}(1999{\natexlab{a}})}]{Kamenev_1999a}%
  \BibitemOpen
  \bibfield  {author} {\bibinfo {author} {\bibfnamefont {A.}~\bibnamefont
  {Kamenev}}\ and\ \bibinfo {author} {\bibfnamefont {M.}~\bibnamefont
  {Mézard}},\ }\bibfield  {title} {\bibinfo {title} {{Wigner-Dyson statistics
  from the replica method}},\ }\href
  {https://doi.org/10.1088/0305-4470/32/24/304} {\bibfield  {journal} {\bibinfo
   {journal} {J. Phys. A}\ }\textbf {\bibinfo {volume} {32}},\ \bibinfo {pages}
  {4373} (\bibinfo {year} {1999}{\natexlab{a}})}\BibitemShut {NoStop}%
\bibitem [{\citenamefont {Kamenev}\ and\ \citenamefont
  {Mézard}(1999{\natexlab{b}})}]{Kamenev_1999b}%
  \BibitemOpen
  \bibfield  {author} {\bibinfo {author} {\bibfnamefont {A.}~\bibnamefont
  {Kamenev}}\ and\ \bibinfo {author} {\bibfnamefont {M.}~\bibnamefont
  {Mézard}},\ }\bibfield  {title} {\bibinfo {title} {Level correlations in
  disordered metals: The replica $\sigma$ model},\ }\href
  {https://doi.org/10.1103/physrevb.60.3944} {\bibfield  {journal} {\bibinfo
  {journal} {Phys. Rev. B}\ }\textbf {\bibinfo {volume} {60}},\ \bibinfo
  {pages} {3944} (\bibinfo {year} {1999}{\natexlab{b}})}\BibitemShut {NoStop}%
\bibitem [{\citenamefont {Yurkevich}\ and\ \citenamefont
  {Lerner}(1999)}]{Yurkevich_Lerner_1999}%
  \BibitemOpen
  \bibfield  {author} {\bibinfo {author} {\bibfnamefont {I.~V.}\ \bibnamefont
  {Yurkevich}}\ and\ \bibinfo {author} {\bibfnamefont {I.~V.}\ \bibnamefont
  {Lerner}},\ }\bibfield  {title} {\bibinfo {title} {{Nonperturbative results
  for level correlations from the replica nonlinear $\ensuremath{\sigma}$
  model}},\ }\href {https://doi.org/10.1103/PhysRevB.60.3955} {\bibfield
  {journal} {\bibinfo  {journal} {Phys. Rev. B}\ }\textbf {\bibinfo {volume}
  {60}},\ \bibinfo {pages} {3955} (\bibinfo {year} {1999})}\BibitemShut
  {NoStop}%
\bibitem [{\citenamefont {Mehta}(2004)}]{mehta2004random}%
  \BibitemOpen
  \bibfield  {author} {\bibinfo {author} {\bibfnamefont {M.~L.}\ \bibnamefont
  {Mehta}},\ }\href@noop {} {\emph {\bibinfo {title} {Random Matrices}}}\
  (\bibinfo  {publisher} {Elsevier},\ \bibinfo {year} {2004})\BibitemShut
  {NoStop}%
\bibitem [{\citenamefont {Chen}\ \emph {et~al.}(2025)\citenamefont {Chen},
  \citenamefont {Kawabata}, \citenamefont {Kulkarni},\ and\ \citenamefont
  {Ryu}}]{CKKR-24}%
  \BibitemOpen
  \bibfield  {author} {\bibinfo {author} {\bibfnamefont {Z.}~\bibnamefont
  {Chen}}, \bibinfo {author} {\bibfnamefont {K.}~\bibnamefont {Kawabata}},
  \bibinfo {author} {\bibfnamefont {A.}~\bibnamefont {Kulkarni}},\ and\
  \bibinfo {author} {\bibfnamefont {S.}~\bibnamefont {Ryu}},\ }\bibfield
  {title} {\bibinfo {title} {{Field theory of non-Hermitian disordered
  systems}},\ }\href {https://doi.org/10.1103/PhysRevB.111.054203} {\bibfield
  {journal} {\bibinfo  {journal} {Phys. Rev. B}\ }\textbf {\bibinfo {volume}
  {111}},\ \bibinfo {pages} {054203} (\bibinfo {year} {2025})}\BibitemShut
  {NoStop}%
\bibitem [{\citenamefont {Nishigaki}(2016)}]{Nishigaki:2016random}%
  \BibitemOpen
  \bibfield  {author} {\bibinfo {author} {\bibfnamefont {S.}~\bibnamefont
  {Nishigaki}},\ }\href
  {https://www.saiensu.co.jp/search/?isbn=978-4-7819-9015-6&y=2024#detail}
  {\emph {\bibinfo {title} {{Random Matrices and Gauge Theory}}}}\ (\bibinfo
  {publisher} {Saiensu-Sha},\ \bibinfo {year} {2016})\BibitemShut {NoStop}%
\bibitem [{\citenamefont {Akemann}\ \emph {et~al.}()\citenamefont {Akemann},
  \citenamefont {Ayg\"un}, \citenamefont {Kieburg},\ and\ \citenamefont
  {P\"aßler}}]{Akemann-24}%
  \BibitemOpen
  \bibfield  {author} {\bibinfo {author} {\bibfnamefont {G.}~\bibnamefont
  {Akemann}}, \bibinfo {author} {\bibfnamefont {N.}~\bibnamefont {Ayg\"un}},
  \bibinfo {author} {\bibfnamefont {M.}~\bibnamefont {Kieburg}},\ and\ \bibinfo
  {author} {\bibfnamefont {P.}~\bibnamefont {P\"aßler}},\ }\bibfield  {title}
  {\bibinfo {title} {{Complex symmetric, self-dual, and Ginibre random
  matrices: Analytical results for three classes of bulk and edge
  statistics}},\ }\Eprint {https://arxiv.org/abs/2410.21032} {arXiv:2410.21032}
  \BibitemShut {NoStop}%
\bibitem [{\citenamefont {Forrester}()}]{Forrester-24}%
  \BibitemOpen
  \bibfield  {author} {\bibinfo {author} {\bibfnamefont {P.~J.}\ \bibnamefont
  {Forrester}},\ }\bibfield  {title} {\bibinfo {title} {{Dualities for
  characteristic polynomial averages of complex symmetric and self dual
  non-Hermitian random matrices}},\ }\Eprint {https://arxiv.org/abs/2411.07356}
  {arXiv:2411.07356} \BibitemShut {NoStop}%
\bibitem [{\citenamefont {Liu}\ and\ \citenamefont
  {Zhang}(2024)}]{Liu:2024xew}%
  \BibitemOpen
  \bibfield  {author} {\bibinfo {author} {\bibfnamefont {D.-Z.}\ \bibnamefont
  {Liu}}\ and\ \bibinfo {author} {\bibfnamefont {L.}~\bibnamefont {Zhang}},\
  }\bibfield  {title} {\bibinfo {title} {{Duality in non-Hermitian random
  matrix theory}},\ }\href {https://doi.org/10.1016/j.nuclphysb.2024.116559}
  {\bibfield  {journal} {\bibinfo  {journal} {Nucl. Phys. B}\ }\textbf
  {\bibinfo {volume} {1004}},\ \bibinfo {pages} {116559} (\bibinfo {year}
  {2024})}\BibitemShut {NoStop}%
\bibitem [{\citenamefont {Akemann}\ \emph {et~al.}(2022)\citenamefont
  {Akemann}, \citenamefont {Mielke},\ and\ \citenamefont
  {P\"a\ss{}ler}}]{Akemann_2022}%
  \BibitemOpen
  \bibfield  {author} {\bibinfo {author} {\bibfnamefont {G.}~\bibnamefont
  {Akemann}}, \bibinfo {author} {\bibfnamefont {A.}~\bibnamefont {Mielke}},\
  and\ \bibinfo {author} {\bibfnamefont {P.}~\bibnamefont {P\"a\ss{}ler}},\
  }\bibfield  {title} {\bibinfo {title} {{Spacing distribution in the
  two-dimensional Coulomb gas: Surmise and symmetry classes of non-Hermitian
  random matrices at noninteger $\ensuremath{\beta}$}},\ }\href
  {https://doi.org/10.1103/PhysRevE.106.014146} {\bibfield  {journal} {\bibinfo
   {journal} {Phys. Rev. E}\ }\textbf {\bibinfo {volume} {106}},\ \bibinfo
  {pages} {014146} (\bibinfo {year} {2022})}\BibitemShut {NoStop}%
\bibitem [{\citenamefont {Verbaarschot}\ and\ \citenamefont
  {Zirnbauer}(1985)}]{Verbaarschot:1985qx}%
  \BibitemOpen
  \bibfield  {author} {\bibinfo {author} {\bibfnamefont {J.~J.~M.}\
  \bibnamefont {Verbaarschot}}\ and\ \bibinfo {author} {\bibfnamefont {M.~R.}\
  \bibnamefont {Zirnbauer}},\ }\bibfield  {title} {\bibinfo {title} {{Critique
  of the replica trick}},\ }\href {https://doi.org/10.1088/0305-4470/18/7/018}
  {\bibfield  {journal} {\bibinfo  {journal} {J. Phys. A}\ }\textbf {\bibinfo
  {volume} {18}},\ \bibinfo {pages} {1093} (\bibinfo {year}
  {1985})}\BibitemShut {NoStop}%
\bibitem [{\citenamefont {Zirnbauer}()}]{zirnbauer1999critiquereplicatrick}%
  \BibitemOpen
  \bibfield  {author} {\bibinfo {author} {\bibfnamefont {M.~R.}\ \bibnamefont
  {Zirnbauer}},\ }\bibfield  {title} {\bibinfo {title} {{Another critique of
  the replica trick}},\ }\Eprint {https://arxiv.org/abs/cond-mat/9903338}
  {arXiv:cond-mat/9903338} \BibitemShut {NoStop}%
\bibitem [{\citenamefont {Kanzieper}(2002)}]{PhysRevLett.89.250201}%
  \BibitemOpen
  \bibfield  {author} {\bibinfo {author} {\bibfnamefont {E.}~\bibnamefont
  {Kanzieper}},\ }\bibfield  {title} {\bibinfo {title} {{Replica Field
  Theories, Painlev\'e Transcendents, and Exact Correlation Functions}},\
  }\href {https://doi.org/10.1103/PhysRevLett.89.250201} {\bibfield  {journal}
  {\bibinfo  {journal} {Phys. Rev. Lett.}\ }\textbf {\bibinfo {volume} {89}},\
  \bibinfo {pages} {250201} (\bibinfo {year} {2002})}\BibitemShut {NoStop}%
\bibitem [{\citenamefont {Splittorff}\ and\ \citenamefont
  {Verbaarschot}(2003)}]{Splittorff:2002eb}%
  \BibitemOpen
  \bibfield  {author} {\bibinfo {author} {\bibfnamefont {K.}~\bibnamefont
  {Splittorff}}\ and\ \bibinfo {author} {\bibfnamefont {J.~J.~M.}\ \bibnamefont
  {Verbaarschot}},\ }\bibfield  {title} {\bibinfo {title} {{Replica Limit of
  the Toda Lattice Equation}},\ }\href
  {https://doi.org/10.1103/PhysRevLett.90.041601} {\bibfield  {journal}
  {\bibinfo  {journal} {Phys. Rev. Lett.}\ }\textbf {\bibinfo {volume} {90}},\
  \bibinfo {pages} {041601} (\bibinfo {year} {2003})}\BibitemShut {NoStop}%
\bibitem [{\citenamefont
  {Kanzieper}()}]{kanzieper2005exactreplicatreatmentnonhermitean}%
  \BibitemOpen
  \bibfield  {author} {\bibinfo {author} {\bibfnamefont {E.}~\bibnamefont
  {Kanzieper}},\ }\bibfield  {title} {\bibinfo {title} {{Exact replica
  treatment of non-Hermitean complex random matrices}},\ }\Eprint
  {https://arxiv.org/abs/cond-mat/0312006} {arXiv:cond-mat/0312006}
  \BibitemShut {NoStop}%
\bibitem [{\citenamefont {Edelman}\ and\ \citenamefont
  {Jeong}(2023)}]{Edelman-23}%
  \BibitemOpen
  \bibfield  {author} {\bibinfo {author} {\bibfnamefont {A.}~\bibnamefont
  {Edelman}}\ and\ \bibinfo {author} {\bibfnamefont {S.}~\bibnamefont
  {Jeong}},\ }\bibfield  {title} {\bibinfo {title} {{Fifty Three Matrix
  Factorizations: A Systematic Approach}},\ }\href
  {https://doi.org/10.1137/21m1416035} {\bibfield  {journal} {\bibinfo
  {journal} {SIAM J. Matrix Anal. Appl.}\ }\textbf {\bibinfo {volume} {44}},\
  \bibinfo {pages} {415} (\bibinfo {year} {2023})}\BibitemShut {NoStop}%
\bibitem [{\citenamefont {Liu}\ \emph {et~al.}(2023)\citenamefont {Liu},
  \citenamefont {Kudler-Flam},\ and\ \citenamefont
  {Kawabata}}]{Liu_Kudler-Flam_Kawabata_2023}%
  \BibitemOpen
  \bibfield  {author} {\bibinfo {author} {\bibfnamefont {Y.}~\bibnamefont
  {Liu}}, \bibinfo {author} {\bibfnamefont {J.}~\bibnamefont {Kudler-Flam}},\
  and\ \bibinfo {author} {\bibfnamefont {K.}~\bibnamefont {Kawabata}},\
  }\bibfield  {title} {\bibinfo {title} {{Symmetry classification of typical
  quantum entanglement}},\ }\href {https://doi.org/10.1103/PhysRevB.108.085109}
  {\bibfield  {journal} {\bibinfo  {journal} {Phys. Rev. B}\ }\textbf {\bibinfo
  {volume} {108}},\ \bibinfo {pages} {085109} (\bibinfo {year}
  {2023})}\BibitemShut {NoStop}%
\bibitem [{Kul()}]{Kulkarni-SCGP}%
  \BibitemOpen
  \href@noop {} {}\bibinfo {note} {{A. K. Kulkarni,
  \href{https://scgp.stonybrook.edu/video_portal/video.php?id=6735}{{\it
  Internal symmetry and dynamical phenomena in open quantum systems}}}
  (Non-Hermitian topology, geometry and symmetry across physical platforms,
  October 7th, 2024).}\BibitemShut {Stop}%
\end{thebibliography}%

\end{document}